\documentclass[aps,twocolumn,nofootinbib]{revtex4-2}

\usepackage{mathtools}
\usepackage{multirow}
\usepackage{booktabs}
\usepackage{adjustbox} 
\usepackage{amssymb}
\usepackage{tikz}

\usepackage{array}
\usepackage{makecell}  
\usepackage{graphicx}
\usepackage{dcolumn}
\usepackage{placeins}
\usepackage{bm}
\usepackage{graphicx}    
\usepackage{subcaption}  
\usepackage{enumitem}   
\usepackage{float}
\usepackage{placeins}
\usepackage{titlesec}
\titlespacing*{\subsubsection}{0pt}{0.6\baselineskip}{0.3\baselineskip}
\usepackage{titlesec}
\titlespacing*{\subsection}
  {0pt}   
  {0.8em} 
  {0.3em} 
\usepackage{hyperref}
\usepackage{wrapfig}
\hypersetup{
    colorlinks,
    citecolor=blue,
    filecolor=blue,
    linkcolor=blue,
    urlcolor=blue,
}
\usepackage{ragged2e}
\renewcommand{\eqref}[1]{Eq.~({\ref{#1}})}

\begin{document}

\author{Kaitlin Gili$^{1}$, Mainak Nistala, Kristen Wendell$^{1}$, and Michael C. Hughes$^{1,\Diamond}$}
\affiliation{$^1$Department of Computer Science, Tufts University, Medford, MA, U.S.A.}
\affiliation{$^2$Departments of Mechanical Engineering and Education, Tufts University, Medford, MA, U.S.A.}

\date{\today}

\title{Locating acts of mechanistic reasoning in student team conversations \\ with mechanistic machine learning}

\begin{abstract}
STEM education researchers are often interested in identifying moments of students' mechanistic reasoning for deeper analysis, but have limited capacity to search through many team conversation transcripts to find segments with a high concentration of such reasoning. We offer a solution in the form of an interpretable machine learning model that outputs time-varying probabilities that individual students are engaging in acts of mechanistic reasoning, leveraging evidence from their own utterances as well as contributions from the rest of the group. Using the toolkit of intentionally-designed probabilistic models, we introduce a specific inductive bias that steers the probabilistic dynamics toward desired, domain-aligned behavior. Experiments compare trained models with and without the inductive bias components, investigating whether their presence improves the desired model behavior on transcripts involving never-before-seen students and a novel discussion context. Our results show that the inductive bias improves generalization -- supporting the claim that interpretability is built into the model for this task rather than imposed post hoc. We conclude with practical recommendations for STEM education researchers seeking to adopt the tool and for ML researchers aiming to extend the model’s design. Overall, we hope this work encourages the development of mechanistically interpretable models that are understandable and controllable for both end users and model designers in STEM education research.
\end{abstract}

\maketitle

\section{Introduction}

We are interested in analyzing students' mechanistic reasoning (MR) in-moment, during collaborative problem-solving conversations. Such conversations are a key activity in the professional practice of science and engineering \cite{vinck2003everyday} and therefore a crucial element of science and engineering education \cite{Dickerson2024CarlaGrumpy,ScherrHammer2009Framing, KoretskyEtAl2023Connected}. To solve technical problems, science and engineering students often need to determine the relationships between different parameters of a natural or designed system. Mechanistic reasoning supports this work because it is a form of thinking focused on characterizing the entities and activities that cause a particular system performance \cite{machamer, vanEck2018Mechanisms}. But to solve difficult technical problems collaboratively, as professional scientists and engineers do, students need not only to engage in mechanistic reasoning as individual thinkers; they must also express and revise their mechanistic reasoning to fellow learners in-the-moment, during conversations. 

The STEM education literature has established that students -- even as young as first graders -- can express mechanistic reasoning during teacher-facilitated and small-group conversations \cite{deAndradeEtAl2022, mech_reason}, but these studies typically analyze at most a handful of conversations due to time constraints. 
In research settings where a large number of student conversations are recorded, it is often resource-intensive to manually identify the segments of text that contain the most evidence of mechanistic reasoning at the individual student level and the collective group level to analyze. Prior work has observed that evidence of mechanistic reasoning is often \emph{dynamic} and \emph{unstable} in student conversations. Students can shift in and out of a mechanistic reasoning frame on the order of minutes \cite{PhysRevPhysEducRes.14.020122}. Thus, having an automated tool that can pinpoint these locations in time-series data that span hours can save STEM education researchers a tremendous amount of time and mental energy in choosing which data segments to prioritize.    

Given the nature of the data task, it is natural to wonder whether advancements in machine learning (ML) can offer a solution. A first consideration might be to offload the task to the context-window of a large-language model (LLM) \cite{ChatGPT_Ed}, which can take text as input and can be prompt-engineered by the user to output references to the most important text segments. While simple sounding, such an approach requires back-and-forth wrestling with a probabilistic token generator \cite{prompt2, prompt3} that lacks inherent interpretability \cite{Rudin2019StopExplaining,Zschech2025Inherently}. Different from a black-box model with a post-hoc explanation for its prediction \cite{Rudin2019StopExplaining}, an inherently interpretable ML model is ``constrained in model form so that it is either useful to someone, or obeys structural knowledge of the domain" \cite{Rudin2019StopExplaining}. In the latter case, predictions arise from inductive biases built in by design. Domain-aligned mathematical mechanisms are explicitly encoded to constrain and guide the model toward the intended behavior. Both model designers and users stand to benefit from this approach. Designers are able to have a clear understanding of the inductive biases built into their implementation, can obtain evidence of their value, and then can remove or refine them as necessary for model improvement. Users, in turn, can place greater confidence in a tool built around domain expertise and tailored to their specific task. Importantly, both designers and users are able to obtain a mechanistic account of the tool’s behavior, enabling more informed decisions about appropriate use.

\begin{figure}[t]
  \includegraphics[width=1\linewidth]{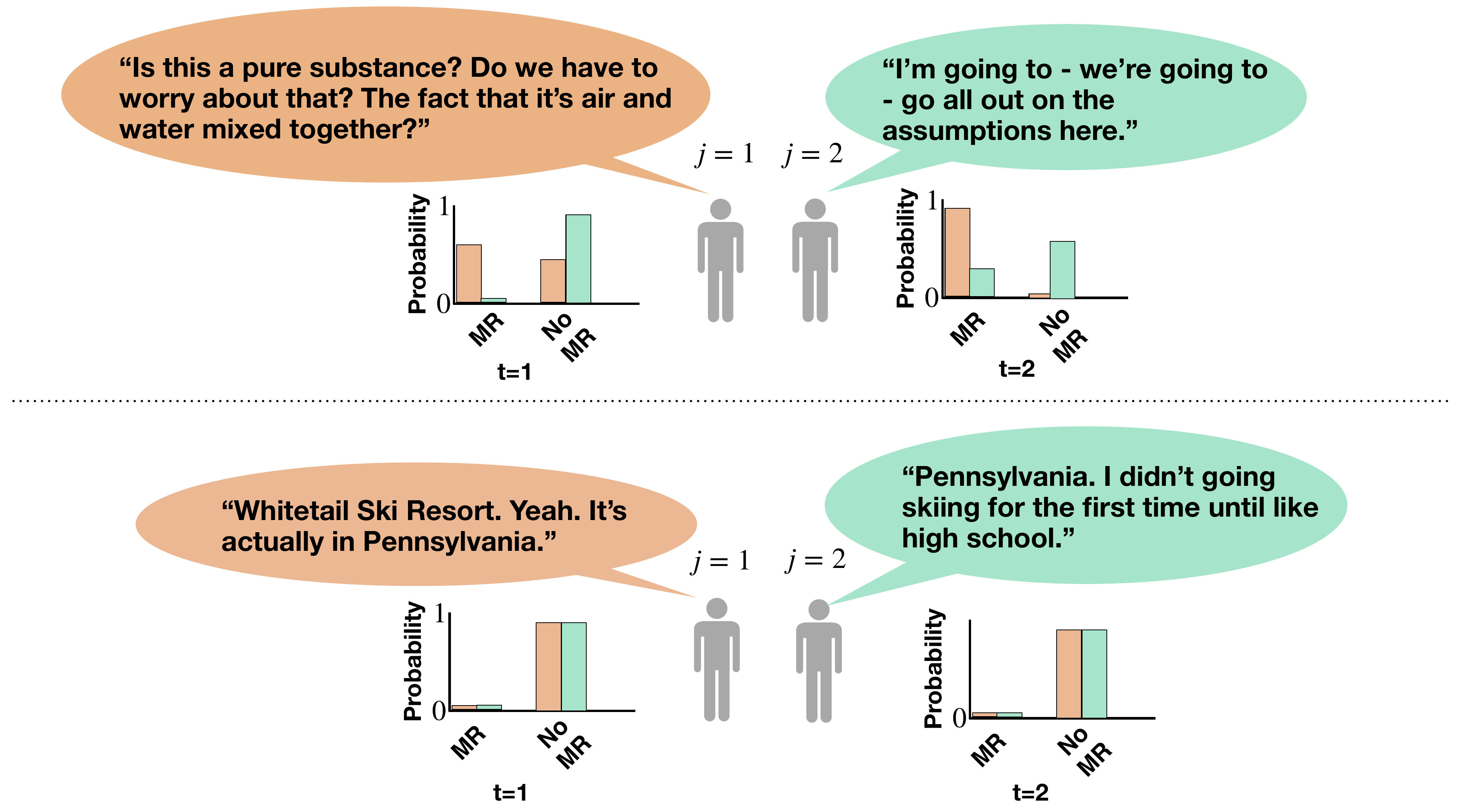}
  \caption{\justifying A toy illustration of our desired ML model behavior, using student utterances from the snowmaking problem. Top: When evidence of mechanistic reasoning (MR) is present at $t=1$, the model sharply increases that student’s latent probability of MR at time $t=2$ and modestly increases the other student’s probability. Bottom: When no MR is observed at $t=1$, the model keeps both students’ latent probabilities low at $t=2$. Comparison: When a student’s observed speech contains mechanistic reasoning evidence at $t=1$, the latent probability of MR is closer to one (top left) than when no evidence is present (bottom left).}
  \label{fig1}
\end{figure}

In this work, we explore an inherently interpretable ML design by leveraging \emph{switching-state} models, which represent the latent state dynamics and their probabilities that drive entity behavior over time \cite{ghahramaniSwitchingState2000, dsarf,wojnowicz2024discovering}. Recent work~\cite{wojnowicz2024discovering} introduces an efficiently trainable ML model containing interpretable mechanisms for how entity observations (e.g~student utterances) influence the probabilities of latent state changes (e.g~evidence of mechanistic reasoning or not) for individual entities as well as the entire group. See Fig.~\ref{fig1} for an intuitive toy illustration of this desired probabilistic model behavior in our student discussion settings. The model in Ref.~\cite{wojnowicz2024discovering} offers a useful starting point for an interpretable ML approach to our data task and motivates the following research questions:

\begin{enumerate}
    \item \emph{RQ1: How might the mechanistic ML model introduced in Ref.~\cite{wojnowicz2024discovering} be adapted with a specialized inductive bias to support the identification of mechanistic reasoning in group discussions of STEM learners?}
    
     \item \emph{RQ2: Does the specialized inductive bias improve the desired ML model behavior when generalizing to discussions involving a novel STEM problem and previously unseen students? And if so, by how much?}
\end{enumerate}

The remainder of this article is organized as follows. Sec.~\ref{Background} provides background with respect to prior work on mechanistic reasoning and  presents examples of mechanistic reasoning evidence from our dataset, which we use to train and test our ML model. We also provide a brief overview of prior work incorporating ML tools to accelerate STEM education research. 

Sec.~\ref{RQ1} describes the model from Ref.~\cite{wojnowicz2024discovering} and how we adapt it to identify regions of high mechanistic reasoning evidence in transcripts of student team conversations. We present our ML modeling decisions in technical detail. First, we outline our probabilistic modeling assumptions and discuss their benefits and limitations for our task. Second, we describe the specialized inductive bias that we bake into the model as this work's contribution.
Namely, we design a feedback mechanism that steers the individual and group latent state probabilities based on the evidence of mechanistic reasoning present in a student utterance. 
Our classifier-based feedback is trained with human annotation supervision, but can generalize to new unannotated transcripts.

In Sec.~\ref{RQ2}, we introduce task-specific metrics for desired model behavior and our hypotheses for these metrics when the feedback mechanism components are present or absent from the model construction. We then examine experimental results for our hypotheses in two data settings.
First, we evaluate predictions on data from previously unseen students discussing a STEM problemthat appears the training data. Second, we evaluate predictions on transcripts that include both unseen students and a new STEM problem. We interpret positive empirical support for each hypothesis as evidence that the model is interpretable by design for the target task. 

In Sec.~\ref{demo_recs}, we present a demonstration of the tool’s output on a transcript segment and provide recommendations for STEM education researchers using our tool. In Sec.~\ref{future_work}, we conclude with an outline of future research directions for ML-experts looking to design tools with inherent interpretability. 

\section{Background}\label{Background}

\subsection{Analyzing students' mechanistic reasoning in conversational data} 
\hspace*{1em} Drawing from depictions of mechanistic reasoning in philosophy of science \cite{machamer}, Russ and colleagues developed a framework for characterizing its presence in classroom science conversations \cite{mech_reason}. It puts forward seven hierarchical elements of mechanistic reasoning that may be evident in student talk. Those categories are (1) describe the \textit{target phenomenon}, (2) identify the \textit{set-up conditions} for the phenomenon, (3) identify the \textit{entities} that play a role in producing the phenomenon, (4-6) identify the \textit{properties, activities}, and \textit{organization} of those entities that affect the outcome of the phenomenon, and (7) \textit{chain} the current state of the entities backward to what happened previously or forward to what will happen next. By describing these categories as hierarchical, we mean both that higher categories represent more sophisticated reasoning and that higher categories generally require lower categories to have already been accomplished. Russ et al.~also proposes two categories of mechanistic reasoning not included in the hierarchy. Reasoning with \textit{analogies} involves making explicit comparisons to other phenomena. Reasoning with \textit{animated models} involves simulating entities and activities through physical gestures or manipulation of objects. We omitted these two categories from our analysis, but they present interesting possibilities for future work. 

To see how the framework's seven hierarchical categories build on each other, consider the following example from Russ and colleagues' original study. First-grade students had observed that a crumpled sheet of paper and a heavy book reached the ground at the same time when dropped from the same height. This result surprised students because when the same sheet of paper had been smooth and flat, it had lagged far behind the falling book. Their teacher gathered the students and asked them to share their thinking about why the crumpled paper and book ``tied". One student pointed out, ``If it’s balled up it’s still not heavy it’s the same size.” Here, her mechanistic reasoning included identifying the paper as a key entity (category \#3), identifying the paper's weight as a key property (category \#4), and identifying the process of crumpling as a key activity (category \#5). She also chained the activity to the property (category \#7) by noting that while the crumpling changed the paper’s shape, it maintained its heaviness. It is important to note that Russ et al's analytic framework does not consider the accuracy of students' claims or the completeness of their explanations. Conversations may exhibit strong evidence of mechanistic reasoning without expressing the ``correct" or ``complete" canonically accepted mechanistic account of a phenomenon.

Since its development, science and engineering education researchers  have used Russ et al.'s framework to characterize the use of mechanistic reasoning by groups of learners. These studies show how mechanistic reasoning in-the-moment contributes to group progress in modeling or designing physical systems \cite{Bachtiar, Wilkerson}. For example, engineering students who engaged in higher levels of mechanistic reasoning as they discussed their physical prototypes tended to make iterations that improved the prototype's performance \cite{Wendell2025MechanisticReasoning}.

\subsection{Evidence of mechanistic reasoning in our dataset} 

\hspace*{1em} The data used in this work was collected by audio recording five homework problem solving sessions carried out by small groups of undergraduate engineering students in a thermal fluid systems course at Tufts university. Each student group was assigned two open-ended problems that involved analyzing ill-defined thermal-fluid systems related to real-world situations, such as making snow for a ski resort or keeping ice cream cold at a summer ice cream stand \cite{Melsky2024}. All groups received one problem that was the same about constructing a resistance heater to raise the temperature of the surrounding gas, and one problem designed around a scenario intended to interest at least one student in the group. For example, the group that received the ski resort snowmaking problem contained a student who was an avid skier. 

We take the snowmaking problem as a central example to show what evidence of mechanistic reasoning looks like in our dataset. The problem statement provided to students is written as: 

\begin{center}
\fbox{\parbox{\linewidth}{
\centering \textbf{Snowmaking problem} \\
\justifying
Choose your favorite ski resort and the desired depth of snow for the best skiing, and use thermodynamics to determine how long it will take to cover the ski trails in that amount of snow. You may assume that one snow gun uses about 100 gallons of water per minute and that the compressor can produce 50 cubic feet per minute  of air. }}\end{center}

Using the seven categories from Russ et al.'s framework, we determined what counts as evidence of each element of mechanistic reasoning within the specific problem space. Here we present Russ' codes applied to the snowmaking problem with examples from our dataset that our coded positively:  \\

\noindent 1.~\textbf{Describe the target phenomenon:}~express the phenomenon that is trying to be explained (e.g.~``covering the ski trail in snow").  

\noindent \emph{Example:} ``Assume that's the surface area [of the resort] that we need to cover [with snow].” \\

\noindent 2.~\textbf{Identify setup conditions:} recognize the initial organization of the environment that is required for the mechanism to run (e.g.~``setting up snowguns around the trail"). 

\noindent \emph{Example:} ``Let's say they have a snow gun - they have like ten on a trail.” \\

\noindent 3.~\textbf{Identify entities:} recognize the things in the environment that interact in cause-effect ways to produce the phenomenon (e.g.~``air", ``water", ``snow", ``ice").   

\noindent \emph{Example:} ``The fact that it's air and water mixed together?” \\

\noindent 4.~\textbf{Identify activities of entities:} recognize the individual entity actions and mutual entity interactions in the mechanism that produce the phenomena (e.g.~``mixing together", ``coming out of the nozzle", ``turning into ice"). Actions can include motions; interactions can include forces. 

\noindent \emph{Example:} ``Alright so we have a volumetric flow rate of water and air, which is gonna be.” \\

\noindent 5.~\textbf{Identify properties of entities:} recognize the entity characteristics or current states that are relevant for the mechanism to run.  (e.g.~``density”, ``compressed”, ``temperature”, ``sparse”).   

\noindent \emph{Example:} ``So we could assume that it's at zero degrees Celsius like just about to freeze, but it is entirely [liquid] water.” \\

\noindent 6.~\textbf{Identify organization of entities:}~recognize the spatial organization of entities in the mechanism (e.g.~``inside of the tube", ``different nozzles").  

\noindent \emph{Example:} ``The fact that it's air and water mixed together?” \\

\noindent 7.~\textbf{Chain backward and forward:} use knowledge about the causal structure of the world to make claims about what must have happened previously to bring about the current state of things (backward) or what will happen next given that certain entities or activities are present now (forward) (e.g.~``expanding air and expanding water are turning into ice").  

\noindent \emph{Example:} ``Outside of the tube, there's expanding air and expanding water which is turning to ice.” \\

The codes for all six problems in our dataset can be found in Appendix \ref{problems}. Our human analysis procedure for code creation and validation, including an inter-rater reliability (IRR) test to show evidence of reproducibility, can be found in Appendix \ref{human_annot}. In Sec.~\ref{RQ1}, we present our ML method for automating mechanistic reasoning detection that is trained and tested on $10$ total transcripts containing student discussion of these problems with human-annotated labels. 

\subsection{Using ML tools to accelerate STEM education research}

\hspace*{1em} There has been a rapid development of ML tools for the purpose of helping STEM education  researchers speed up their data analysis procedures. Given that the majority of data studied in STEM education research is qualitative -- such as written responses and transcripts from digital recordings -- improvements in large language models (LLMs) \cite{BERT, touvron_llama_2023, jiang2023mistral7b} have enabled the large-scale automation of data annotation \cite{PhysRevPhysEducRes.20.010116, PhysRevPhysEducRes.18.010141, 2023.EDM-posters.59, PhysRevPhysEducRes.21.010128, gili2025, ChatGPT_Ed, automated_reflections, harpreet_ml}. Automated methods for data labeling include both in-context learning (ICL) and transfer learning with the possibility of foundation model supervised fine-tuning. In ICL, an engineered prompt for the task along with a few human labeled examples are fed directly into a chatbot window (e.g. ChatGPT) and the model then generates labels for the rest of the user provided data. The chatbot window is LLM-powered, but does not require any user interaction with the pre-trained LLM components, making it accessible to non-ML experts. Under the assumption that off-the-shelf LLMs can be guided with few data to perform well on a general set of language tasks, ICL is extremely data efficient. However, it also happens to be sensitive to the provided prompt \cite{prompting, jiangHowCanWe2020} and the selection and order of data shown \cite{liuWhatMakesGood2022}. Computing reliable statistics across models with different initializations (i.e.~different prompts) is a challenge given that even the same exact input prompt may lead to different outputs due to the probabilistic nature of chatbots \cite{stochastic_parrots}. Lastly, ICL methods lack inherent interpretability for their predictions. 

Rather than feed text directly into a context-window, one can consider a transfer learning approach. In transfer learning, one leverages an LLM as an encoder that transforms text into numerical vectors (i.e.~text embeddings) that can be used to train a downstream classification model for a particular task. Under the assumption that off-the-shelf text-embeddings are representationally meaningful, this approach removes the need to re-train the foundation LLM model and reduces training costs only to the downstream model. To drop the assumption, one can fine-tune the pre-trained LLM (i.e.~update the foundation model weights) to produce embeddings that are more aligned with the downstream classification task. In recent work, Gili et al.~\cite{gili2025} show that for the downstream task of identifying students’ mechanistic reasoning in written explanations, fine-tuning yields only slightly better performance than using off-the-shelf LLM embeddings, while requiring nearly $23\times$ more training time. Overall, transfer learning offers more inherent interpretability than in-context learning (ICL) because it produces text embeddings that can be directly analyzed and allows researchers to select an interpretable downstream classifier (e.g.~a linear model).

The majority of ICL and transfer learning methods explored for data annotation purposes in STEM education has been focused around annotating static data -- isolated student responses that are assumed to be time independent. For example, Fussell et al.~\cite{PhysRevPhysEducRes.21.010128} provide an evaluation of different LLMs for identifying specific skills in students’ typed lab notes. In Ref.~\cite{org_chem_ml}, Watts and her co-authors separate Russ' original categories for identifying mechanistic reasoning into binary (present or not present) codes specific to organic chemistry problems. They then compare the performance differences between neural network model variants that are trained to classify each code on independent student explanations. In Ref.~\cite{gili2025}, the authors also adapt Russ' original categories into binary codes, but specific to classical mechanics problems and with a PER-ML co-design approach that takes the ML model decisions into account when refining the qualitative coding framework. The authors train probabilistic classifiers with different LLM encoders, and evaluate the trade-off between performance and computational resource use. Our work is similar in that we initially adapt Russ’s original categories into binary codes, but we do so for time-varying conversational data from a thermal-fluid systems course and use the resulting signals to estimate the probability of mechanistic reasoning, rather than producing a binary yes/no label. 

The latter point separates our work from that of Ref.~\cite{moreau_pernet2024_talkmoves}, which uses LLMs to classify talk moves in transcripts of tutoring sessions. They report classification results from two types of prompt-engineering ICL methods with GPT-3.5-turbo and transfer learning without fine-tuning. In two of the scenarios, they treat each student utterance in the transcript as independent, despite the underlying sequential nature of the data. In the third scenario, they prompt GPT-3.5-turbo to reproduce the entire transcript with annotations line-by-line, allowing for the context of all surrounding lines to influence the annotations. Our work focuses on constructing an ML model that is inherently interpretable so that the user can understand the context mechanism -- how evidence present in a single student utterance influences their state and that of all others. This enables users to search for locations of high mechanistic evidence at the level of an individual student or the entire group, using probabilities calibrated to the evidence present rather than a binary label. 

\vspace{-2mm}

\section{RQ1: Specializing ML Inductive Biases to Locate Mechanistic Reasoning}\label{RQ1}

In this section, we describe the design of an inherently interpretable ML model for the task of locating evidence of mechanistic reasoning in student conversations. Our model is a task-specific adaptation of the hierarchical switching-state recurrent dynamical model (HSRDM), a previously proposed~\citep{wojnowicz2024discovering} general framework for modeling interacting entities over time.

We specialize this model in two ways: (1) to model text transcripts of conversations between students, and (2) so latent states at both team-level and student-level directly correspond to notations of mechanistic reasoning.
The subsections below describe the core modeling framework as well as specialized components that provide desired inductive bias for our goals. More detail, especially about model training, can be found in Appendix \ref{ml_method_details} for the interested reader. 

\subsection{Data representation}
\label{sec:data_repr}

We observe $N$ sequences of observed data from $J$ interacting entities.
In our setting, each sequence, indexed by $i \in \{1, ..., N\}$, represents a separate conversation between students on a problem-solving team. Each of the $J$ students, indexed by $j \in \{1,.., J\}$, is an ``entity'' \footnote{Note that the usage of the term ``entity'' here is distinct from the usage of the term in Sec.~\ref{Background} when discussing the framework for mechanistic reasoning evidence.}.
In our data, some of the $N$ conversations involve different unique students.

We represent each turn in the conversation as a discrete timestep, indexed $t \in \{0,.., T\}$.
The data sequence $x^{i, j}_{0:T} = [x^{i,j}_0, x^{i,j}_1, \ldots, x^{i,j}_T]$ denotes the text observed from the $j$-th student in conversation $i$ over time. The turn-by-turn nature of conversations means that each $x^{i,j}_t$ represents either a contiguous utterance of speech if $j$ is talking, or complete silence if $j$ does not talk at turn $t$. The length of each conversation may vary, but to keep notation simple we denote length $T$ as constant across $i$.

Each actual utterance from the audio recording of the conversation is transcribed to written text by human researchers. We  set $x^{i,j}_{t}$ to the $D$-dim. vector embedding of that text via the recent open-weight EmbeddingGemma encoder-only model \cite{embedding_gemma}. The EmbeddingGemma encoder is selected due to its state-of-the-art performance\footnote{EmbeddingGemma was the highest ranked on the on the Massive Multilingual Text Embedding Benchmark (MMTEB) at the time of conducting this work.}, while using relatively few parameters (308M parameters) compared to those of decoder-only models (e.g.~Mistral-7B \cite{jiang2023mistral7b}: 7.3 billion parameters). A primary benefit of the EmbeddingGemma encoder is that it allows a relatively compact embeddings size due to Matryoshka representation learning~\citep{kusupati2022matryoshka}. We use $D{=}128$-dimensional embeddings. This reduced dimensionality helps keep training runtime efficient. If turn $t$ lacks an utterance for speaker $j$, we set the observed vector $x^{i,j}_t$ to the vector embedding of the word ``silence" via EmbeddingGemma. At each turn $t$, there is \emph{always} just one speaker with a non-silence embedding, whereas the rest of the group has this silence embedding.

\subsection{Probabilistic model}
\label{probabilistic_assumptions}

As done in \citep{wojnowicz2024discovering}, we model each sequence $i$ as an i.i.d.~draw from a switching-state space model with two levels of latent variables. First, a shared ``team-level'' discrete state sequence $s^{i}_{0:T}$ represents the time-evolving state of reasoning for the entire group of $J$ students. We assume $L$ possible team-level states.
Second, an ``entity-level'' discrete state sequence $z_{0:T}^{i,j}$ describes student-specific state over time. We assume $K$ possible states at each entity, indexed $k \in \{1, 2, \ldots K\}$.

Our model defines a joint distribution over all random variables: $p(s^{1:N}_{0:T}, z^{1:N,1:J}_{0:T}, x^{1:N,1:J}_{0:T}, \theta)$. Throughout, we denote essential model parameters shared by all sequences as $\theta$. We assume this joint distribution factorizes as a product over i.i.d.~sequences. Each sequence $i$'s variables are generated by moving forward in time.

First, for the very first timestep $t{=}0$, we draw \emph{initial} values:
\begin{align}
    &s^i_0 \sim \text{Cat}_L( \theta_{s-ini} )
    \label{eq:gen_model_for_t_eq_0}
    \\ \notag
    &\text{for~$j$~in~} 1, 2, \ldots J:
    \\ \notag 
    &\qquad z^{i,j}_0 \sim \text{Cat}_K( \theta_{z-ini} )
    \\ \notag 
    &\qquad x^{i,j}_0 \sim \mathcal{N}_D( \mu_{x-ini}, \Sigma_{x-ini} )
\end{align}

Then, each subsequent time $t \in 1, 2, \ldots T$ depends only on variables from the previous or current time:
\begin{align}
    &s^{i}_t \sim \text{Cat}_L (~ G( s^{i}_{t-1}, x^{i,1:J}_{t-1}, \theta_{ss} )~ )
    \label{eq:gen_model_for_t_geq_1}
    \\ \notag 
    &\text{for~ $j$ ~in~} 1, 2, \ldots J:
    \\ \notag 
    &\qquad z^{i,j}_t \sim \text{Cat}_K(~ F(s^{i}_t, z^{i,j}_{t-1}, x^{i,j}_{t-1}, \theta_{se}) ~)
    \\ \notag 
    &\qquad x^{i,j}_t | z^{i,j}_t {=} k \sim \mathcal{N}_D( A_k x^{i,j}_{t-1} + b_k, Q_k)
\end{align}
Here, $\text{Cat}_L$ denotes a Categorical distribution over $L$ possible discrete values, parameterized by an $L$-dim.~vector of probabilities. Similarly, $\text{Cat}_K$ denotes a Categorical distribution over $K$ possible discrete values, parameterized by a $K$-dim.~vector of probabilities. $\mathcal{N}_D$ denotes a multivariate Normal which takes a mean vector in $\mathbb{R}^D$ and a $D\times D$ covariance matrix.
The key parameters here for initialization ($\{\theta_{s-ini}, \theta_{z-ini}, \mu_{x-ini}, \Sigma_{x-ini}\}$), latent state transitions ($\{\theta_{ss}, \theta_{se}\}$) and data emissions ($ \{A_k, b_k, Q_k\}_{k=1}^K$) are all contained in $\theta$.

\begin{figure}[!t]
  \includegraphics[width=1\linewidth]{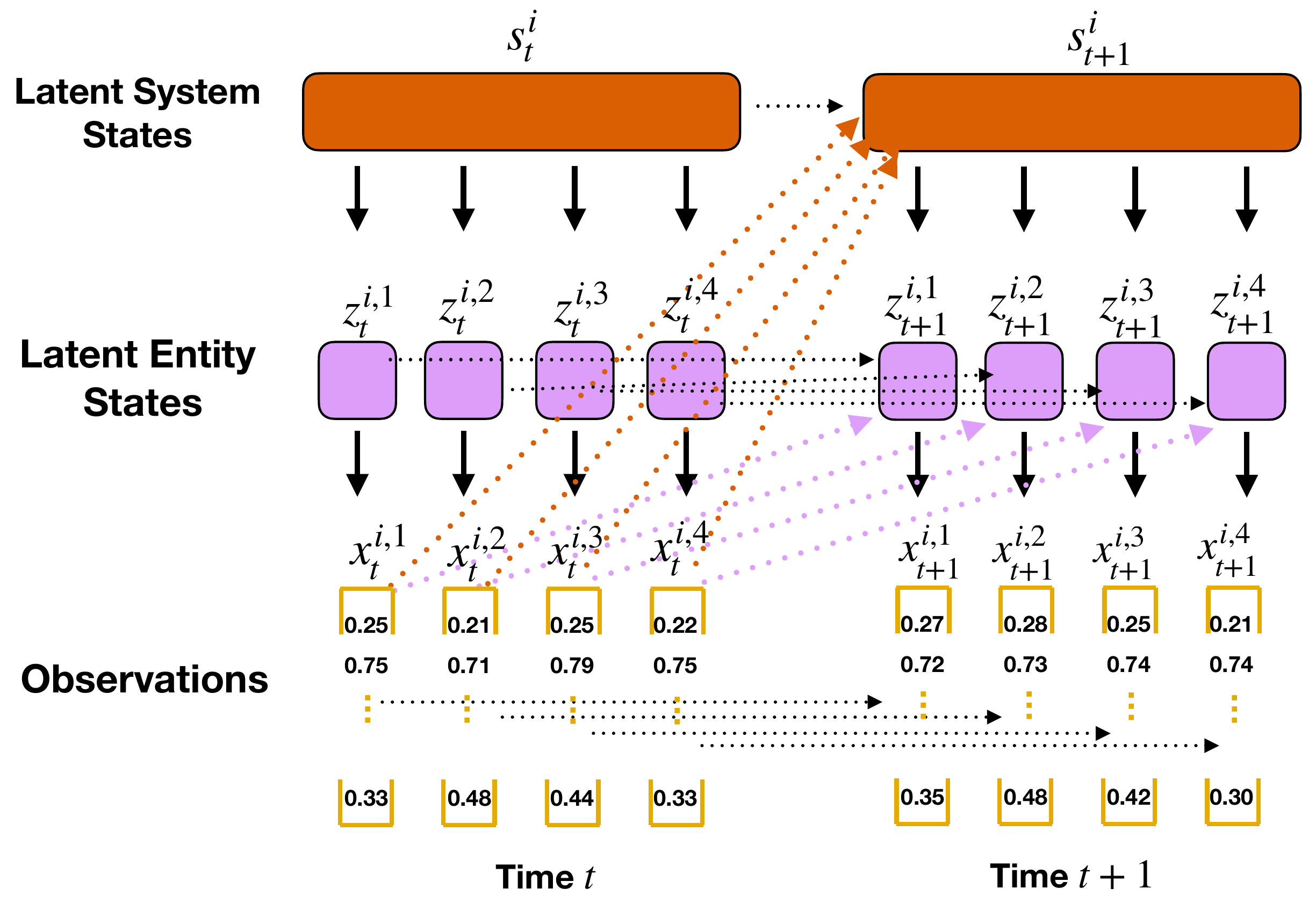}
  \caption{\justifying A graphical illustration of the HSRDM introduced in Ref.~\cite{wojnowicz2024discovering}, which we adapt for our task of locating evidence of mechanistic reasoning at the individual and student level. This representation visualizes our first three primary probabilistic modeling assumptions: (1) each observation and latent variable is Markovian (indicated by the black dashed lines), (2) the entity observations and latent states only interact via a shared system state (solid black lines), and (3) the system and latent entity variables are influenced by the observations at the previous timestep through the feedback mechanisms (dashed colored lines).}
  \label{fig:model_diagram}
\end{figure}

We encourage readers to first understand the dependencies between the variables, as visualized in Fig.~\ref{fig:model_diagram} for adjacent times $t-1,t$. Later bold-headed paragraphs provide more detail about each sampling step, such as the form of functions $F$ and $G$ that produce $L$-dim.~and $K$-dim.~vectors of probabilities. These individual sampling steps are obtained from further factorizing the joint distribution over the random variables in each sequence $i$ via statistical assumptions that enable computational tractability, efficiency, and domain alignment. A mathematical formulation of how these assumptions are incorporated into the joint distribution is provided in Appendix~\ref{factorization_details} for interested readers. Here, we present the four key modeling assumptions more descriptively: (1) the Markovian assumption that all variables at time $t$ depend only on variables from time $t-1$, (2) the conditional independence of per-entity variables given the shared team state $s_t$, (3) the presence of data-to-latent feedback that informs both team and entity-level variables, and (4) parameters controlling transitions and emissions are shared by all entities, not learned to be entity-specific. 

The Markovian assumption (1) makes computation tractable, but also aligns with \emph{domain knowledge}. 
Each turn in conversation will most strongly depend on the previous turn.
For example, if a student spoke at turn $t-1$, by our construction their observation at turn $t$ is going to be silence. Our generative model can account for this.
More importantly, Markovian time dynamics can model the latent trends in mechanistic reasoning we expect to see over time. Since reasoning mechanistically is typically a method to make sense of a particular phenomenon \cite{definition_sensemaking}, if a student--or the group--is already engaged in this practice, they are likely to continue until ambiguity has been resolved. While this is not \emph{always} true, we consider this a reasonable assumption. The feedback mechanism can account for moments when mechanistic-reasoning evidence abruptly disappears from the conversation. 

Assumption (2), the conditional independence of entity states given system states, is primarily for \emph{efficiency}. 
Because entity observations and latent states only influence one another through the system latent variable, the cost of generating data from the model or evaluating the probability of data under the model remains \emph{linear in $J$} as the number of entities grows. While our present analysis looks at small student teams of just 2-4 students, this model could scale easily to classroom-wide discussions of 50 or more students. We do acknowledge that using a single variable to channel the aggregate influence from all entities may be limited in flexibility compared to direct pairwise modeling of entity–entity interactions. 

Assumption (3), using feedback from previous data to influence the next system-level state and entity-level state, is crucial for modeling our chosen student conversation data. Our conception is that students have their own internal states that drive distinct utterances, but their individual states can be influenced by the rest of the group. 
Feedback mechanisms allow an individual's utterance to change both their own trajectory and the group's.

Finally, Assumption 4 assumes all students are governed by the same transition probabilities and emission probabilities in the generative model, rather than specialized parameters for individual students. This is a necessity given the limited overall number of sequences available in use case and the need to generalize at ``test time'' to students we have never seen before. 
Individualized parameters are well-suited for a context where one desires to model the same entities over many sequences, and when there is a large amount of data available for each entity. 
The original HSRDM work~\citep{wojnowicz2024discovering} showed how modeling the same 5 basketball players over a season with player-specific parameters could be beneficial.
We don't yet have enough student-specific data to build student-specific models; even in longer conversations, some students barely speak.

\newcommand{\SSN}{\mathsf{S0}}
\newcommand{\SSY}{\mathsf{S1}}
\newcommand{\TTN}{\mathsf{T0}}
\newcommand{\TTY}{\mathsf{T1}}

\newcommand*\circled[1]{\tikz[baseline=(char.base)]{
            \node[shape=circle,draw,inner sep=1pt] (char) {#1};}}
\newcommand{\sysY}{\circled{$\mathsf{1}$}}
\newcommand{\sysN}{\circled{$\mathsf{0}$}}

\medskip

\noindent \textbf{System-level states.} Each $s_t^i$ is a latent system state indicator for sequence $i$ at time $t$. Given our overall modeling goals, we consider just $L=2$ possible states, denoted as $\sysN$ and $\sysY$. State $\sysY$ indicates the team is in a collective mechanistic reasoning state $l=1$, while $\sysN$ suggests the opposite. In our model, the system state is designed to mediate how entities influence one another. 
For example, if the probability of system state $\sysY$ is high, the desired model behavior is that each entity has a higher probability of producing text that indicates mechanistic reasoning.

\medskip
\noindent \textbf{Entity-level states.} Each  $z_t^{i,j}$ is a latent entity state indicator for student $j$ and sequence $i$ at time $t$. By design, we define $K{=}4$ possible entity-level states, denoted as $\SSN, \SSY, \TTN,$ or $\TTY$. We intend states with prefix $\mathsf{S}$  correspond to the emission of \emph{silence}, whereas states with prefix $\mathsf{T}$ correspond to that entity \emph{talking} at time $t$. States with suffix $\mathsf{0}$ indicate no evidence of mechanistic reasoning in the $t-1$ utterance, while suffix $\mathsf{1}$ indicates present evidence in the $t-1$ utterance. 

The exact latent state for a given observation may not always be clear. For example, if there is a speaking observation with evidence of mechanistic reasoning, but no evidence was present in the previous speaker, should the entity latent be the state $\TTN$ or $\TTY$? We prioritize estimating a \emph{probability distribution} over these states rather forcing the model to choose just one state. Thus, the desired model behavior in this instance would be to place the highest probability on $\TTY$, and second-highest on $\TTN$, with zero probability on any $\mathsf{S}$ states.

\medskip 
\noindent \textbf{System-level starting distribution.}
We assume a simple uniform probability over the 2 system states, so $\theta_{s-ini} = [0.5, 0.5]$ in~\eqref{eq:gen_model_for_t_eq_0}.

\medskip
\noindent \textbf{System-level transition distributions.} To define the transitions between system-level latent states in~\eqref{eq:gen_model_for_t_geq_1}, the function $G$ needs to take in the previous state $s^i_{t-1}$ and recent data from all $J$ entities $x^{i,1:J}_{t-1}$, and produce a probability vector defining a valid distribution over the $L{=}2$ system states.
\begin{align}
        G(s^i_{t-1}, x^{i,1:J}_{t-1}, \theta_{ss}) = \textsc{sm}(
        \tilde{\Pi}^{\mathsf T} e_{s^i_{t-1}} + \Lambda g( x_{t-1}^{i, 1:J}))
\label{eq:sys_transition_probas_G}
\end{align}
Here, $\textsc{sm}$ is the softmax function, $e_{s^i_{t-1}}$ is a one-hot vector of length $L$,
and $\tilde{\Pi}$ is a $L\times L$ matrix where $\tilde{\Pi}_{l,m}$ gives a \emph{log} prior probability of transition from $l$ to $m$. 
The feedback function $g$ produces features useful for determining transition probabilities, and $\Lambda$ is a coefficient matrix that maps the features from $g$ to a vector of size $L$.
This feedback mechanism can aggregate from all $J$ entities.
The definition of $g$ is one of our deliberate inductive bias components that will be discussed in detail in Sec.~\ref{feedback_mech}, alongside a fixed setting for $\Lambda$. 
Only $\tilde{\Pi}$ is a learnable parameter in $\theta_{ss}$.

\medskip
\noindent \textbf{Entity-level starting distribution.} We assume a uniform distribution over all $K$ entity states at time $t=0$, so $\theta_{z-ini} = [\frac{1}{4}, \frac{1}{4}, \frac{1}{4}, \frac{1}{4}]$ in~\eqref{eq:gen_model_for_t_eq_0}.

\medskip 
\noindent \textbf{Entity-level transition distribution.} The function $F$ in~\eqref{eq:gen_model_for_t_geq_1} produces $K$-dim.~probabilities defining the chance of the next entity-level state for student $j$. We construct $F$ in a similar manner as $G$ above, via the definition:
\begin{align}
    F(s^i_t, z^{i,j}_{t-1}, x^{i,j}_{t-1}, \theta_{es}) = \textsc{sm}(
    \tilde{P}_{s^i_t}^{\mathsf T} e_{z^{i,j}_{t-1}} + \Psi_{s^i_t} f(x^{i,j}_{t-1}) )
    \label{eq:transition_probas_F}
\end{align}
Here, a key difference between $F$ and the system function $G$ is that only data feedback from the focal entity $j$ is needed for $F$. No other entities are directly involved, as system state mediates all entity-to-entity interaction.
The $K \times K$ matrix $\tilde{P}_{s^i_t}$ defines log transition probabilities specific to the current system state, while $e_k$ here denotes a $K$-dim.~one-hot vector.
The feedback feature function $f$ and its coefficient matrix $\Psi_{s^i_t}$ are deliberately fixed as a way to control inductive bias. Details will be provided in Sec.~\ref{feedback_mech}.
Overall, the only learnable parameters for $F$ are $\theta_{se} = \{ \tilde{P}_{\ell} \}_{\ell=1}^L$.

\medskip
\noindent \textbf{Emission at $t{=}0$.} 
For our initial observations $x^{i,j}_0$, we model the emission distributions for when $k$ is $\TTN$ or $\TTY$ as multi-variate gaussian distributions with zero mean vectors and an identity covariance. Thus, $\mu_{x-ini} = 0_D$ and $\Sigma_{x-ini} = \mathcal{I}_D$ in~\eqref{eq:gen_model_for_t_geq_1}. We model the emission distributions for when $k$ is $\SSN$ or $\SSY$ as multi-variate gaussian distributions with fixed parameters: $A_k = 0$, $Q_k = 0.001 I_D$, and $b_k$ to the embedding of ``silence''. To further ensure a good starting point for each sequence, we penalize the density emissions of all non-$k{=}0$ states for the non-speakers. As in, all non-speakers occupy the $k{=}0$ state given that there is no previous observation for evidence of mechanistic reasoning to occur.

\medskip
\noindent \textbf{Emission for $t{\geq}1$.} As defined in~\eqref{eq:gen_model_for_t_geq_1}, to generate each $x^{i,j}_t$ the current entity state indicator $z^{i,j}_t \in \{\SSN, \SSY, \TTN, \TTY\}$ defines a state-specific autoregressive Gaussian emission model. The state ID as an integer $k$ provides the state-specific coefficient matrix $A_k \in \mathbb{R}^D$, mean offset $b_k \in \mathbb{R}^D$, and $D\times D$ covariance matrix $Q_k$. 

For our task, we intend for state IDs with prefix $\mathsf{S}$ to generate only silent (non-talking) utterances. Text transcripts contain no interesting time-varying data for such turns $t$. Thus, instead of learning these parameters, we fix $A_k = 0$, fix $Q_k = 0.001 I_D$, and set $b_k$ to the embedding of ``silence'' for $k \in \{\SSN, \SSY\}$. This leaves only the $A_k, b_k, Q_k$ for talking states as learnable parameters in $\theta_{ee}$. 

We use Gaussian distributions for the emissions because they accommodate continuous data and have proven empirically versatile. Exploring alternative distributions better suited to modeling language embeddings, however, remains an important direction for future work.

\medskip
\noindent \textbf{Priors on Parameters.}
Whereas the previous work \cite{wojnowicz2024discovering} incorporated a sticky Dirichlet prior over the system transition parameters to make remaining in the same latent state more likely than transitioning, we do not include this bias in our work. Given that we lack knowledge around how many turns of talk we expect prior to a change of state, we use uninformative priors on all transition parameters -- both at the system and entity level. We also use uninformative priors over the multi-variate gaussian parameters that govern student utterances. We highlight that incorporating informed priors to enhance the inductive bias of the modeling is an important part of future work. 

\subsection{Training procedure}\label{unsupervised_training}
\hspace*{1em} To train our adapted HSRDM given a dataset of $N$ sequences, we follow the fully unsupervised procedure introduced by~\citet{wojnowicz2024discovering}.  
Their algorithm simultaneously produces a maximum a-posteriori point estimate of the parameter vector $\theta$ and an estimated posterior distribution over \emph{all} state sequences $\{ s^i_{0:T}, z^{i,1:J}_{0:T} \}_{i=1}^N$.
Practically, using these results can query the probability that $s^i_t = \sysY$ at any time $t$, or the probability that $z^{i,j}_t = \TTN$ or $z^{i,j}_t = \SSY$ at any time $t$.
We should understand these as posterior probabilities, because they condition on the entire data sequence $x^{i,1:J}_{0:T}$.
The underlying algorithm uses Bayesian variational methods
\cite{beal2003variational, Blei_2017} for posterior estimation, with structured approximations rather than naive ``mean-field'' independence assumptions.
Further details for this algorithm are provided in App.~\ref{vi_alg_overview}.

The learned posterior distributions over the entity and system-level latent variables are our primary quantities of interest for identifying students’ mechanistic reasoning over time in a conversation. In the next section, we introduce the specialized inductive bias we use to steer these posterior probabilities based on evidence in the conversation.

\subsection{Specialized inductive bias via data feedback}
\label{inductive_bias}
\label{feedback_mech}

Our probabilistic model in~\eqref{eq:gen_model_for_t_geq_1} uses previous data to inform next state probabilities for both system and entity-level states, via the function $ g(x_{t-1}^{i, 1:J})$ in~\eqref{eq:sys_transition_probas_G} and function $f(x_{t-1}^{i,j})$ in~\eqref{eq:transition_probas_F}.
\citet{wojnowicz2024discovering} introduced such feedback functions in a generic fashion.
In our work, we develop concrete custom functions as a primary inductive bias to guide our model behavior. 

Specifically, we train a classifier that takes EmbeddingGemma encoded utterance vectors $x_i \in \mathbb{R}^{128}$ as input and predicts the maximum mechanistic-evidence class $y_i \in \{0, 1, 2 ,...7\}$ under Russ’s taxonomy (Sec.~\ref{Background}). The zeroth class is for no-evidence, whereas classes $1-7$ quantify how much mechanistic reasoning evidence is present in the utterance. 
Denote the ultimate classifier as a function $c$ that, given an $x^{i,j}_t$, produces a 8-dim.~vector of probabilities for the 8 classes in Russ' framework.

Training labels for the classifier come from a subset of our human expert annotations of the data. We highlight that these annotations are utterance independent -- meaning that each utterance is annotated as a stand-alone segment of text, rather than incorporating inferred meaning from the surrounding utterances. Further details on the human annotation analysis and evidence for code reproducibility can be found in Appendix \ref{human_annot}. 
Our use of ``ground-truth'' labels from a subset of data to inform the feedback mechanism makes our adapted-HSRDM a \emph{semi-supervised method}, rather than fully-unsupervised. We are careful to make sure the data used to train this classifier is a subset of data used to train the adapted-HSDRM overall, and has no overlap with later ``test'' data used to evaluate ability to generalize.

\medskip
\noindent \textbf{Classifier model.}
For the classifier $c$, we utilize a 2-layer feed-forward neural network (NN) with a ReLu activation function and a hidden dimension size equal to the size of our text embeddings ($D = 128$).
We made this choice after evaluating a range of pre-trained and train-from-scratch options. This choice reflects a favorable trade-off between task alignment and deployment cost (e.g.~parameter storage for inference-time compute). We detail this rationale in the Appendix \ref{classifier_selection}.

\medskip
\noindent \textbf{Classifier training.}
We fit this classifier to all embedding-label pairs from $N=4$ sequences of data, a subset of our training set. 
To avoid overfitting, we minimize a loss that includes a prediction-quality term (cross entropy) and an L2 penalty on the weights of the NN. This penalty is often called ``
weight decay'' in deep learning. 

Standard practice to select the strength hyperparameter for the weight decay requires a grid search, where at each candidate value we fit a model on one subset of the total dataset and then selected based on performance on a validation set--a separate subset of the total data. In our setting, we cannot afford a separate validation set. We have limited data and would like to reserve as much as possible for the overall HSRDM training; especially because we also need a separate held-out test for assessing generalization performance. Training the feedback classifier on the same exact dataset as the overall HSRDM training runs the risk of overfitting the HSRDM.

To avoid needing a validation set for our classifier, we utilize a recent Bayesian model selection strategy from~\citet{harvey2025}. This paper suggests a way to use the marginal likelihood of the training set, which has a well-known ability to avoid over-fitting \cite{10.1162/neco.1992.4.3.415}, via a practical approximation in the case of deep neural networks where the marginal likelihood itself is not tractable. We provide technical details of the classifier training, including our training data selection, implementation decisions, and results in Appendix \ref{feedback_app}.

\medskip
\noindent \textbf{System-level feedback $g$.}
For the system recurrence function, we set $g(x^{i,1:J}_t) = \textsc{OH}(c(x^{i,j^*_t}_t))$, where $j^*_t$ indicates the \emph{speaking} entity at time $t$ and $\textsc{OH}$ produces a one-hot vector indicating the maximum value of its input vector.
As other entities remain silent by construction, the speaker is the only entity whose speech can have evidence of mechanistic reasoning.

We then set the coefficient matrix $\Lambda$, which is multiplied by the 8-dim.~vector from $g$ in~\eqref{eq:sys_transition_probas_G},~to be: 
\[
\Lambda \;=\;
\begin{bmatrix}
2 & 0 & 0 & 0 & 0 & 0 & 0 & 0 \\
0 & 2 & 3 & 4 & 5 & 6 & 7 & 8
\end{bmatrix},
\]
where the ordering of the rows corresponds to system state $\sysN$ and then $\sysY$. Using these coefficients, as the classifier $c$ predicts an increasing amount of mechanistic reasoning is present (mass moving rightwards), the probability of transitioning to $\sysY$ becomes higher. 

\medskip 
\noindent \textbf{Entity-specific feedback $f$.}
The recurrence function $f$ for entity $j$ takes only the one-hot prediction from that entity as input: $f(x^{i,j}_t) = \textsc{OH}( c(x^{i,j}_t) )$. We utilize one-hot predictions rather than predictive probabilities from the classifier to avoid the extra step of analyzing probability-prediction calibration quality. Using the one-hot predictions is a common practice in machine learning; however we acknowledge that more investigation could be done to understand which output is more suitable given the context. 

We then set the system-state-specific coefficient matrix $\Psi$ from~\eqref{eq:transition_probas_F} to one of two values:
\begin{align} \notag 
\Psi_{{\tiny \sysN}} {=}
    \begin{bmatrix}
2   & 0 & 0 & 0 & 0 & 0 & 0 & 0 \\
0   & 0 & 0 & 0 & 0 & 0 & 0 & 0 \\
\frac{1}{2} & 0 & 0 & 0 & 0 & 0 & 0 & 0 \\
0   & 0 & 0 & 0 & 0 & 0 & 0 & 0
\end{bmatrix},
\Psi_{{\tiny \sysY}} {=}
    \begin{bmatrix}
 0 & 0 & 0 & 0 & 0 & 0 & 0 & 0 \\
 2 & 3 & 4 & 5 & 6 & 7 & 8 & 9 \\
 0 & 0 & 0 & 0 & 0 & 0 & 0 & 0 \\
 2 & 0 & 0 & 0 & 0 & 0 & 0 & 0
\end{bmatrix}.
\end{align}

Here, the ordering of rows corresponds to entity states $\SSN, \SSY, \TTN, \TTY$.
When $s^i_t{=}\sysN$, the left matrix above will increase the probability of the non-reasoning entity states, $\SSN, \TTN$, given that this matrix interacts with only one-hot vectors that have a $1$ in the zeroth index. The integer shift is stronger for the $\SSN$ state than the $\TTN$, as only one student is able to transition to talk independent of the number of students in the group. These integer values are found empirically from implementing the model on the training data containing students of group size $4$ and human-optimizing for the combination of integer values that provide quality training performance. We provide the model results on training data in Appendix \ref{ML_results}. 

When $s^i_t{=}\sysY$, interactions between the \emph{speaker} one-hot vector from the classifier and the right matrix above instead increase the probabilities of $\SSY$, with the strength of this increase determined by how much evidence is expressed in the utterance. Since the one-hot vectors for all silent observations have a $1$ in the zeroth index, there is a $2$ in the zeroth index for the $k \in \{\SSY, \TTY\}$ states, so that silent observations also shift probability mass toward these mechanistic reasoning states, but more weakly than the speaker does.

Together, the supervised classifier underlying the feedback functions $f,g$ and the fixed settings of coefficients above implement a feedback mechanism that instills a desired inductive bias, pushing the model toward higher probabilities of mechanistic reasoning states when previous student utterances contain relevant evidence.

\subsection{Informed initialization procedure}\label{gauss_init}

The training optimization problem for the HSRDM is non-convex. Thus, smart parameter initialization can substantially improve the quality of solutions delivered by our algorithm.
Here we describe a way to initialize the parameters of the emission model,
so that it can generate embedding vectors containing evidence of mechanistic reasoning when $k = \TTY$, and embedding vectors without evidence of such reasoning when $k = \TTN$. Recall these are the only states with learnable parameters in $\theta$; Sec.~\ref{probabilistic_assumptions} describes how we fix $A_k, b_k, Q_k$ for $k \in \SSN, \SSY$. 

Concretely, we use true reasoning labels $y$ provided by our human annotators to aid initialization. We utilize the same labels as those used to train the classifier discussed in Sec.\ref{feedback_mech}. From this data, we create two subsets: $\mathcal{S}_{\mathsf{1}}$ contains adjacent-time pairs $x^{i,j}_{t-1}, x^{i,j}_{t}$ with reasoning present in $y_{t-1}$ and $j$ talking at $t$, and $\mathcal{S}_{\mathsf{0}}$ contains pairs with reasoning absent yet $j$ is talking.
We then estimate $A_k, b_k$ by minimizing squared error $\sum_{t \in \mathcal{S}_k} (x_t - A_k x_{t-1} - b_k)^2$. The initial covariance $Q_k$ is computed from the residuals, enforcing a variance floor of $0.0001$ to prevent too small values along the diagonal.

This mechanism, when combined with the recurrent feedback, enables the model to balance desired transition and emission probabilities. When the previous utterance contains mechanistic evidence, the feedback mechanism biases the next speaker toward the $\TTY$ state. If that speaker's utterance subsequently contains no mechanistic evidence, the Gaussian emission instead pulls the latent state toward $\TTN$. The intent is for these competing influences to balance probability mass between the two states, so that neither state becomes overwhelmingly likely.

\section{RQ2: How Inductive Biases Impact Generalization}\label{RQ2}
In this section, we introduce our chosen metrics for model evaluation, and investigate if and how much the inductive bias via the data feedback (introduced in Sec.\ref{inductive_bias}) impacts the performance of the adapted-HSRDM on unseen data. Our aim is to obtain evidence (or lack of) that the adapted-HSRDM is interpretable by design for the task of identifying evidence of student's mechanistic reasoning in time-evolving conversational data.
Given the relatively small amount of data we have, we stress this investigation is preliminary and exploratory.

\subsection{Metrics and hypotheses}\label{evalaution_metrics} 
\hspace*{1em} We introduce task-specific metrics tailored to desired model behavior in this setting. By no means do these metrics capture all desired model behavior; however, they offer a reasonable starting point for the primary behaviors we both desire and \emph{expect} to observe empirically based on the theoretical inductive bias baked into the model. Thus, we frame our investigation around \emph{hypotheses} of the following form: when the inductive bias is present, metric \emph{X} should be \emph{Y} relative to when it is absent or weaker. Positive support for these hypotheses is evidence that our adapted-HSRDM is inherently interpretable for the target task. Below, we define each hypothesis and provide the rationale for its inclusion in our exploratory evaluation.

\medskip 
\noindent \textbf{H(i): The \emph{correlation} between a talker doing MR and being in S1 at the next turn should be \emph{stronger}.} 
We compute the Pearson correlation between the hand-coded mechanistic-reasoning strength of an entity's observation at time $t$ and the model's posterior probability that the \emph{same} entity occupies latent state $\SSY$ at time $t{+}1$. Pearson's $r$ captures the direction and magnitude of a linear association. We focus on the $t \rightarrow t{+}1$ relationship because, in our data, an entity who speaks at time $t$ always transitions to silence at the next time step. Thus, we evaluate correlations specifically for the subsequent silent state after a speaking turn. A large positive correlation in this setting signals that higher mechanistic evidence strength is associated with greater posterior mass on $\SSY$ at the next turn. Thus, a stronger correlation when the inductive bias is present supports our hypothesis that the the mechanism is operating as intended. 

\medskip
\noindent \textbf{H(ii): The \emph{mean probability gap} in being in S1 at the next turn between talkers who reason and those who do not should be \emph{larger}.} Across all time steps $t$, we examine the human annotation $y_t$ and assign to one of two sets: the times $\mathcal{T}_Y$ with present evidence of mechanistic reasoning, and the times with no evidence $\mathcal{T}_N$. Denote the talking entity at each time $t$ as $j^*_t$; we know that they will move to a silent state at time $t+1$. Concretely, we'd like to compare the mean posterior probability of the target state across the two sets:
\begin{align}
    \frac{1}{|\mathcal{T}_Y|} \sum_{t \in \mathcal{T}_Y} q( z^{i,j^*_t}_{t+1} = \mathsf{S1} ) 
\gg 
\frac{1}{|\mathcal{T}_N|}\sum_{t \in \mathcal{T}_N} q( z^{i,j^*_t}_{t+1} = \mathsf{S1} )
\end{align}

 This metric offers more context to the one above, as we \emph{compare} how likely the speaker is to occupy the mechanistic reasoning state after their utterance contained evidence vs.~when their previous utterance did not. We report standard deviation as well as mean to help readers understand variation. A larger mean probability gap when the inductive bias is present` supports our hypothesis that the the mechanism is operating as intended.

\medskip
\noindent \textbf{H(iii): The \emph{mean probability gap} in being in T1 at the next turn between non-talkers who hear MR and non-talkers who do not should be \emph{larger}.} Using the same $\mathcal{T}_Y$ and $\mathcal{T}_N$ defined above, we now focus on those who did not talk at $t$ but may have \emph{heard} MR at that time. Denote $j^*_{t+1}$ as the speaker at $t+1$, who we know was silent at time $t$. Concretely, we'd like to compare the following: 
\begin{align} \notag 
    \frac{1}{|\mathcal{T}_Y|} \sum_{t \in \mathcal{T}_Y} q( z^{i,j^*_{t+1}}_{t+1} = \mathsf{T1} ) 
\gg 
\frac{1}{|\mathcal{T}_N|}\sum_{t \in \mathcal{T}_N} q( z^{i,j^*_{t+1}}_{t+1} = \mathsf{T1} )
\end{align}
This metric informs how well mechanistic reasoning in the recent past influences the group member who talks next. 
Given that this metric provides information about a non-speaker at time $t$, we do not expect their increase in posterior mass at the next timestep to be as large as the speaker's increase, so the delta from $\mathcal{T}_Y$ to $\mathcal{T}_N$ will be smaller than H(ii). We report standard deviation as well as mean to help readers understand variation. A larger mean probability gap when the inductive bias is present` supports our hypothesis that the the mechanism is operating as intended.

It is also important to separately measure how well mechanistic reasoning in the recent past influences the group members who remain silent. By tweaking the construction above, we compute the mean posterior probabilities of all entities who transition from silent to silent at all time steps in each of the two sets $\mathcal{T}_Y$, $\mathcal{T}_N$. A larger likelihood gap when the inductive bias is present` supports our hypothesis that the the mechanism is operating as intended. We report this metric for our training data in the Appendix, but do not use it for the results in the main text as our test data only contains $J{=}2$ students (i.e~no students are transitioning from silent to silent).

\subsection{Inductive bias ablations}

\hspace*{1em} To explore the hypotheses put forth in Sec.~\ref{evalaution_metrics},  we run variants of the model that either contain, weaken, or completely remove the inductive bias introduced in Sec.~\ref{inductive_bias}. To weaken or remove the bias, we ablate targeted components of the model while keeping the others intact. We describe these model variants below: 

\medskip
\noindent \textbf{Human feedback.}
As a ``gold standard'', we consider replacing the classifier $c$ with the true labels provided by humans for the 8 levels of MR evidence at each time $t$. This helps assess a \emph{ceiling} for performance of the feedback mechanism. Naturally, such labels are rarely readily available at test time due to the expensive annotation effort required.

\medskip
\noindent \textbf{Classifier feedback.} We use the classifier $c$ as the feedback mechanism described in Sec.~\ref{inductive_bias}. Thus, this model variant is the adapted-HSRDM with the specialized inductive bias for our task and serves as the out-of-the-box method proposed in this work.

\medskip
\noindent \textbf{Weak classifier feedback (no system state).} We consider a version of the model that removes the top-level system state variables entirely of the adapted-HSRDM. Removing this while keeping the classifier feedback largely weakens the inductive bias introduced in Sec.~\ref{inductive_bias} as \emph{only} the entity level feedback is utilized and the entity feedback matrix remains stationary as $\Psi_{{\tiny \sysN}}$. Additionally, ablating the system state provides evidence that the hierarchical group coordination structure in the model is inherently useful for the task.

\medskip
\noindent \textbf{No feedback.} We consider a version of the model that removed the feedback mechanism completely -- removing the $f$ term in~\eqref{eq:sys_transition_probas_G} and the $f$ term in~\eqref{eq:transition_probas_F}. Thus, this model represents the complete removal of the inductive bias introduced in Sec.~\ref{inductive_bias}. 

\begin{table*}[!t]
\centering
\centering\begin{tabular}{l | c r r r r | c r r r r r}
\toprule
&  \multicolumn{5}{c}{Seen Problem}
& \multicolumn{4}{c}{New Problem}
\\
Feedback 
    & Entity & $\SSN$ & $\boldsymbol{\SSY}$ & $\TTN$ & $\TTY$
    & $\SSN$ & $\boldsymbol{\SSY}$ & $\TTN$ & $\TTY$ \\
\midrule 
\multirow[c]{2}{*}{{Human}}
& 0 & -0.069 & \textbf{0.548} & -0.099 & -0.062 
& -0.086 & \textbf{0.869} & -0.247 & -0.087 \\
    & 1 & -0.100 & \textbf{0.780} & -0.166 & -0.087 
  & -0.090 & \textbf{0.894} & -0.283 & -0.110 \\
\addlinespace
\multirow[c]{2}{*}{Classifier} 
    & 0 & -0.087 & \textbf{0.257} & -0.097 & -0.065 
& -0.019 & \textbf{0.484} & -0.244 & -0.089 \\
    & 1 & -0.137 & \textbf{0.436} & -0.160 & -0.091 
& -0.005 & \textbf{0.467} & -0.276 & -0.120 \\
\addlinespace
\multirow[c]{2}{*}{No System} 
& 0 & 0.007 & \textbf{0.147} & -0.100 & -0.061 
& 0.097 & \textbf{0.298} & -0.255 & -0.076 \\
     & 1 & -0.004 & \textbf{0.233} & -0.173 & -0.078 
& 0.058 & \textbf{0.349} & -0.293 & -0.100 \\
\addlinespace
\multirow[c]{2}{*}{None} 
    & 0 & 0.140 & \textbf{0.135} & -0.097 & -0.064 
& 0.288 & \textbf{0.286} & -0.246 & -0.088 \\
& 1 & 0.221 & \textbf{0.212} & -0.160 & -0.091 
& 0.340 & \textbf{0.334} & -0.279 & -0.117 \\
\bottomrule
\end{tabular}
\caption{\justifying \textbf{H(i): Pearson $r$ correlation experiments on the seen (left) and unseen (right) test problems.}
For each model ablation, we examine turns when an entity speaks, 
and assess correlation between the posterior probability mass of the next state (columns) and the human-annotated evidence strength present in the utterance.
For good models, the desired behavior is a strong positive correlation for state $\SSY$ (bold). We observe the correlation decrease as we ablate inductive bias components in the model architecture. 
}
\label{seen_speaker_correlation}
\label{unseen_speaker_correlation}
\end{table*}
\subsection{Experimental setup details} 
\hspace*{1em} We divide our total dataset of thermo-fluids student conversations into a training set of $N{=}8$ transcripts and a test set of $N{=}2$ transcripts. Details about training and test data selection are provided in Appendix \ref{FM training details}.

Each training sequence has $J{=}4$ students on a team. 
A subset of $N{=}4$ sequences from the training set was used to develop the classifier feedback mechanism. The test set contains two transcripts from previously unseen students discussing two different thermo-fluids problems. One of these problems also appears in the training data; we refer to it as the seen problem and to the other as the unseen problem. Importantly, the test transcripts are different from that of the training in that they only contain two students in the group rather than four. In a less data scarce setting, we would evaluate model performance on groups matching the number of students seen during training, as well as on smaller and larger groups. Focusing evaluation on $J{=}2$ student transcripts is a limitation of this study, which we discuss further in Sec.~\ref{future_work}.

To train the model parameters, we utilized closed-form updates for the gaussian emission parameters and gradient-ascent on the ELBO objective for all others. We adopted the same low-level configuration settings as Ref.~\cite{wojnowicz2024discovering}: after initialization, each M-step used $50$ gradient-ascent iterations; the initialization procedure consisted of a single E-step and M-step with $5$ gradient-ascent iterations in the M-step. All gradient-ascent was computed with an Adam optimizer with a learning rate $0.01$. We trained the model with $15$ total CAVI iterations. We present the training ELBO plot for our best model in Appendix \ref{ML_results}. 

Rather than using the ELBO as the main guidepost to end model training, we attended to the correlation between the system level state occupying $\sysY$ and human-annotated evidence of mechanistic reasoning being present in the previous time step. We observed empirically that the correlation stabilized to two decimal places by iteration $15$. Given the metrics we are interested in (see Sec.~\ref{evalaution_metrics}), we believed this to be a more appropriate metric than monitoring the ELBO to determine convergence. 

We thus used the human annotations in the evaluation of the optimization procedure for all trained model feedback ablations. Each set of training parameters per ablation is used for that same ablation during test. We include all trained model parameters in our codebase \cite{tufts_ml_mech4mech_2026} and the training results can be found in Appendix \ref{ML_results}. Additionally, the appendix includes training results for some model variants initialized with an uninformed, label-free k-means procedure rather than the informed procedure described in Sec.~\ref{gauss_init}. These results stem from related experimentation on the initialization procedure and are deferred to the appendix because they are not central to our main research questions.

No parameters in the adapted-HSRDM are randomly initialized and no randomness is introduced in the gradient-ascent algorithm (each step uses all data, not a random minibatch). Thus, our training stability is not sensitive to any random seeds. As described in Sec.~\ref{RQ1}, the adapted-HSRDM is completely deterministic given a fixed training set.

\begin{table*}[!t]
\centering
\centering\begin{tabular}{l l  | c c | c c}
\toprule
& &  \multicolumn{2}{c}{\hspace{0.5cm} Seen Problem \hspace{0.5cm}}
& \multicolumn{2}{c}{\hspace{0.5cm} New Problem \hspace{0.5cm}}
\\
Feedback & Evidence 
    & Mean Proba. & Std.~Dev.
    & Mean Proba. & Std.~Dev.
\\
\midrule
\multirow[c]{2}{*}{Human}
  & Yes & \textbf{0.791} & 0.105
& \textbf{0.649} & 0.270 \\
  & No & \textbf{0.151} & 0.167 
& \textbf{0.096} & 0.124 \\
\addlinespace
\multirow[c]{2}{*}{Classifier} 
 & Yes & \textbf{0.973} & 0.017 
& \textbf{0.769} & 0.230 \\
  & No & \textbf{0.543} & 0.292 
& \textbf{0.476} & 0.280 \\
\addlinespace
\multirow[c]{2}{*}{No System} 
  & Yes & 0.964 & 0.011 
& 0.963 & 0.047 \\
  & No & 0.926 & 0.052
& 0.937 & 0.039 \\
\addlinespace
\multirow[c]{2}{*}{None} 
  & Yes & 0.495 & 0.004
& 0.501 & 0.004 \\
  & No & 0.500 & 0.004
& 0.502 & 0.004 \\
\addlinespace
\bottomrule
\end{tabular}
\caption{\justifying 
\textbf{H(ii): Mean probability of entity state $\SSY$, for turns with and without evidence of MR, on the seen (center) and unseen (right) problems.}
For all model ablations, we report the mean and standard deviation of the previous speaker’s posterior probability mass assigned to state $\SSY$, conditioned on whether their utterance contains evidence of mechanistic reasoning.
We expect for this state to have a larger probability when evidence is present than not, and we highlight in bold cases where the delta is at least 0.25.}
\label{seen_speaker_mean}
\label{unseen_speaker_mean}
\end{table*}

\begin{table*}[t]
\centering
\centering\begin{tabular}{l l  | c c | c c}
\toprule
& &  \multicolumn{2}{c}{\hspace{0.5cm} Seen Problem \hspace{0.5cm}}
& \multicolumn{2}{c}{\hspace{0.5cm} New Problem \hspace{0.5cm}}
\\
Feedback & Evidence 
    & Mean Proba. & Std.~Dev.
    & Mean Proba. & Std.~Dev.
\\
\midrule
\multirow[c]{2}{*}{{Human}}
& Yes & 0.143 & 0.350 
& 0.199 & 0.373 \\
& No & 0.309 & 0.450 
& 0.164 & 0.360 \\
\addlinespace
\multirow[c]{2}{*}{Classifier} 
& Yes & 0.143 & 0.350
& \textbf{0.230} & 0.403 \\
& No & 0.335 & 0.461
& \textbf{0.179} & 0.371 \\
\addlinespace
\multirow[c]{2}{*}{No System} 
& Yes & 0.143 & 0.350
& 0.148 & 0.342 \\
& No & 0.282 & 0.442 
& 0.134 & 0.326 \\
\addlinespace
\multirow[c]{2}{*}{None} 
& Yes & 0.143 & 0.350 
& 0.218 & 0.391 \\
& No & 0.334 & 0.464 
& 0.171 & 0.363 \\
\addlinespace
\bottomrule
\end{tabular}
\caption{\justifying 
\textbf{ H(iii): Mean probability of entity state $\TTY$, for turns with and without evidence of MR, on the seen (center) and unseen (right) problems.}
For all model ablations, we report the mean and standard deviation of the current speaker’s posterior probability mass assigned to state $\TTY$, conditioned on whether the previous speaker's utterance contains evidence of mechanistic reasoning.
We expect for this state to have \emph{noticeably} higher probability when evidence is present than not, and this, we highlight in bold cases where the delta is at least 0.05.
}
\label{seen_nonspeaker_mean}
\label{unseen_nonspeaker_mean}
\end{table*}

 \subsection{H(i) Results: Correlation}

In Table~\ref{seen_speaker_correlation}, we report metric results for all inductive bias ablations. 
Focusing on the seen problem setting, the column most relevant is $\SSY$, representing the correlation between the human-annotated evidence of mechanistic reasoning and the posterior probability mass of the $\SSY$ occupancy. We observe the highest average correlation values across both entities when there is human feedback. When there is classifier evidence, we observe the correlation decrease as expected, but still remain higher than when there is weak feedback and no feedback. Relative to no feedback, classifier feedback produces about a $2\times$ increase in correlation strength. In non-$\SSY$ columns for both human and classifier evidence feedback, we observe small evidence of an anti-correlation, which is evidence of the feedback working as expected. Whereas, for no feedback we observe small evidence of a correlation in the $\SSN$ column. This is evidence that the inductive bias removes unwanted correlations. Overall, there is strong evidence to support the hypothesis that the classifier feedback is a useful inductive bias. 

In Table \ref{unseen_speaker_correlation} (right), we report metric results for all ablations in the unseen-problem setting. As before, the column most relevant is $\SSY$, representing the correlation between the human-annotated evidence of mechanistic reasoning and the posterior probability mass of the $\SSY$ occupancy. We observe consistently higher correlation values across all the ablations relative to the seen-problem setting. The highest average correlation values across both entities are observed when there is human feedback. When classifier evidence is used, the model performs worse as expected. However, when the classifier evidence or the system state is removed, the correlation across both entities decreases further. Compared to the no-feedback setting, incorporating classifier feedback produces an approximately $1.5\times$ stronger correlation. This supports the hypothesis that the classifier feedback induces a useful inductive bias. In the $\TTN$ and $\TTY$ columns, we observe consistent negative correlations. 

\subsection{H(ii) Results: MR Effect on Talker Next State}

In Table \ref{seen_speaker_mean}, we report metric results for all ablations on the seen problem. We observe the largest increase in the $\SSY$ state occupancy posterior mean -- conditioned on mechanistic reasoning evidence -- with the human feedback mechanism. As expected, the classifier feedback performs worse than the human feedback, but greatly outperforms removing the system state and the feedback altogether. When we ablate the system state from the model or take away the evidence feedback mechanism completely, we see there there is essentially \emph{no difference} in the $\SSY$ state occupancy posterior mean. Compared to the no-feedback setting, incorporating classifier feedback produces an approximately $86\times$ larger mean probability gap. This is strong evidence to support the hypothesis that the classifier feedback induces a useful inductive bias. 

In Table \ref{unseen_speaker_mean} (right columns), we focus on the unseen problem.  As before, we observe the largest increase in the $\SSY$ state occupancy posterior mean with the human feedback mechanism. Classifier evidence performs worse human evidence and performance decreases further when the system state is ablated or the evidence feedback mechanism is removed. Relative to when no feedback is present, incorporating classifier feedback produces an approximately $313\times$ larger mean probability gap. These results are consistent with the hypothesis that classifier feedback induces a useful inductive bias.

\subsection{H(iii) Results: MR Effect on Non-talker Next State}

In Table \ref{seen_nonspeaker_mean}, we report metric results for all ablations for the seen problem. We observe all results run counter to our expectations: the average posterior mass on state $\TTY$ is higher when no mechanistic reasoning evidence is present at the previous time step than when evidence is present. One possible reason for this poor generalization is that there are only two students in the group rather than four in our training data. However, we believe a stronger hypothesis is that in this particular transcript, there are very few examples of mechanistic reasoning. Given the small sample size (seven instances in this transcript), learning how the current speaker’s evidence of mechanistic reasoning should affect the next speaker's posterior may be difficult, and likely harder than the previous two metrics. We include the number of positive label counts by problem in Appendix Table \ref{tab:positive_label_counts}. 

\begin{figure*}[!t]
  \centering
  \includegraphics[width=\textwidth]{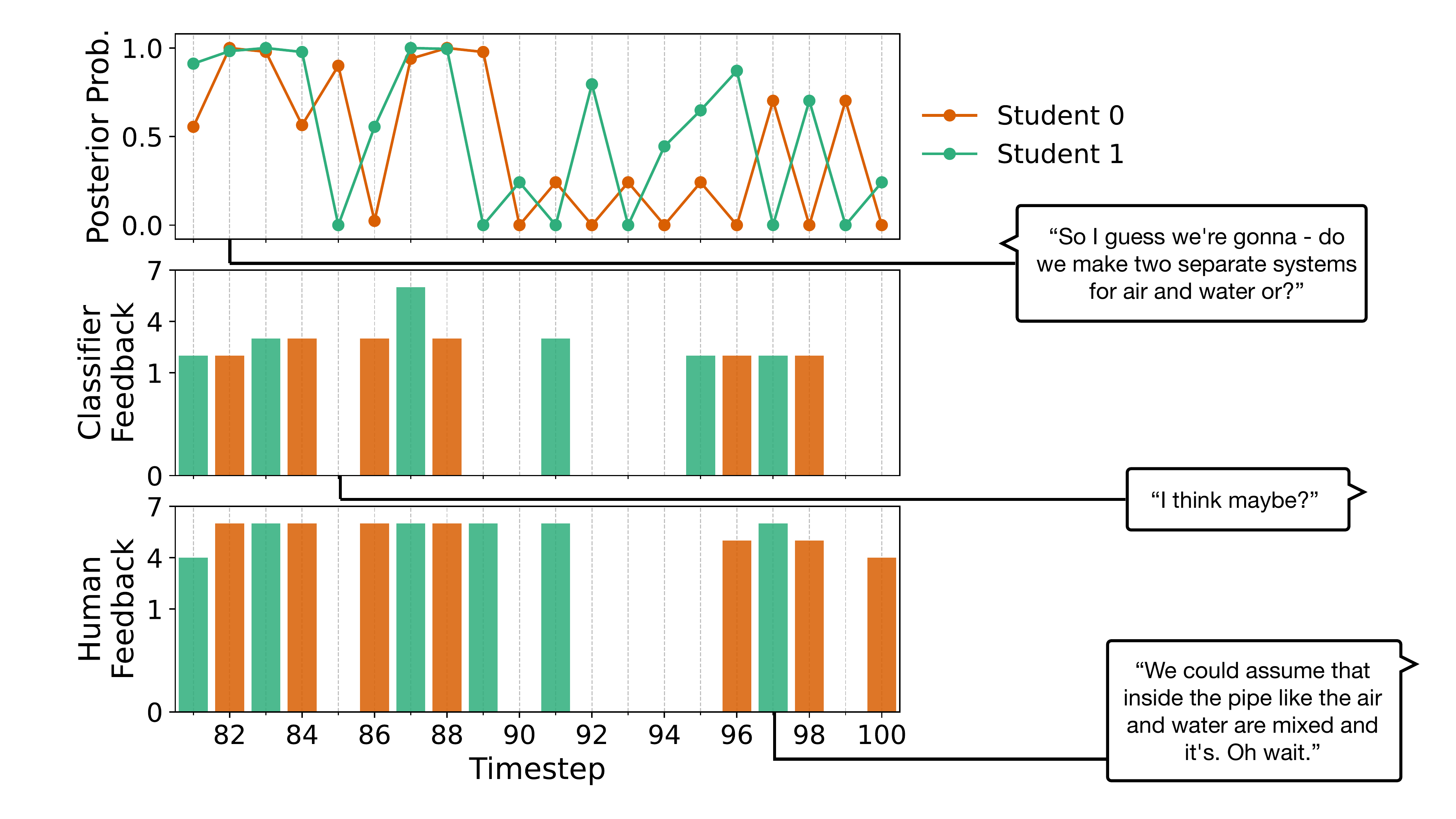}
  \caption{\justifying \textbf{The tool’s output for a segment of the
unseen test problem transcript.} Top: 
  the posterior probabilities that each student is engaging in mechanistic reasoning throughout the segment of conversation. These probabilities are direct outputs from the adapted-HSRDM with classifier feedback -- selected as the region of 20 time steps that had the highest average human feedback. Middle: the classifier feedback predictions at aligned timesteps in the conversation. Bottom: the human feedback annotations at aligned timesteps in the conversation.
  Our main motivation for this figure is to demonstrate a visual output that a user can generate with our tool. We also observe visually that the human feedback strength and the posterior probabilities at the next timestep for a speaker are strongly positively correlated (as evidenced in our numerical results). We also observe that classifier mistakes clearly influence the adapted-HSRDM's behavior. For example, in step 89, where the classifier predicts no evidence of mechanistic reasoning and the posterior probabilities remain low.  }
  \label{fig3}
\end{figure*}

In Table \ref{unseen_nonspeaker_mean}'s right columns, we report metric results for all ablations on the seen problem. For this metric, we do not expect that there be such large differences in the posterior means based on the conditional; rather, we expect that there be noticeable differences, which we do see here. Surprisingly, we observe the largest increase in the $\TTY$ state occupancy posterior mean with the classifier feedback mechanism. When we ablate the system state, performance drops significantly. It makes sense that ablating the system state would cause the most disturbance given that the metric validates the influence of an individual on the entire group.  However, no evidence performs similarly to classifier evidence, suggesting that the feedback mechanism does not have a large affect on the $\TTY$ state occupancy posterior mean. The performance on the unseen-problem setting is better relative to the performance on the seen-problem setting which we hypothesize is because the unseen problem contains 75 instances of mechanistic reasoning, compared to only 7 instances in the seen problem. We include the number of positive label counts by problem in Appendix Table \ref{tab:positive_label_counts}. In summary, we have little evidence from this metric to conclude that the classifier feedback mechanism offers a useful inductive bias. However, we do have evidence to support that the system state is an integral inductive bias component.

\section{Tool demonstration and user recommendations}\label{demo_recs} In this manuscript, we explore an inherently interpretable ML tool to automate the identification of students' mechanistic reasoning in time-series data. In Fig.~\ref{fig3}, we show the adapted-HSRDM's output for a segment of the unseen test problem transcript. We took a moving average of human labels of mechanistic reasoning for 20 timesteps, found the max, and then selected those 20 timesteps as our region. The first row displays the posterior probabilities that each student is engaging in mechanistic reasoning throughout the segment of conversation. These probabilities are direct outputs from the adapted-HSRDM with \emph{classifier feedback}. The second and third rows show the classifier feedback predictions and the human feedback annotations, respectively at aligned timesteps in the conversation. Our main motivation for including Fig.~\ref{fig3} is to demonstrate a visual output of a high density region of mechanistic reasoning that a user can generate with our tool. Fig.~\ref{fig3} is \emph{not meant} to represent a ground truth comparison of the posterior probabilities to the labels from the two feedback mechanisms. Recall that both the human and classifier feedback are labels of \emph{independent} utterances that do not include information from surrounding utterances. Nonetheless, as a compliment to the evidence shown in Table \ref{unseen_speaker_correlation} that our model outputs desired behavior, we see visually that the strength of the human feedback and the posterior probabilities of the speaker at the next timestep are strongly positively correlated. Additionally, the human feedback is a ground truth for the classifier feedback, and we can compare these predictions directly. We observe that classifier mistakes clearly influence the adapted-HSRDM's behavior. For example, in step 89, where the classifier predicts no evidence of mechanistic reasoning and the posterior probabilities remain low. Overall, although the human and classifier annotations exhibit substantial variance in their exact numerical agreement, they align well on the binary distinction of whether evidence of mechanistic reasoning is present.

The adapted-HSRDM with instructions for use can be found in the public Github repository \cite{tufts_ml_mech4mech_2026}, under the project name ``Mech4Mech''. We provide clear guidelines for users to run the pre-trained classifier feedback model on their own data. We recommend using the classifier variant for new data without human feedback annotations, since it is most appropriate to deploy a model under the same information conditions available at test time as during training. That said, we provide the parameters for all model variants in the codebase, so users can easily switch to the human-feedback variant if they wish to test their data with it.

Users can choose to generate a posterior probability plot over a specified time window for the region exhibiting the highest density of mechanistic reasoning (e.g., similar to the top row of Fig.~\ref{fig3}). The highest density region is computed by taking a moving average across the system state probabilities for $s^i_t{=}\sysY$ for a specified user time window. Users can also choose to output a csv file of the per-student max posterior probability with the associated state over time. To enable others' explorations, we provide open-access embeddings of the raw student text \footnote{The raw student data used in this project is not able to made publicly available as dictated by the original IRB documentation and authorization.} and an open version of the code. 

When using this model on new student transcripts, we ask the user to be cautious of the following contextual variables: (1) varying group sizes, (2) non open-ended problems, (3) what counts as evidence of mechanistic reasoning, and (4) small amounts of evidence present in the data. 

As for (1), our training set contained transcripts of student groups of four and our test set contained transcripts of student groups of two. While the generalization results support that the model can generalize to student groups of new sizes, we do not have enough evidence to conclude that we would see good performance on groups of other sizes, or much larger sizes. With respect to (2), all of our transcripts contained student discourse about open-ended thermo-fluids problems, which biases our model's capability towards dialogue that would be generated from such prompts. We caution a user against applying this tool on problems that would overly constrain a student's thinking towards a specific answer. Although we have not validated the model beyond thermo-fluids contexts, the diversity of thermo-fluids problems used in training and evaluation gives us reasonable confidence that it may transfer to discussions of other open-ended science and engineering problems. Of course, we are most confident that the model can generalize to new student transcripts of the same problems from the training set. Given that questions which are too constraining may be less likely to lead to evidence of mechanistic reasoning, this point also aligns with (3). The generalization results on a transcript containing very little evidence of mechanistic reasoning show that it is important for a user to consider whether their data might have enough evidence of mechanistic reasoning for the tool to be used properly. Without looking at the data, this judgment is best informed by the user’s familiarity with the problem prompts and the data-collection procedure. As a quick sanity check, we recommend for the user to annotate a small sequence of data and evaluate the model according to the metrics used in this work. Lastly, it is important for any user to determine whether their model of what counts as evidence of mechanistic reasoning aligns with that of the one used in this work. We encourage a user to review our process for human-annotation in the Appendix \ref{human_annot} and decide whether the labels taken as ``ground truth" in this work align with their description of mechanistic reasoning evidence.

\section{Research opportunities and limitations}\label{future_work} 

\hspace*{1em} We outline several future directions for ML experts aiming to build on this tool for STEM education research, especially through addressing its current limitations. An initial direction is to further characterize the variational posteriors at different time points, and how they change with respect to the inductive bias being present. For example, we did not investigate if and when the posteriors exhibit very extreme probability values (i.e.~very close to zero or one) and how this might impact recommendations for tool usage. Given that our test experiments were conducted on only two transcripts, an important next step is to evaluate the model on a broader range of data that varies in problem type, student group size, and student populations. This would further inform probabilistic priors that we can introduce as additional inductive biases for the task.

At present, one might reasonably ask why the classifier itself should not simply be used as a proxy method to identify text regions with the highest density of mechanistic reasoning. This question can be pursued systematically if provided ground truth human annotations that capture context, rather than assigning labels only at the level of acts of mechanistic reasoning in individual utterances. Generating context-sensitive human annotations poses a greater challenge when coding transcript data line by line using Russ’ framework. If obtained, these annotations would enable systematic comparisons between our model’s learned latent states, independent classifier predictions, and uninterpretable in-context learning outputs. Doing this human coding may also reveal new human-decision making elements that make useful inductive biases to incorporate into the modeling. As an alternative approach to more human annotation, all three methods could be used to identify regions of highest mechanistic reasoning density, which education researchers could then qualitatively evaluate for their usefulness. This would answer the question of which tool effectively sends education researchers to the most relevant transcript segments. Independent of the approach taken, we believe systematic comparisons are an important next step for understanding trade-offs among interpretability, usability, sustainability, and performance.

As a separate direction, we encourage others to expand upon the adapted-HSRDM to introduce personalized features for individual entities, inductive biases that allow for shuffled student groups, and multi-modal observations to include data beyond language. Especially with respect to the latter, we can include more informed evidence of mechanistic reasoning (e.g.~student drawings, hand-gestures) that will strengthen the tool's utility. We encourage the construction of similar models for identifying other disciplinary practices, behaviors, and emotional affects of students in conversation -- such as engaging in confusion \cite{hammer_uncertainty}, analogical reasoning \cite{bergeron2025reasoning}, and meta-cognition. 

Upon reflection, there are a set of smaller improvements that could be made to the adapted-HSRDM to enhance performance. For one, we could take advantage of the talk-silent knowledge in the data and add prior distributions to the system and entity transition probabilities reflecting this knowledge. Furthermore, we could generate a more mathematically principled silent vector than simply embedding the word ``silence". For example, we could generate an embedding that is maximally orthogonal to the mean of the talk embeddings. 

We hope that this work inspires an interpretable ML approach to the design and investigation of tools for STEM education research. While we acknowledge that interpretable methods do not always offer the largest short-term efficiency win, we believe that tools that can be reasoned about mechanistically provide longer-term advantages for both designers and users. Understanding the underlying mechanisms offers designers more principled control and evidence that can support or nullify a hypothesis, which in turn, offers users greater confidence in the tool and clearer guidance of when it should be applied. We emphasize that tools of this kind are especially valuable for fostering \emph{convergence} between tool designers and end users, and we encourage deeper collaboration at the intersection of ML and STEM education.
\vspace{-1em}
\begin{acknowledgments}
We are grateful for financial support from the U.S. National Science Foundation via awards GCR $\#2428640$ and CAREER $\#2338962$. We wish to acknowledge other members of the Hughes lab research group, namely Ethan Harvey for help training the feedback classifier with a data-emphasized variational ELBO approach. We also thank Erica Kemmerling, Katie Melsky, and Isabella Stuopis for their work creating the original data set.
 \end{acknowledgments}

\bibliography{ml_ed.bib,refs_from_zotero.bib}

\begin{thebibliography}{50}%
\makeatletter
\providecommand \@ifxundefined [1]{%
 \@ifx{#1\undefined}
}%
\providecommand \@ifnum [1]{%
 \ifnum #1\expandafter \@firstoftwo
 \else \expandafter \@secondoftwo
 \fi
}%
\providecommand \@ifx [1]{%
 \ifx #1\expandafter \@firstoftwo
 \else \expandafter \@secondoftwo
 \fi
}%
\providecommand \natexlab [1]{#1}%
\providecommand \enquote  [1]{``#1''}%
\providecommand \bibnamefont  [1]{#1}%
\providecommand \bibfnamefont [1]{#1}%
\providecommand \citenamefont [1]{#1}%
\providecommand \href@noop [0]{\@secondoftwo}%
\providecommand \href [0]{\begingroup \@sanitize@url \@href}%
\providecommand \@href[1]{\@@startlink{#1}\@@href}%
\providecommand \@@href[1]{\endgroup#1\@@endlink}%
\providecommand \@sanitize@url [0]{\catcode `\\12\catcode `\$12\catcode `\&12\catcode `\#12\catcode `\^12\catcode `\_12\catcode `\%12\relax}%
\providecommand \@@startlink[1]{}%
\providecommand \@@endlink[0]{}%
\providecommand \url  [0]{\begingroup\@sanitize@url \@url }%
\providecommand \@url [1]{\endgroup\@href {#1}{\urlprefix }}%
\providecommand \urlprefix  [0]{URL }%
\providecommand \Eprint [0]{\href }%
\providecommand \doibase [0]{https://doi.org/}%
\providecommand \selectlanguage [0]{\@gobble}%
\providecommand \bibinfo  [0]{\@secondoftwo}%
\providecommand \bibfield  [0]{\@secondoftwo}%
\providecommand \translation [1]{[#1]}%
\providecommand \BibitemOpen [0]{}%
\providecommand \bibitemStop [0]{}%
\providecommand \bibitemNoStop [0]{.\EOS\space}%
\providecommand \EOS [0]{\spacefactor3000\relax}%
\providecommand \BibitemShut  [1]{\csname bibitem#1\endcsname}%
\let\auto@bib@innerbib\@empty
\bibitem [{\citenamefont {Vinck}\ and\ \citenamefont {Blanco}(2003)}]{vinck2003everyday}%
  \BibitemOpen
  \bibfield  {author} {\bibinfo {author} {\bibfnamefont {D.}~\bibnamefont {Vinck}}\ and\ \bibinfo {author} {\bibfnamefont {{\'E}.}~\bibnamefont {Blanco}},\ }\href {https://doi.org/10.7551/mitpress/2862.001.0001} {\emph {\bibinfo {title} {Everyday Engineering: An Ethnography of Design and Innovation}}}\ (\bibinfo  {publisher} {The MIT Press},\ \bibinfo {address} {Cambridge, MA},\ \bibinfo {year} {2003})\BibitemShut {NoStop}%
\bibitem [{\citenamefont {Dickerson}\ \emph {et~al.}(2024)\citenamefont {Dickerson}, \citenamefont {Masta}, \citenamefont {Ohland},\ and\ \citenamefont {Pawley}}]{Dickerson2024CarlaGrumpy}%
  \BibitemOpen
  \bibfield  {author} {\bibinfo {author} {\bibfnamefont {D.~A.}\ \bibnamefont {Dickerson}}, \bibinfo {author} {\bibfnamefont {S.}~\bibnamefont {Masta}}, \bibinfo {author} {\bibfnamefont {M.~W.}\ \bibnamefont {Ohland}},\ and\ \bibinfo {author} {\bibfnamefont {A.~L.}\ \bibnamefont {Pawley}},\ }\bibfield  {title} {\bibinfo {title} {Is carla grumpy? analysis of peer evaluations to explore microaggressions and other marginalizing behaviors in engineering student teams},\ }\href {https://doi.org/10.1002/jee.20606} {\bibfield  {journal} {\bibinfo  {journal} {Journal of Engineering Education}\ }\textbf {\bibinfo {volume} {113}},\ \bibinfo {pages} {603} (\bibinfo {year} {2024})}\BibitemShut {NoStop}%
\bibitem [{\citenamefont {Scherr}\ and\ \citenamefont {Hammer}(2009)}]{ScherrHammer2009Framing}%
  \BibitemOpen
  \bibfield  {author} {\bibinfo {author} {\bibfnamefont {R.~E.}\ \bibnamefont {Scherr}}\ and\ \bibinfo {author} {\bibfnamefont {D.}~\bibnamefont {Hammer}},\ }\bibfield  {title} {\bibinfo {title} {Student behavior and epistemological framing: Examples from collaborative active-learning activities in physics},\ }\href {https://doi.org/10.1080/07370000902797379} {\bibfield  {journal} {\bibinfo  {journal} {Cognition and Instruction}\ }\textbf {\bibinfo {volume} {27}},\ \bibinfo {pages} {147} (\bibinfo {year} {2009})}\BibitemShut {NoStop}%
\bibitem [{\citenamefont {Koretsky}\ \emph {et~al.}(2023)\citenamefont {Koretsky}, \citenamefont {Nefcy}, \citenamefont {Nolen},\ and\ \citenamefont {Champagne}}]{KoretskyEtAl2023Connected}%
  \BibitemOpen
  \bibfield  {author} {\bibinfo {author} {\bibfnamefont {M.~D.}\ \bibnamefont {Koretsky}}, \bibinfo {author} {\bibfnamefont {E.~J.}\ \bibnamefont {Nefcy}}, \bibinfo {author} {\bibfnamefont {S.~B.}\ \bibnamefont {Nolen}},\ and\ \bibinfo {author} {\bibfnamefont {A.~B.}\ \bibnamefont {Champagne}},\ }\bibfield  {title} {\bibinfo {title} {Connected epistemic practices in laboratory-based engineering design projects for large-course instruction},\ }\href {https://doi.org/10.1002/sce.21769} {\bibfield  {journal} {\bibinfo  {journal} {Science Education}\ }\textbf {\bibinfo {volume} {107}},\ \bibinfo {pages} {510} (\bibinfo {year} {2023})}\BibitemShut {NoStop}%
\bibitem [{\citenamefont {Machamer}\ \emph {et~al.}(2000)\citenamefont {Machamer}, \citenamefont {Darden},\ and\ \citenamefont {Craver}}]{machamer}%
  \BibitemOpen
  \bibfield  {author} {\bibinfo {author} {\bibfnamefont {P.}~\bibnamefont {Machamer}}, \bibinfo {author} {\bibfnamefont {L.}~\bibnamefont {Darden}},\ and\ \bibinfo {author} {\bibfnamefont {C.~F.}\ \bibnamefont {Craver}},\ }\bibfield  {title} {\bibinfo {title} {Thinking about mechanisms},\ }\href {http://www.jstor.org/stable/188611} {\bibfield  {journal} {\bibinfo  {journal} {Philosophy of Science}\ }\textbf {\bibinfo {volume} {67}},\ \bibinfo {pages} {1} (\bibinfo {year} {2000})}\BibitemShut {NoStop}%
\bibitem [{\citenamefont {van Eck}(2018)}]{vanEck2018Mechanisms}%
  \BibitemOpen
  \bibfield  {author} {\bibinfo {author} {\bibfnamefont {D.}~\bibnamefont {van Eck}},\ }\bibfield  {title} {\bibinfo {title} {Mechanisms and engineering science},\ }in\ \href@noop {} {\emph {\bibinfo {booktitle} {The Routledge Handbook of Mechanisms and Mechanical Philosophy}}},\ \bibinfo {editor} {edited by\ \bibinfo {editor} {\bibfnamefont {S.}~\bibnamefont {Glennan}}\ and\ \bibinfo {editor} {\bibfnamefont {P.}~\bibnamefont {Illari}}}\ (\bibinfo  {publisher} {Routledge},\ \bibinfo {year} {2018})\ pp.\ \bibinfo {pages} {447--461}\BibitemShut {NoStop}%
\bibitem [{\citenamefont {de~Andrade}\ \emph {et~al.}(2022)\citenamefont {de~Andrade}, \citenamefont {Shwartz}, \citenamefont {Freire},\ and\ \citenamefont {Baptista}}]{deAndradeEtAl2022}%
  \BibitemOpen
  \bibfield  {author} {\bibinfo {author} {\bibfnamefont {V.}~\bibnamefont {de~Andrade}}, \bibinfo {author} {\bibfnamefont {Y.}~\bibnamefont {Shwartz}}, \bibinfo {author} {\bibfnamefont {S.}~\bibnamefont {Freire}},\ and\ \bibinfo {author} {\bibfnamefont {M.}~\bibnamefont {Baptista}},\ }\bibfield  {title} {\bibinfo {title} {Students' mechanistic reasoning in practice: Enabling functions of drawing, gestures and talk},\ }\href {https://doi.org/10.1002/sce.21685} {\bibfield  {journal} {\bibinfo  {journal} {Science Education}\ }\textbf {\bibinfo {volume} {106}},\ \bibinfo {pages} {199} (\bibinfo {year} {2022})}\BibitemShut {NoStop}%
\bibitem [{\citenamefont {Russ}\ \emph {et~al.}(2008)\citenamefont {Russ}, \citenamefont {Scherr}, \citenamefont {Hammer},\ and\ \citenamefont {Mikeska}}]{mech_reason}%
  \BibitemOpen
  \bibfield  {author} {\bibinfo {author} {\bibfnamefont {R.~S.}\ \bibnamefont {Russ}}, \bibinfo {author} {\bibfnamefont {R.~E.}\ \bibnamefont {Scherr}}, \bibinfo {author} {\bibfnamefont {D.}~\bibnamefont {Hammer}},\ and\ \bibinfo {author} {\bibfnamefont {J.}~\bibnamefont {Mikeska}},\ }\bibfield  {title} {\bibinfo {title} {Recognizing mechanistic reasoning in student scientific inquiry: A framework for discourse analysis developed from philosophy of science},\ }\href {https://doi.org/https://doi.org/10.1002/sce.20264} {\bibfield  {journal} {\bibinfo  {journal} {Science Education}\ }\textbf {\bibinfo {volume} {92}},\ \bibinfo {pages} {499} (\bibinfo {year} {2008})}\BibitemShut {NoStop}%
\bibitem [{\citenamefont {Odden}\ and\ \citenamefont {Russ}(2018)}]{PhysRevPhysEducRes.14.020122}%
  \BibitemOpen
  \bibfield  {author} {\bibinfo {author} {\bibfnamefont {T.~O.~B.}\ \bibnamefont {Odden}}\ and\ \bibinfo {author} {\bibfnamefont {R.~S.}\ \bibnamefont {Russ}},\ }\bibfield  {title} {\bibinfo {title} {Sensemaking epistemic game: A model of student sensemaking processes in introductory physics},\ }\href {https://doi.org/10.1103/PhysRevPhysEducRes.14.020122} {\bibfield  {journal} {\bibinfo  {journal} {Phys. Rev. Phys. Educ. Res.}\ }\textbf {\bibinfo {volume} {14}},\ \bibinfo {pages} {020122} (\bibinfo {year} {2018})}\BibitemShut {NoStop}%
\bibitem [{\citenamefont {Long}\ \emph {et~al.}(2024)\citenamefont {Long}, \citenamefont {Luo},\ and\ \citenamefont {Zhang}}]{ChatGPT_Ed}%
  \BibitemOpen
  \bibfield  {author} {\bibinfo {author} {\bibfnamefont {Y.}~\bibnamefont {Long}}, \bibinfo {author} {\bibfnamefont {H.}~\bibnamefont {Luo}},\ and\ \bibinfo {author} {\bibfnamefont {Y.}~\bibnamefont {Zhang}},\ }\bibfield  {title} {\bibinfo {title} {Evaluating large language models in analysing classroom dialogue.},\ }\href {https://doi.org/10.1038/s41539-024-00273-3} {\bibfield  {journal} {\bibinfo  {journal} {npj Sci. Learn.}\ } (\bibinfo {year} {2024})}\BibitemShut {NoStop}%
\bibitem [{\citenamefont {He}\ \emph {et~al.}(2025)\citenamefont {He}, \citenamefont {Naphade},\ and\ \citenamefont {Huang}}]{prompt2}%
  \BibitemOpen
  \bibfield  {author} {\bibinfo {author} {\bibfnamefont {Z.}~\bibnamefont {He}}, \bibinfo {author} {\bibfnamefont {S.}~\bibnamefont {Naphade}},\ and\ \bibinfo {author} {\bibfnamefont {T.-H.~K.}\ \bibnamefont {Huang}},\ }\bibfield  {title} {\bibinfo {title} {Prompting in the dark: Assessing human performance in prompt engineering for data labeling when gold labels are absent},\ }in\ \href {https://doi.org/10.1145/3706598.3714319} {\emph {\bibinfo {booktitle} {Proceedings of the 2025 CHI Conference on Human Factors in Computing Systems}}},\ \bibinfo {series and number} {CHI '25}\ (\bibinfo  {publisher} {Association for Computing Machinery},\ \bibinfo {address} {New York, NY, USA},\ \bibinfo {year} {2025})\BibitemShut {NoStop}%
\bibitem [{\citenamefont {Subramonyam}\ \emph {et~al.}(2024)\citenamefont {Subramonyam}, \citenamefont {Pea}, \citenamefont {Pondoc}, \citenamefont {Agrawala},\ and\ \citenamefont {Seifert}}]{prompt3}%
  \BibitemOpen
  \bibfield  {author} {\bibinfo {author} {\bibfnamefont {H.}~\bibnamefont {Subramonyam}}, \bibinfo {author} {\bibfnamefont {R.}~\bibnamefont {Pea}}, \bibinfo {author} {\bibfnamefont {C.}~\bibnamefont {Pondoc}}, \bibinfo {author} {\bibfnamefont {M.}~\bibnamefont {Agrawala}},\ and\ \bibinfo {author} {\bibfnamefont {C.}~\bibnamefont {Seifert}},\ }\bibfield  {title} {\bibinfo {title} {Bridging the gulf of envisioning: Cognitive challenges in prompt based interactions with llms},\ }in\ \href {https://doi.org/10.1145/3613904.3642754} {\emph {\bibinfo {booktitle} {Proceedings of the 2024 CHI Conference on Human Factors in Computing Systems}}},\ \bibinfo {series and number} {CHI '24}\ (\bibinfo  {publisher} {Association for Computing Machinery},\ \bibinfo {address} {New York, NY, USA},\ \bibinfo {year} {2024})\BibitemShut {NoStop}%
\bibitem [{\citenamefont {Rudin}(2019)}]{Rudin2019StopExplaining}%
  \BibitemOpen
  \bibfield  {author} {\bibinfo {author} {\bibfnamefont {C.}~\bibnamefont {Rudin}},\ }\bibfield  {title} {\bibinfo {title} {Stop explaining black box machine learning models for high stakes decisions and use interpretable models instead},\ }\href {https://doi.org/10.1038/s42256-019-0048-x} {\bibfield  {journal} {\bibinfo  {journal} {Nature Machine Intelligence}\ }\textbf {\bibinfo {volume} {1}},\ \bibinfo {pages} {206} (\bibinfo {year} {2019})},\ \bibinfo {note} {received 30 Dec 2018; accepted 26 Mar 2019; published 13 May 2019}\BibitemShut {NoStop}%
\bibitem [{\citenamefont {Zschech}\ \emph {et~al.}(2025)\citenamefont {Zschech}, \citenamefont {Weinzierl},\ and\ \citenamefont {Kraus}}]{Zschech2025Inherently}%
  \BibitemOpen
  \bibfield  {author} {\bibinfo {author} {\bibfnamefont {P.}~\bibnamefont {Zschech}}, \bibinfo {author} {\bibfnamefont {S.}~\bibnamefont {Weinzierl}},\ and\ \bibinfo {author} {\bibfnamefont {M.}~\bibnamefont {Kraus}},\ }\bibfield  {title} {\bibinfo {title} {Inherently interpretable machine learning: A contrasting paradigm to post-hoc explainable ai},\ }\href {https://doi.org/10.1007/s12599-025-00964-0} {\bibfield  {journal} {\bibinfo  {journal} {Business \& Information Systems Engineering}\ ,\ \bibinfo {pages} {1}} (\bibinfo {year} {2025})},\ \bibinfo {note} {received 17 Dec 2024; accepted 24 Jul 2025; published 15 Sep 2025}\BibitemShut {NoStop}%
\bibitem [{\citenamefont {Ghahramani}\ and\ \citenamefont {Hinton}(2000)}]{ghahramaniSwitchingState2000}%
  \BibitemOpen
  \bibfield  {author} {\bibinfo {author} {\bibfnamefont {Z.}~\bibnamefont {Ghahramani}}\ and\ \bibinfo {author} {\bibfnamefont {G.~E.}\ \bibnamefont {Hinton}},\ }\bibfield  {title} {\bibinfo {title} {Variational learning for switching state-space models},\ }\href@noop {} {\bibfield  {journal} {\bibinfo  {journal} {Neural computation}\ }\textbf {\bibinfo {volume} {12}} (\bibinfo {year} {2000})}\BibitemShut {NoStop}%
\bibitem [{\citenamefont {Farnoosh}\ \emph {et~al.}(2021)\citenamefont {Farnoosh}, \citenamefont {Azari},\ and\ \citenamefont {Ostadabbas}}]{dsarf}%
  \BibitemOpen
  \bibfield  {author} {\bibinfo {author} {\bibfnamefont {A.}~\bibnamefont {Farnoosh}}, \bibinfo {author} {\bibfnamefont {B.}~\bibnamefont {Azari}},\ and\ \bibinfo {author} {\bibfnamefont {S.}~\bibnamefont {Ostadabbas}},\ }\bibfield  {title} {\bibinfo {title} {Deep switching auto-regressive factorization: Application to time series forecasting},\ }\href {https://doi.org/10.1609/aaai.v35i8.16907} {\bibfield  {journal} {\bibinfo  {journal} {Proceedings of the AAAI Conference on Artificial Intelligence}\ }\textbf {\bibinfo {volume} {35}},\ \bibinfo {pages} {7394} (\bibinfo {year} {2021})}\BibitemShut {NoStop}%
\bibitem [{\citenamefont {Wojnowicz}\ \emph {et~al.}(2024)\citenamefont {Wojnowicz}, \citenamefont {Gili}, \citenamefont {Rath}, \citenamefont {Miller}, \citenamefont {Miller}, \citenamefont {Hancock}, \citenamefont {O'Donovan}, \citenamefont {Elkin-Frankston}, \citenamefont {Brunyé},\ and\ \citenamefont {Hughes}}]{wojnowicz2024discovering}%
  \BibitemOpen
  \bibfield  {author} {\bibinfo {author} {\bibfnamefont {M.~T.}\ \bibnamefont {Wojnowicz}}, \bibinfo {author} {\bibfnamefont {K.}~\bibnamefont {Gili}}, \bibinfo {author} {\bibfnamefont {P.}~\bibnamefont {Rath}}, \bibinfo {author} {\bibfnamefont {E.}~\bibnamefont {Miller}}, \bibinfo {author} {\bibfnamefont {J.}~\bibnamefont {Miller}}, \bibinfo {author} {\bibfnamefont {C.}~\bibnamefont {Hancock}}, \bibinfo {author} {\bibfnamefont {M.}~\bibnamefont {O'Donovan}}, \bibinfo {author} {\bibfnamefont {S.}~\bibnamefont {Elkin-Frankston}}, \bibinfo {author} {\bibfnamefont {T.~T.}\ \bibnamefont {Brunyé}},\ and\ \bibinfo {author} {\bibfnamefont {M.~C.}\ \bibnamefont {Hughes}},\ }\href {https://arxiv.org/abs/2401.14973} {\bibinfo {title} {Discovering group dynamics in coordinated time series via hierarchical recurrent switching-state models}} (\bibinfo {year} {2024})\BibitemShut {NoStop}%
\bibitem [{\citenamefont {Bachtiar}\ \emph {et~al.}(2021)\citenamefont {Bachtiar}, \citenamefont {Meulenbroeks},\ and\ \citenamefont {van Joolingen}}]{Bachtiar}%
  \BibitemOpen
  \bibfield  {author} {\bibinfo {author} {\bibfnamefont {R.~W.}\ \bibnamefont {Bachtiar}}, \bibinfo {author} {\bibfnamefont {R.~F.~G.}\ \bibnamefont {Meulenbroeks}},\ and\ \bibinfo {author} {\bibfnamefont {W.~R.}\ \bibnamefont {van Joolingen}},\ }\bibfield  {title} {\bibinfo {title} {Stimulating mechanistic reasoning in physics using student-constructed stop-motion animations},\ }\href {https://doi.org/10.1007/s10956-021-09918-z} {\bibfield  {journal} {\bibinfo  {journal} {Journal of Science Education and Technology}\ }\textbf {\bibinfo {volume} {30}},\ \bibinfo {pages} {777} (\bibinfo {year} {2021})},\ \bibinfo {note} {published 28 Apr 2021; issue date Dec 2021}\BibitemShut {NoStop}%
\bibitem [{\citenamefont {Wilkerson-Jerde}\ \emph {et~al.}(2015)\citenamefont {Wilkerson-Jerde}, \citenamefont {Gravel},\ and\ \citenamefont {Macrander}}]{Wilkerson}%
  \BibitemOpen
  \bibfield  {author} {\bibinfo {author} {\bibfnamefont {M.~H.}\ \bibnamefont {Wilkerson-Jerde}}, \bibinfo {author} {\bibfnamefont {B.~E.}\ \bibnamefont {Gravel}},\ and\ \bibinfo {author} {\bibfnamefont {C.~A.}\ \bibnamefont {Macrander}},\ }\bibfield  {title} {\bibinfo {title} {Exploring shifts in middle school learners’ modeling activity while generating drawings, animations, and computational simulations of molecular diffusion},\ }\href {https://doi.org/10.1007/s10956-014-9497-5} {\bibfield  {journal} {\bibinfo  {journal} {Journal of Science Education and Technology}\ }\textbf {\bibinfo {volume} {24}},\ \bibinfo {pages} {396} (\bibinfo {year} {2015})}\BibitemShut {NoStop}%
\bibitem [{\citenamefont {Wendell}\ \emph {et~al.}(2025)\citenamefont {Wendell}, \citenamefont {Top{\c{c}}u},\ and\ \citenamefont {Andrews}}]{Wendell2025MechanisticReasoning}%
  \BibitemOpen
  \bibfield  {author} {\bibinfo {author} {\bibfnamefont {K.}~\bibnamefont {Wendell}}, \bibinfo {author} {\bibfnamefont {M.~S.}\ \bibnamefont {Top{\c{c}}u}},\ and\ \bibinfo {author} {\bibfnamefont {C.}~\bibnamefont {Andrews}},\ }\bibfield  {title} {\bibinfo {title} {Mechanistic reasoning: How cause-and-effect thinking supports engineering design problem-solving},\ }\href {https://doi.org/10.1080/00368148.2025.2504136} {\bibfield  {journal} {\bibinfo  {journal} {Science and Children}\ }\textbf {\bibinfo {volume} {62}},\ \bibinfo {pages} {44} (\bibinfo {year} {2025})}\BibitemShut {NoStop}%
\bibitem [{\citenamefont {Melsky}\ \emph {et~al.}(2024)\citenamefont {Melsky}, \citenamefont {Stuopis}, \citenamefont {Wendell},\ and\ \citenamefont {Kemmerling}}]{Melsky2024}%
  \BibitemOpen
  \bibfield  {author} {\bibinfo {author} {\bibfnamefont {K.}~\bibnamefont {Melsky}}, \bibinfo {author} {\bibfnamefont {I.}~\bibnamefont {Stuopis}}, \bibinfo {author} {\bibfnamefont {K.}~\bibnamefont {Wendell}},\ and\ \bibinfo {author} {\bibfnamefont {E.~C.}\ \bibnamefont {Kemmerling}},\ }\bibfield  {title} {\bibinfo {title} {Personalized problems and student discourse in thermal fluid transport courses},\ }\href {https://doi.org/10.1177/03064190231195609} {\bibfield  {journal} {\bibinfo  {journal} {International Journal of Mechanical Engineering Education}\ }\textbf {\bibinfo {volume} {52}},\ \bibinfo {pages} {457} (\bibinfo {year} {2024})}\BibitemShut {NoStop}%
\bibitem [{\citenamefont {Devlin}\ \emph {et~al.}(2019)\citenamefont {Devlin}, \citenamefont {Chang}, \citenamefont {Lee},\ and\ \citenamefont {Toutanova}}]{BERT}%
  \BibitemOpen
  \bibfield  {author} {\bibinfo {author} {\bibfnamefont {J.}~\bibnamefont {Devlin}}, \bibinfo {author} {\bibfnamefont {M.-W.}\ \bibnamefont {Chang}}, \bibinfo {author} {\bibfnamefont {K.}~\bibnamefont {Lee}},\ and\ \bibinfo {author} {\bibfnamefont {K.}~\bibnamefont {Toutanova}},\ }\href {https://arxiv.org/abs/1810.04805} {\bibinfo {title} {Bert: Pre-training of deep bidirectional transformers for language understanding}} (\bibinfo {year} {2019})\BibitemShut {NoStop}%
\bibitem [{\citenamefont {Touvron}\ \emph {et~al.}(2023)\citenamefont {Touvron}, \citenamefont {Lavril}, \citenamefont {Izacard}, \citenamefont {Martinet}, \citenamefont {Lachaux}, \citenamefont {Lacroix}, \citenamefont {Rozière}, \citenamefont {Goyal}, \citenamefont {Hambro}, \citenamefont {Azhar}, \citenamefont {Rodriguez}, \citenamefont {Joulin}, \citenamefont {Grave},\ and\ \citenamefont {Lample}}]{touvron_llama_2023}%
  \BibitemOpen
  \bibfield  {author} {\bibinfo {author} {\bibfnamefont {H.}~\bibnamefont {Touvron}}, \bibinfo {author} {\bibfnamefont {T.}~\bibnamefont {Lavril}}, \bibinfo {author} {\bibfnamefont {G.}~\bibnamefont {Izacard}}, \bibinfo {author} {\bibfnamefont {X.}~\bibnamefont {Martinet}}, \bibinfo {author} {\bibfnamefont {M.-A.}\ \bibnamefont {Lachaux}}, \bibinfo {author} {\bibfnamefont {T.}~\bibnamefont {Lacroix}}, \bibinfo {author} {\bibfnamefont {B.}~\bibnamefont {Rozière}}, \bibinfo {author} {\bibfnamefont {N.}~\bibnamefont {Goyal}}, \bibinfo {author} {\bibfnamefont {E.}~\bibnamefont {Hambro}}, \bibinfo {author} {\bibfnamefont {F.}~\bibnamefont {Azhar}}, \bibinfo {author} {\bibfnamefont {A.}~\bibnamefont {Rodriguez}}, \bibinfo {author} {\bibfnamefont {A.}~\bibnamefont {Joulin}}, \bibinfo {author} {\bibfnamefont {E.}~\bibnamefont {Grave}},\ and\ \bibinfo {author} {\bibfnamefont {G.}~\bibnamefont {Lample}},\ }\href {http://arxiv.org/abs/2302.13971} {\bibinfo {title} {Llama: Open and efficient foundation language
  models}},\ \bibinfo {howpublished} {arXiv preprint arXiv:2302.13971} (\bibinfo {year} {2023})\BibitemShut {NoStop}%
\bibitem [{\citenamefont {Jiang}\ \emph {et~al.}(2023)\citenamefont {Jiang}, \citenamefont {Sablayrolles}, \citenamefont {Mensch}, \citenamefont {Bamford}, \citenamefont {Chaplot}, \citenamefont {de~las Casas}, \citenamefont {Bressand}, \citenamefont {Lengyel}, \citenamefont {Lample}, \citenamefont {Saulnier}, \citenamefont {Lavaud}, \citenamefont {Lachaux}, \citenamefont {Stock}, \citenamefont {Scao}, \citenamefont {Lavril}, \citenamefont {Wang}, \citenamefont {Lacroix},\ and\ \citenamefont {Sayed}}]{jiang2023mistral7b}%
  \BibitemOpen
  \bibfield  {author} {\bibinfo {author} {\bibfnamefont {A.~Q.}\ \bibnamefont {Jiang}}, \bibinfo {author} {\bibfnamefont {A.}~\bibnamefont {Sablayrolles}}, \bibinfo {author} {\bibfnamefont {A.}~\bibnamefont {Mensch}}, \bibinfo {author} {\bibfnamefont {C.}~\bibnamefont {Bamford}}, \bibinfo {author} {\bibfnamefont {D.~S.}\ \bibnamefont {Chaplot}}, \bibinfo {author} {\bibfnamefont {D.}~\bibnamefont {de~las Casas}}, \bibinfo {author} {\bibfnamefont {F.}~\bibnamefont {Bressand}}, \bibinfo {author} {\bibfnamefont {G.}~\bibnamefont {Lengyel}}, \bibinfo {author} {\bibfnamefont {G.}~\bibnamefont {Lample}}, \bibinfo {author} {\bibfnamefont {L.}~\bibnamefont {Saulnier}}, \bibinfo {author} {\bibfnamefont {L.~R.}\ \bibnamefont {Lavaud}}, \bibinfo {author} {\bibfnamefont {M.-A.}\ \bibnamefont {Lachaux}}, \bibinfo {author} {\bibfnamefont {P.}~\bibnamefont {Stock}}, \bibinfo {author} {\bibfnamefont {T.~L.}\ \bibnamefont {Scao}}, \bibinfo {author} {\bibfnamefont {T.}~\bibnamefont {Lavril}}, \bibinfo {author} {\bibfnamefont
  {T.}~\bibnamefont {Wang}}, \bibinfo {author} {\bibfnamefont {T.}~\bibnamefont {Lacroix}},\ and\ \bibinfo {author} {\bibfnamefont {W.~E.}\ \bibnamefont {Sayed}},\ }\href {https://arxiv.org/abs/2310.06825} {\bibinfo {title} {Mistral 7b}} (\bibinfo {year} {2023})\BibitemShut {NoStop}%
\bibitem [{\citenamefont {Campbell}\ \emph {et~al.}(2024)\citenamefont {Campbell}, \citenamefont {Ansell},\ and\ \citenamefont {Stelzer}}]{PhysRevPhysEducRes.20.010116}%
  \BibitemOpen
  \bibfield  {author} {\bibinfo {author} {\bibfnamefont {J.}~\bibnamefont {Campbell}}, \bibinfo {author} {\bibfnamefont {K.}~\bibnamefont {Ansell}},\ and\ \bibinfo {author} {\bibfnamefont {T.}~\bibnamefont {Stelzer}},\ }\bibfield  {title} {\bibinfo {title} {Evaluating ibm's watson natural language processing artificial intelligence as a short-answer categorization tool for physics education research},\ }\href {https://doi.org/10.1103/PhysRevPhysEducRes.20.010116} {\bibfield  {journal} {\bibinfo  {journal} {Phys. Rev. Phys. Educ. Res.}\ }\textbf {\bibinfo {volume} {20}},\ \bibinfo {pages} {010116} (\bibinfo {year} {2024})}\BibitemShut {NoStop}%
\bibitem [{\citenamefont {Wilson}\ \emph {et~al.}(2022)\citenamefont {Wilson}, \citenamefont {Pollard}, \citenamefont {Aiken}, \citenamefont {Caballero},\ and\ \citenamefont {Lewandowski}}]{PhysRevPhysEducRes.18.010141}%
  \BibitemOpen
  \bibfield  {author} {\bibinfo {author} {\bibfnamefont {J.}~\bibnamefont {Wilson}}, \bibinfo {author} {\bibfnamefont {B.}~\bibnamefont {Pollard}}, \bibinfo {author} {\bibfnamefont {J.~M.}\ \bibnamefont {Aiken}}, \bibinfo {author} {\bibfnamefont {M.~D.}\ \bibnamefont {Caballero}},\ and\ \bibinfo {author} {\bibfnamefont {H.~J.}\ \bibnamefont {Lewandowski}},\ }\bibfield  {title} {\bibinfo {title} {Classification of open-ended responses to a research-based assessment using natural language processing},\ }\href {https://doi.org/10.1103/PhysRevPhysEducRes.18.010141} {\bibfield  {journal} {\bibinfo  {journal} {Phys. Rev. Phys. Educ. Res.}\ }\textbf {\bibinfo {volume} {18}},\ \bibinfo {pages} {010141} (\bibinfo {year} {2022})}\BibitemShut {NoStop}%
\bibitem [{\citenamefont {Wang}\ \emph {et~al.}(2023)\citenamefont {Wang}, \citenamefont {Shan}, \citenamefont {Zheng}, \citenamefont {Guo}, \citenamefont {Chen},\ and\ \citenamefont {Lu}}]{2023.EDM-posters.59}%
  \BibitemOpen
  \bibfield  {author} {\bibinfo {author} {\bibfnamefont {D.}~\bibnamefont {Wang}}, \bibinfo {author} {\bibfnamefont {D.}~\bibnamefont {Shan}}, \bibinfo {author} {\bibfnamefont {Y.}~\bibnamefont {Zheng}}, \bibinfo {author} {\bibfnamefont {K.}~\bibnamefont {Guo}}, \bibinfo {author} {\bibfnamefont {G.}~\bibnamefont {Chen}},\ and\ \bibinfo {author} {\bibfnamefont {Y.}~\bibnamefont {Lu}},\ }\bibfield  {title} {\bibinfo {title} {Can chatgpt detect student talk moves in classroom discourse? a preliminary comparison with bert},\ }in\ \href {https://doi.org/10.5281/zenodo.8115772} {\emph {\bibinfo {booktitle} {Proceedings of the 16th International Conference on Educational Data Mining}}},\ \bibinfo {editor} {edited by\ \bibinfo {editor} {\bibfnamefont {M.}~\bibnamefont {Feng}}, \bibinfo {editor} {\bibfnamefont {T.}~\bibnamefont {KÃ¤ser}},\ and\ \bibinfo {editor} {\bibfnamefont {P.}~\bibnamefont {Talukdar}}}\ (\bibinfo  {publisher} {International Educational Data Mining Society},\ \bibinfo {address} {Bengaluru,
  India},\ \bibinfo {year} {2023})\ pp.\ \bibinfo {pages} {515--519}\BibitemShut {NoStop}%
\bibitem [{\citenamefont {Fussell}\ \emph {et~al.}(2025)\citenamefont {Fussell}, \citenamefont {Flynn}, \citenamefont {Damle}, \citenamefont {Fox},\ and\ \citenamefont {Holmes}}]{PhysRevPhysEducRes.21.010128}%
  \BibitemOpen
  \bibfield  {author} {\bibinfo {author} {\bibfnamefont {R.~K.}\ \bibnamefont {Fussell}}, \bibinfo {author} {\bibfnamefont {M.}~\bibnamefont {Flynn}}, \bibinfo {author} {\bibfnamefont {A.}~\bibnamefont {Damle}}, \bibinfo {author} {\bibfnamefont {M.~F.~J.}\ \bibnamefont {Fox}},\ and\ \bibinfo {author} {\bibfnamefont {N.~G.}\ \bibnamefont {Holmes}},\ }\bibfield  {title} {\bibinfo {title} {Comparing large language models for supervised analysis of students' lab notes},\ }\href {https://doi.org/10.1103/PhysRevPhysEducRes.21.010128} {\bibfield  {journal} {\bibinfo  {journal} {Phys. Rev. Phys. Educ. Res.}\ }\textbf {\bibinfo {volume} {21}},\ \bibinfo {pages} {010128} (\bibinfo {year} {2025})}\BibitemShut {NoStop}%
\bibitem [{\citenamefont {Gili}\ \emph {et~al.}(2025)\citenamefont {Gili}, \citenamefont {Heuton}, \citenamefont {Shah}, \citenamefont {Hammer},\ and\ \citenamefont {Hughes}}]{gili2025}%
  \BibitemOpen
  \bibfield  {author} {\bibinfo {author} {\bibfnamefont {K.}~\bibnamefont {Gili}}, \bibinfo {author} {\bibfnamefont {K.}~\bibnamefont {Heuton}}, \bibinfo {author} {\bibfnamefont {A.}~\bibnamefont {Shah}}, \bibinfo {author} {\bibfnamefont {D.}~\bibnamefont {Hammer}},\ and\ \bibinfo {author} {\bibfnamefont {M.~C.}\ \bibnamefont {Hughes}},\ }\href {https://arxiv.org/abs/2503.15638} {\bibinfo {title} {Combining physics education and machine learning research to measure evidence of students' mechanistic sensemaking}} (\bibinfo {year} {2025}),\ \Eprint {https://arxiv.org/abs/2503.15638} {arXiv:2503.15638 [physics.ed-ph]} \BibitemShut {NoStop}%
\bibitem [{\citenamefont {Ullmann}(2019)}]{automated_reflections}%
  \BibitemOpen
  \bibfield  {author} {\bibinfo {author} {\bibfnamefont {T.}~\bibnamefont {Ullmann}},\ }\bibfield  {title} {\bibinfo {title} {Automated analysis of reflection in writing: Validating machine learning approaches},\ }\href {https://doi.org/10.1007/s40593-019-00174-2} {\bibfield  {journal} {\bibinfo  {journal} {Int J Artif Intell Educ}\ }\textbf {\bibinfo {volume} {29}},\ \bibinfo {pages} {217–257} (\bibinfo {year} {2019})}\BibitemShut {NoStop}%
\bibitem [{\citenamefont {Auby}\ \emph {et~al.}(2025)\citenamefont {Auby}, \citenamefont {Shivagunde}, \citenamefont {Deshpande}, \citenamefont {Rumshisky},\ and\ \citenamefont {Koretsky}}]{harpreet_ml}%
  \BibitemOpen
  \bibfield  {author} {\bibinfo {author} {\bibfnamefont {H.}~\bibnamefont {Auby}}, \bibinfo {author} {\bibfnamefont {N.}~\bibnamefont {Shivagunde}}, \bibinfo {author} {\bibfnamefont {V.}~\bibnamefont {Deshpande}}, \bibinfo {author} {\bibfnamefont {A.}~\bibnamefont {Rumshisky}},\ and\ \bibinfo {author} {\bibfnamefont {M.~D.}\ \bibnamefont {Koretsky}},\ }\bibfield  {title} {\bibinfo {title} {Analysis of student understanding in short-answer explanations to concept questions using a human-centered ai approach},\ }\href {https://doi.org/https://doi.org/10.1002/jee.70032} {\bibfield  {journal} {\bibinfo  {journal} {Journal of Engineering Education}\ }\textbf {\bibinfo {volume} {114}},\ \bibinfo {pages} {e70032} (\bibinfo {year} {2025})},\ \Eprint {https://arxiv.org/abs/https://onlinelibrary.wiley.com/doi/pdf/10.1002/jee.70032} {https://onlinelibrary.wiley.com/doi/pdf/10.1002/jee.70032} \BibitemShut {NoStop}%
\bibitem [{\citenamefont {Zamfirescu-Pereira}\ \emph {et~al.}(2023)\citenamefont {Zamfirescu-Pereira}, \citenamefont {Wong}, \citenamefont {Hartmann},\ and\ \citenamefont {Yang}}]{prompting}%
  \BibitemOpen
  \bibfield  {author} {\bibinfo {author} {\bibfnamefont {J.}~\bibnamefont {Zamfirescu-Pereira}}, \bibinfo {author} {\bibfnamefont {R.~Y.}\ \bibnamefont {Wong}}, \bibinfo {author} {\bibfnamefont {B.}~\bibnamefont {Hartmann}},\ and\ \bibinfo {author} {\bibfnamefont {Q.}~\bibnamefont {Yang}},\ }\bibfield  {title} {\bibinfo {title} {Why johnny can’t prompt: How non-ai experts try (and fail) to design llm prompts}\ }(\bibinfo  {publisher} {Association for Computing Machinery},\ \bibinfo {address} {New York, NY, USA},\ \bibinfo {year} {2023})\BibitemShut {NoStop}%
\bibitem [{\citenamefont {Jiang}\ \emph {et~al.}(2020)\citenamefont {Jiang}, \citenamefont {Xu}, \citenamefont {Araki},\ and\ \citenamefont {Neubig}}]{jiangHowCanWe2020}%
  \BibitemOpen
  \bibfield  {author} {\bibinfo {author} {\bibfnamefont {Z.}~\bibnamefont {Jiang}}, \bibinfo {author} {\bibfnamefont {F.~F.}\ \bibnamefont {Xu}}, \bibinfo {author} {\bibfnamefont {J.}~\bibnamefont {Araki}},\ and\ \bibinfo {author} {\bibfnamefont {G.}~\bibnamefont {Neubig}},\ }\bibfield  {title} {\bibinfo {title} {How {{Can We Know What Language Models Know}}?},\ }\href {https://aclanthology.org/2020.tacl-1.28/} {\bibfield  {journal} {\bibinfo  {journal} {Transactions of the Association for Computational Linguistics}\ }\textbf {\bibinfo {volume} {8}},\ \bibinfo {pages} {423} (\bibinfo {year} {2020})}\BibitemShut {NoStop}%
\bibitem [{\citenamefont {Liu}\ \emph {et~al.}(2022)\citenamefont {Liu}, \citenamefont {Shen}, \citenamefont {Zhang}, \citenamefont {Dolan}, \citenamefont {Carin},\ and\ \citenamefont {Chen}}]{liuWhatMakesGood2022}%
  \BibitemOpen
  \bibfield  {author} {\bibinfo {author} {\bibfnamefont {J.}~\bibnamefont {Liu}}, \bibinfo {author} {\bibfnamefont {D.}~\bibnamefont {Shen}}, \bibinfo {author} {\bibfnamefont {Y.}~\bibnamefont {Zhang}}, \bibinfo {author} {\bibfnamefont {B.}~\bibnamefont {Dolan}}, \bibinfo {author} {\bibfnamefont {L.}~\bibnamefont {Carin}},\ and\ \bibinfo {author} {\bibfnamefont {W.}~\bibnamefont {Chen}},\ }\bibfield  {title} {\bibinfo {title} {What {{Makes Good In-Context Examples}} for {{GPT-3}}?},\ }in\ \href {https://aclanthology.org/2022.deelio-1.10/} {\emph {\bibinfo {booktitle} {Proceedings of {{Deep Learning Inside Out}} ({{DeeLIO}} 2022): {{The}} 3rd {{Workshop}} on {{Knowledge Extraction}} and {{Integration}} for {{Deep Learning Architectures}}}}},\ \bibinfo {editor} {edited by\ \bibinfo {editor} {\bibfnamefont {E.}~\bibnamefont {Agirre}}, \bibinfo {editor} {\bibfnamefont {M.}~\bibnamefont {Apidianaki}},\ and\ \bibinfo {editor} {\bibfnamefont {I.}~\bibnamefont {Vuli{\'c}}}}\ (\bibinfo  {publisher} {Association for
  Computational Linguistics},\ \bibinfo {address} {Dublin, Ireland and Online},\ \bibinfo {year} {2022})\ pp.\ \bibinfo {pages} {100--114}\BibitemShut {NoStop}%
\bibitem [{\citenamefont {Bender}\ \emph {et~al.}(2021)\citenamefont {Bender}, \citenamefont {Gebru}, \citenamefont {McMillan-Major},\ and\ \citenamefont {Shmitchell}}]{stochastic_parrots}%
  \BibitemOpen
  \bibfield  {author} {\bibinfo {author} {\bibfnamefont {E.~M.}\ \bibnamefont {Bender}}, \bibinfo {author} {\bibfnamefont {T.}~\bibnamefont {Gebru}}, \bibinfo {author} {\bibfnamefont {A.}~\bibnamefont {McMillan-Major}},\ and\ \bibinfo {author} {\bibfnamefont {S.}~\bibnamefont {Shmitchell}},\ }\bibfield  {title} {\bibinfo {title} {On the dangers of stochastic parrots: Can language models be too big?},\ }in\ \href {https://doi.org/10.1145/3442188.3445922} {\emph {\bibinfo {booktitle} {Proceedings of the 2021 ACM Conference on Fairness, Accountability, and Transparency}}},\ \bibinfo {series and number} {FAccT '21}\ (\bibinfo  {publisher} {Association for Computing Machinery},\ \bibinfo {address} {New York, NY, USA},\ \bibinfo {year} {2021})\ p.\ \bibinfo {pages} {610–623}\BibitemShut {NoStop}%
\bibitem [{\citenamefont {Watts}\ \emph {et~al.}(2022)\citenamefont {Watts}, \citenamefont {Dood},\ and\ \citenamefont {Shultz}}]{org_chem_ml}%
  \BibitemOpen
  \bibfield  {author} {\bibinfo {author} {\bibfnamefont {F.~M.}\ \bibnamefont {Watts}}, \bibinfo {author} {\bibfnamefont {A.~J.}\ \bibnamefont {Dood}},\ and\ \bibinfo {author} {\bibfnamefont {G.~V.}\ \bibnamefont {Shultz}},\ }\bibfield  {title} {\bibinfo {title} {Developing machine learning models for automated analysis of organic chemistry students’ written descriptions of organic reaction mechanisms},\ }in\ \href {https://doi.org/10.1039/9781839167782-00285} {\emph {\bibinfo {booktitle} {Student Reasoning in Organic Chemistry}}}\ (\bibinfo  {publisher} {The Royal Society of Chemistry},\ \bibinfo {year} {2022})\BibitemShut {NoStop}%
\bibitem [{\citenamefont {Moreau-Pernet}\ \emph {et~al.}(2024)\citenamefont {Moreau-Pernet}, \citenamefont {Tian}, \citenamefont {Sawaya}, \citenamefont {Foltz}, \citenamefont {Cao}, \citenamefont {Milne},\ and\ \citenamefont {Christie}}]{moreau_pernet2024_talkmoves}%
  \BibitemOpen
  \bibfield  {author} {\bibinfo {author} {\bibfnamefont {B.}~\bibnamefont {Moreau-Pernet}}, \bibinfo {author} {\bibfnamefont {Y.}~\bibnamefont {Tian}}, \bibinfo {author} {\bibfnamefont {S.}~\bibnamefont {Sawaya}}, \bibinfo {author} {\bibfnamefont {P.}~\bibnamefont {Foltz}}, \bibinfo {author} {\bibfnamefont {J.}~\bibnamefont {Cao}}, \bibinfo {author} {\bibfnamefont {B.}~\bibnamefont {Milne}},\ and\ \bibinfo {author} {\bibfnamefont {T.}~\bibnamefont {Christie}},\ }\bibfield  {title} {\bibinfo {title} {Classifying tutor discursive moves at scale in mathematics classrooms with large language models},\ }in\ \href {https://doi.org/10.1145/3657604.3664664} {\emph {\bibinfo {booktitle} {Proceedings of the Eleventh ACM Conference on Learning @ Scale (L@S '24)}}}\ (\bibinfo  {publisher} {Association for Computing Machinery},\ \bibinfo {address} {Atlanta, GA, USA},\ \bibinfo {year} {2024})\ pp.\ \bibinfo {pages} {361--365}\BibitemShut {NoStop}%
\bibitem [{\citenamefont {Schechter}\ \emph {et~al.}(2025)\citenamefont {Schechter}, \citenamefont {Dua}, \citenamefont {Dua}, \citenamefont {Zhang}, \citenamefont {Salz}, \citenamefont {Mullins}, \citenamefont {Panyam}, \citenamefont {Smoot}, \citenamefont {Naim}, \citenamefont {Zou}, \citenamefont {Chen}, \citenamefont {Cer}, \citenamefont {Lisak}, \citenamefont {Choi}, \citenamefont {Gonzalez}, \citenamefont {Sanseviero}, \citenamefont {Cameron}, \citenamefont {Ballantyne}, \citenamefont {Black}, \citenamefont {Chen}, \citenamefont {Wang}, \citenamefont {Li}, \citenamefont {Martins}, \citenamefont {Lee}, \citenamefont {Sherwood}, \citenamefont {Ji}, \citenamefont {Wu}, \citenamefont {Zheng}, \citenamefont {Singh}, \citenamefont {Sharma}, \citenamefont {Sreepat}, \citenamefont {Jain}, \citenamefont {Elarabawy}, \citenamefont {Co}, \citenamefont {Doumanoglou}, \citenamefont {Samari}, \citenamefont {Hora}, \citenamefont {Potetz}, \citenamefont {Kim}, \citenamefont {Alfonseca}, \citenamefont {Moiseev},
  \citenamefont {Gomez},\ and\ \citenamefont {Hernández}}]{embedding_gemma}%
  \BibitemOpen
  \bibfield  {author} {\bibinfo {author} {\bibfnamefont {V.}~\bibnamefont {Schechter}}, \bibinfo {author} {\bibfnamefont {H.}~\bibnamefont {Dua}}, \bibinfo {author} {\bibfnamefont {S.}~\bibnamefont {Dua}}, \bibinfo {author} {\bibfnamefont {B.}~\bibnamefont {Zhang}}, \bibinfo {author} {\bibfnamefont {D.}~\bibnamefont {Salz}}, \bibinfo {author} {\bibfnamefont {R.}~\bibnamefont {Mullins}}, \bibinfo {author} {\bibfnamefont {S.~R.}\ \bibnamefont {Panyam}}, \bibinfo {author} {\bibfnamefont {S.}~\bibnamefont {Smoot}}, \bibinfo {author} {\bibfnamefont {I.}~\bibnamefont {Naim}}, \bibinfo {author} {\bibfnamefont {J.}~\bibnamefont {Zou}}, \bibinfo {author} {\bibfnamefont {F.}~\bibnamefont {Chen}}, \bibinfo {author} {\bibfnamefont {D.}~\bibnamefont {Cer}}, \bibinfo {author} {\bibfnamefont {A.}~\bibnamefont {Lisak}}, \bibinfo {author} {\bibfnamefont {M.}~\bibnamefont {Choi}}, \bibinfo {author} {\bibfnamefont {L.}~\bibnamefont {Gonzalez}}, \bibinfo {author} {\bibfnamefont {O.}~\bibnamefont {Sanseviero}}, \bibinfo {author}
  {\bibfnamefont {G.}~\bibnamefont {Cameron}}, \bibinfo {author} {\bibfnamefont {I.}~\bibnamefont {Ballantyne}}, \bibinfo {author} {\bibfnamefont {K.}~\bibnamefont {Black}}, \bibinfo {author} {\bibfnamefont {K.}~\bibnamefont {Chen}}, \bibinfo {author} {\bibfnamefont {W.}~\bibnamefont {Wang}}, \bibinfo {author} {\bibfnamefont {Z.}~\bibnamefont {Li}}, \bibinfo {author} {\bibfnamefont {G.}~\bibnamefont {Martins}}, \bibinfo {author} {\bibfnamefont {J.}~\bibnamefont {Lee}}, \bibinfo {author} {\bibfnamefont {M.}~\bibnamefont {Sherwood}}, \bibinfo {author} {\bibfnamefont {J.}~\bibnamefont {Ji}}, \bibinfo {author} {\bibfnamefont {R.}~\bibnamefont {Wu}}, \bibinfo {author} {\bibfnamefont {J.}~\bibnamefont {Zheng}}, \bibinfo {author} {\bibfnamefont {J.}~\bibnamefont {Singh}}, \bibinfo {author} {\bibfnamefont {A.}~\bibnamefont {Sharma}}, \bibinfo {author} {\bibfnamefont {D.}~\bibnamefont {Sreepat}}, \bibinfo {author} {\bibfnamefont {A.}~\bibnamefont {Jain}}, \bibinfo {author} {\bibfnamefont {A.}~\bibnamefont
  {Elarabawy}}, \bibinfo {author} {\bibfnamefont {A.}~\bibnamefont {Co}}, \bibinfo {author} {\bibfnamefont {A.}~\bibnamefont {Doumanoglou}}, \bibinfo {author} {\bibfnamefont {B.}~\bibnamefont {Samari}}, \bibinfo {author} {\bibfnamefont {B.}~\bibnamefont {Hora}}, \bibinfo {author} {\bibfnamefont {B.}~\bibnamefont {Potetz}}, \bibinfo {author} {\bibfnamefont {D.}~\bibnamefont {Kim}}, \bibinfo {author} {\bibfnamefont {E.}~\bibnamefont {Alfonseca}}, \bibinfo {author} {\bibfnamefont {F.}~\bibnamefont {Moiseev}}, \bibinfo {author} {\bibfnamefont {F.~P.}\ \bibnamefont {Gomez}},\ and\ \bibinfo {author} {\bibfnamefont {G.}~\bibnamefont {Hernández}},\ }\href {https://arxiv.org/abs/2509.20354} {\bibinfo {title} {Embeddinggemma: Powerful and lightweight text representations}} (\bibinfo {year} {2025})\BibitemShut {NoStop}%
\bibitem [{\citenamefont {Kusupati}\ \emph {et~al.}(2022)\citenamefont {Kusupati}, \citenamefont {Bhatt}, \citenamefont {Rege}, \citenamefont {Wallingford}, \citenamefont {Sinha}, \citenamefont {Ramanujan}, \citenamefont {Howard-Snyder}, \citenamefont {Chen}, \citenamefont {Kakade}, \citenamefont {Jain},\ and\ \citenamefont {Farhadi}}]{kusupati2022matryoshka}%
  \BibitemOpen
  \bibfield  {author} {\bibinfo {author} {\bibfnamefont {A.}~\bibnamefont {Kusupati}}, \bibinfo {author} {\bibfnamefont {G.}~\bibnamefont {Bhatt}}, \bibinfo {author} {\bibfnamefont {A.}~\bibnamefont {Rege}}, \bibinfo {author} {\bibfnamefont {M.}~\bibnamefont {Wallingford}}, \bibinfo {author} {\bibfnamefont {A.}~\bibnamefont {Sinha}}, \bibinfo {author} {\bibfnamefont {V.}~\bibnamefont {Ramanujan}}, \bibinfo {author} {\bibfnamefont {W.}~\bibnamefont {Howard-Snyder}}, \bibinfo {author} {\bibfnamefont {K.}~\bibnamefont {Chen}}, \bibinfo {author} {\bibfnamefont {S.}~\bibnamefont {Kakade}}, \bibinfo {author} {\bibfnamefont {P.}~\bibnamefont {Jain}},\ and\ \bibinfo {author} {\bibfnamefont {A.}~\bibnamefont {Farhadi}},\ }\bibfield  {title} {\bibinfo {title} {Matryoshka representation learning},\ }in\ \href {https://proceedings.neurips.cc/paper_files/paper/2022/file/c32319f4868da7613d78af9993100e42-Paper-Conference.pdf} {\emph {\bibinfo {booktitle} {Advances in Neural Information Processing Systems (NeurIPS)}}},\
  Vol.~\bibinfo {volume} {35}\ (\bibinfo {year} {2022})\ pp.\ \bibinfo {pages} {36005--36018}\BibitemShut {NoStop}%
\bibitem [{\citenamefont {Odden}\ and\ \citenamefont {Russ}(2019)}]{definition_sensemaking}%
  \BibitemOpen
  \bibfield  {author} {\bibinfo {author} {\bibfnamefont {T.~O.~B.}\ \bibnamefont {Odden}}\ and\ \bibinfo {author} {\bibfnamefont {R.~S.}\ \bibnamefont {Russ}},\ }\bibfield  {title} {\bibinfo {title} {Defining sensemaking: Bringing clarity to a fragmented theoretical construct},\ }\href {https://doi.org/https://doi.org/10.1002/sce.21452} {\bibfield  {journal} {\bibinfo  {journal} {Science Education}\ }\textbf {\bibinfo {volume} {103}},\ \bibinfo {pages} {187} (\bibinfo {year} {2019})}\BibitemShut {NoStop}%
\bibitem [{\citenamefont {Beal}(2003)}]{beal2003variational}%
  \BibitemOpen
  \bibfield  {author} {\bibinfo {author} {\bibfnamefont {M.~J.}\ \bibnamefont {Beal}},\ }\href {https://discovery.ucl.ac.uk/id/eprint/10101435} {\emph {\bibinfo {title} {Variational algorithms for approximate Bayesian inference}}}\ (\bibinfo  {publisher} {University of London, University College London (United Kingdom)},\ \bibinfo {year} {2003})\BibitemShut {NoStop}%
\bibitem [{\citenamefont {Blei}\ \emph {et~al.}(2017)\citenamefont {Blei}, \citenamefont {Kucukelbir},\ and\ \citenamefont {McAuliffe}}]{Blei_2017}%
  \BibitemOpen
  \bibfield  {author} {\bibinfo {author} {\bibfnamefont {D.~M.}\ \bibnamefont {Blei}}, \bibinfo {author} {\bibfnamefont {A.}~\bibnamefont {Kucukelbir}},\ and\ \bibinfo {author} {\bibfnamefont {J.~D.}\ \bibnamefont {McAuliffe}},\ }\bibfield  {title} {\bibinfo {title} {Variational inference: A review for statisticians},\ }\href {https://doi.org/10.1080/01621459.2017.1285773} {\bibfield  {journal} {\bibinfo  {journal} {Journal of the American Statistical Association}\ }\textbf {\bibinfo {volume} {112}},\ \bibinfo {pages} {859–877} (\bibinfo {year} {2017})}\BibitemShut {NoStop}%
\bibitem [{\citenamefont {Harvey}\ \emph {et~al.}(2025)\citenamefont {Harvey}, \citenamefont {Petrov},\ and\ \citenamefont {Hughes}}]{harvey2025}%
  \BibitemOpen
  \bibfield  {author} {\bibinfo {author} {\bibfnamefont {E.}~\bibnamefont {Harvey}}, \bibinfo {author} {\bibfnamefont {M.}~\bibnamefont {Petrov}},\ and\ \bibinfo {author} {\bibfnamefont {M.~C.}\ \bibnamefont {Hughes}},\ }\href {https://arxiv.org/abs/2502.01861} {\bibinfo {title} {Learning hyperparameters via a data-emphasized variational objective}} (\bibinfo {year} {2025}),\ \Eprint {https://arxiv.org/abs/2502.01861} {arXiv:2502.01861 [cs.LG]} \BibitemShut {NoStop}%
\bibitem [{\citenamefont {MacKay}(1992)}]{10.1162/neco.1992.4.3.415}%
  \BibitemOpen
  \bibfield  {author} {\bibinfo {author} {\bibfnamefont {D.~J.~C.}\ \bibnamefont {MacKay}},\ }\bibfield  {title} {\bibinfo {title} {Bayesian interpolation},\ }\href {https://doi.org/10.1162/neco.1992.4.3.415} {\bibfield  {journal} {\bibinfo  {journal} {Neural Computation}\ }\textbf {\bibinfo {volume} {4}},\ \bibinfo {pages} {415} (\bibinfo {year} {1992})},\ \Eprint {https://arxiv.org/abs/https://direct.mit.edu/neco/article-pdf/4/3/415/812340/neco.1992.4.3.415.pdf} {https://direct.mit.edu/neco/article-pdf/4/3/415/812340/neco.1992.4.3.415.pdf} \BibitemShut {NoStop}%
\bibitem [{\citenamefont {{tufts-ml}}(2026)}]{tufts_ml_mech4mech_2026}%
  \BibitemOpen
  \bibfield  {author} {\bibinfo {author} {\bibnamefont {{tufts-ml}}},\ }\href {https://github.com/tufts-ml/Mech4Mech} {\bibinfo {title} {Mech4mech}} (\bibinfo {year} {2026}),\ \bibinfo {note} {gitHub repository. Accessed: 2026-04-03}\BibitemShut {NoStop}%
\bibitem [{\citenamefont {Jennifer~Radoff}\ and\ \citenamefont {Hammer}(2019)}]{hammer_uncertainty}%
  \BibitemOpen
  \bibfield  {author} {\bibinfo {author} {\bibfnamefont {L.~Z.~J.}\ \bibnamefont {Jennifer~Radoff}}\ and\ \bibinfo {author} {\bibfnamefont {D.}~\bibnamefont {Hammer}},\ }\bibfield  {title} {\bibinfo {title} {“it’s scary but it’s also exciting”: Evidence of meta-affective learning in science},\ }\href {https://doi.org/10.1080/07370008.2018.1539737} {\bibfield  {journal} {\bibinfo  {journal} {Cognition and Instruction}\ }\textbf {\bibinfo {volume} {37}},\ \bibinfo {pages} {73} (\bibinfo {year} {2019})}\BibitemShut {NoStop}%
\bibitem [{\citenamefont {Bergeron}\ \emph {et~al.}(2025)\citenamefont {Bergeron}, \citenamefont {Pamuk~Turner},\ and\ \citenamefont {Hammer}}]{bergeron2025reasoning}%
  \BibitemOpen
  \bibfield  {author} {\bibinfo {author} {\bibfnamefont {K.}~\bibnamefont {Bergeron}}, \bibinfo {author} {\bibfnamefont {D.}~\bibnamefont {Pamuk~Turner}},\ and\ \bibinfo {author} {\bibfnamefont {D.}~\bibnamefont {Hammer}},\ }\bibfield  {title} {\bibinfo {title} {Reasoning through uncertainty: expert chemists' analogical thinking on a novel problem},\ }\href {https://doi.org/10.1039/D5RP00102A} {\bibfield  {journal} {\bibinfo  {journal} {Chemistry Education Research and Practice}\ }\textbf {\bibinfo {volume} {26}},\ \bibinfo {pages} {1031} (\bibinfo {year} {2025})}\BibitemShut {NoStop}%
\bibitem [{\citenamefont {Weller}\ \emph {et~al.}(2025)\citenamefont {Weller}, \citenamefont {Ricci}, \citenamefont {Marone}, \citenamefont {Chaffin}, \citenamefont {Lawrie},\ and\ \citenamefont {Durme}}]{weller2025seqvsseqopen}%
  \BibitemOpen
  \bibfield  {author} {\bibinfo {author} {\bibfnamefont {O.}~\bibnamefont {Weller}}, \bibinfo {author} {\bibfnamefont {K.}~\bibnamefont {Ricci}}, \bibinfo {author} {\bibfnamefont {M.}~\bibnamefont {Marone}}, \bibinfo {author} {\bibfnamefont {A.}~\bibnamefont {Chaffin}}, \bibinfo {author} {\bibfnamefont {D.}~\bibnamefont {Lawrie}},\ and\ \bibinfo {author} {\bibfnamefont {B.~V.}\ \bibnamefont {Durme}},\ }\href {https://arxiv.org/abs/2507.11412} {\bibinfo {title} {Seq vs seq: An open suite of paired encoders and decoders}} (\bibinfo {year} {2025}),\ \Eprint {https://arxiv.org/abs/2507.11412} {arXiv:2507.11412 [cs.CL]} \BibitemShut {NoStop}%
\bibitem [{\citenamefont {Rabiner}(1989)}]{rabinerTutorialHiddenMarkov1989}%
  \BibitemOpen
  \bibfield  {author} {\bibinfo {author} {\bibfnamefont {L.~R.}\ \bibnamefont {Rabiner}},\ }\bibfield  {title} {\bibinfo {title} {A {{Tutorial}} on {{Hidden Markov Models}} and {{Selected Applications}} in {{Speech Recognition}}},\ }\href@noop {} {\bibfield  {journal} {\bibinfo  {journal} {Proc. of the IEEE}\ }\textbf {\bibinfo {volume} {77}},\ \bibinfo {pages} {257} (\bibinfo {year} {1989})}\BibitemShut {NoStop}%
\bibitem [{\citenamefont {Pedregosa}\ \emph {et~al.}(2011)\citenamefont {Pedregosa}, \citenamefont {Varoquaux}, \citenamefont {Gramfort}, \citenamefont {Michel}, \citenamefont {Thirion}, \citenamefont {Grisel}, \citenamefont {Blondel}, \citenamefont {Prettenhofer}, \citenamefont {Weiss}, \citenamefont {Dubourg}, \citenamefont {Vanderplas}, \citenamefont {Passos}, \citenamefont {Cournapeau}, \citenamefont {Brucher}, \citenamefont {Perrot},\ and\ \citenamefont {Duchesnay}}]{scikit-learn}%
  \BibitemOpen
  \bibfield  {author} {\bibinfo {author} {\bibfnamefont {F.}~\bibnamefont {Pedregosa}}, \bibinfo {author} {\bibfnamefont {G.}~\bibnamefont {Varoquaux}}, \bibinfo {author} {\bibfnamefont {A.}~\bibnamefont {Gramfort}}, \bibinfo {author} {\bibfnamefont {V.}~\bibnamefont {Michel}}, \bibinfo {author} {\bibfnamefont {B.}~\bibnamefont {Thirion}}, \bibinfo {author} {\bibfnamefont {O.}~\bibnamefont {Grisel}}, \bibinfo {author} {\bibfnamefont {M.}~\bibnamefont {Blondel}}, \bibinfo {author} {\bibfnamefont {P.}~\bibnamefont {Prettenhofer}}, \bibinfo {author} {\bibfnamefont {R.}~\bibnamefont {Weiss}}, \bibinfo {author} {\bibfnamefont {V.}~\bibnamefont {Dubourg}}, \bibinfo {author} {\bibfnamefont {J.}~\bibnamefont {Vanderplas}}, \bibinfo {author} {\bibfnamefont {A.}~\bibnamefont {Passos}}, \bibinfo {author} {\bibfnamefont {D.}~\bibnamefont {Cournapeau}}, \bibinfo {author} {\bibfnamefont {M.}~\bibnamefont {Brucher}}, \bibinfo {author} {\bibfnamefont {M.}~\bibnamefont {Perrot}},\ and\ \bibinfo {author} {\bibfnamefont
  {E.}~\bibnamefont {Duchesnay}},\ }\bibfield  {title} {\bibinfo {title} {Scikit-learn: Machine learning in {P}ython},\ }\href@noop {} {\bibfield  {journal} {\bibinfo  {journal} {Journal of Machine Learning Research}\ }\textbf {\bibinfo {volume} {12}},\ \bibinfo {pages} {2825} (\bibinfo {year} {2011})}\BibitemShut {NoStop}%
\end{thebibliography}%
\appendix 
\section{Thermo-Fluid mechanics problems \& codes}\label{problems}
Below, we present all six thermo-fluid mechanics problems given to students; their responses comprise the dataset used to train the adapted-HSRDM models. The resistance heater problem was the only problem administered to all five student groups, so we denote it as problem B whereas all other problems are denoted by A and the group number (e.g.~1A). For each problem, we report the problem-specific codebook produced by applying Russ’ framework to that problem instance. Codes for which we found no evidence in students’ responses are marked “[Not mentioned by the students]”.
\begin{center}
\fbox{\parbox{\linewidth}{
\centering \textbf{Problem 1A: Ice Cream Shop} \\
\raggedright
\textbf{Problem:} The building of Farfar’s Danish Ice Cream Shop in Duxbury, MA is somewhat old and thus does not seem to have a great cooling system. As a result, sometimes the ice cream gets a bit melty even when it’s still in the freezer. The temperature in the ice cream shop is to be maintained at 55°F. Estimate the dimensions of the building, use thermodynamics principles to determine the maximum heat loss the shop can have, and suggest a method for minimizing this heat loss.  \\
\textbf{Target phenomena:} Ice cream shop maintained at 55 degrees fahrenheit, where the ice cream does not melt. \\
\textbf{Set-up conditions:} Dimensions of the ice cream shop, environmental conditions outside of the shop, refrigerator and/or air conditioner inside of the shop. \\
\textbf{Entities:} Ice cream, air, refrigerant, air conditioner coil, compressor, condenser. \\
\textbf{Activities:} Air moving through the coils, air mixing with the refrigerant, the compressor pressurizing and raising the temperature of the air. [Not mentioned by the students] \\
\textbf{Properties:} Temperature of the air, temperature of the ice cream, mass of the ice cream, density of the air. [Not mentioned by the students] \\
\textbf{Organization:} Hot air traveling from the inside to the outside of the building. [Not mentioned by the students] \\
\textbf{Chaining:} Hot air travels through the coils and is absorbed by the refrigerant, and flows to the outside of the building, where it is compressed into a higher temperature and pressure gas before being absorbed by the surrounding air. [Not mentioned by the students]}}
\end{center}

\begin{center}
\fbox{\parbox{\linewidth}{
\centering \textbf{Problem B: Resistance Heater} \\
\raggedright
\textbf{Problem:} Design an experiment complete with instrumentation to determine the specific heats of a gas using a resistance heater. Discuss how the experiment will be conducted, what measurements need to be taken, and how the specific heats will be determined. What are the sources of error in your system? How can you minimize the experimental error? \\
\textbf{Target phenomena:} Gas inside the container is heated to a different measured temperature. [Not mentioned by the students] \\
\textbf{Set-up conditions:} Setting up the container conditions, the gas, the resistor, the power sources, and the sources of measurement. \\
\textbf{Entities:} electrons, atoms (inside of the resistor), molecules, atoms (inside of the gas). [Not mentioned by the students] \\
\textbf{Activities:} Electrons moving through the wire, atoms colliding with electrons to resist their movement, gas atoms moving quickly. [Not mentioned by the students] 
\textbf{Properties:} Charges, masses of the electrons or atoms  [Not mentioned by the students] \\
\textbf{Organization:} Electrons in the resistor, atoms in the gas, spread of the atoms in the container. [Not mentioned by the students] \\
\textbf{Chaining:} The electrons move through the wire and collide with atoms in the resistor, which causes the electrons to lose energy and the resistor atoms to gain energy/heat up. [Not mentioned by the students] }}
\end{center}

\begin{center}
\fbox{\parbox{\linewidth}{
\centering \textbf{Problem 2A: Snowmaking} \\
\raggedright
\textbf{Problem:} Most ski resorts in the U.S. use snow guns to make additional snow to supplement natural snow. These machines use water and compressed air. The air forces the water to form tiny droplets, which are then expelled from the nozzle and form ice crystals, which then fall to the ground as snow. Compressed air cools as it expands, which assists with converting the water droplets into snow. Choose your favorite ski resort and the desired depth of snow for the best skiing, and use thermodynamics to determine how long it will take to cover the ski trails in that amount of snow. You may assume that one snow gun uses about 100 gallons of water per minute and that the compressor can produce 50 cfm (cubic feet per minute) of air. \\
\textbf{Target phenomena:} Covering the ski trail in snow. \\
\textbf{Set-up conditions:} Setting up snowguns around the trail. \\
\textbf{Entities:} Air, water, snow, ice. \\
\textbf{Activities:} Mixing together, combing out of the novel, turning into ice. \\
\textbf{Properties:} Density, compressed, temperature, sparse. \\
\textbf{Organization:} Inside of the tube, different novels or systems. \\
\textbf{Chaining:} expanding air and expanding water are turning into ice. }}
\end{center}

\begin{center}

\fbox{\parbox{\linewidth}{
\centering \textbf{Problem 3A: Bus} \\
\raggedright
\textbf{Problem:} About a decade ago, Stanford University successfully tried using waste vegetable oil from the dining halls as fuel for campus shuttles $(https://news.stanford.edu/news/2006/january25/biodiesel-012506.html)$. What if our campus tried to do this? Plan a useful bus route around our campus and specify the volume of fuel needed for the bus to travel this route without having to refuel. You may assume the energy density of vegetable oil is 42.20 MJ/kg or 30.53 MJ/L.  \\
\textbf{Target phenomena:} Bus travels a route on vegetable oil fuel without needing to re-fuel. \\
\textbf{Set-up conditions:} Route distance, road type, bus, bus weight.  \\
\textbf{Entities:} Fuel, vegetable oil, air molecules,crankshaft (of piston), valves (of piston), camshaft (of piston). \\
\textbf{Activities:} Fuel is released into the combustion chamber, fuel and air are compressed, crankshaft turns [Not mentioned by the students]. \\
\textbf{Properties:} Amount of space taken up by the fuel or the air. [Not mentioned by the students] \\
\textbf{Organization:} Fuel mixes with air, the crankshaft provides energy to the rest of the vehicle engine. [Not mentioned by the students]. \\
\textbf{Chaining:} The extension rod of the piston provides a force on the crankshaft, such that it turns and powers the rest of the engine as well as the camshaft, which opens the valves that let fuel into the chamber for compression and combustion. [Not mentioned by the students]. }}
\end{center}

\begin{center}
\fbox{\parbox{\linewidth}{
\centering \textbf{Problem 4A: High-altitude plane} \\
\raggedright
\textbf{Problem:} Gas turbine engines used in airplanes consist of a fan followed by a compressor, diffuser, combustor, turbine, and sometimes an afterburner. You are designing the engine for a high- altitude airplane. Normally, commercial planes operate best around 35,000 ft above sea level, but your plane should operate optimally at around 100,000 ft. Because of the high altitude, there will be a lower concentration of oxygen than normal, and the air entering the engine will be colder. Design a protocol for getting the oxygen up to the appropriate temperature and pressure needed for combustion. Keep in mind your solution has to be relatively light. \\
\textbf{Target phenomena:} Plane is flying at 100 thousand feet with high enough temperature and pressure oxygen to keep the plane in the air. \\
\textbf{Set-up conditions:} Type of plane, atmospheric conditions at high altitude.\\
\textbf{Entities:} Air, diffuser, compressor, combustor, turbine, afterburner, oxygen, fuel, exhaust, nozzle. \\ 
\textbf{Activities:} Air slowing down, oxygen burning, turbine spinning. \\
\textbf{Properties:} Air temperature, air pressure, air volume, compressor or diffuser material. \\
\textbf{Organization:} air moving through diffuser and into compressor and into combustor, energy coming out of the combustor. \\
\textbf{Chaining:} high energy gas leaving the combustor to spin the turbine, which manually controls the compressor.}}
\end{center} 

\begin{center}
\fbox{\parbox{\linewidth}{
\centering\textbf{Problem 5A: Rocket} \\
\raggedright
\textbf{Problem:} Hybrid rockets use a combination of solid and liquid or gaseous propellants. In hybrid rockets, a stable oxidizer is used with a solid fuel. In order to be used, the fuel needs to be vaporized. The primary difficulty with hybrids is with mixing the propellants during the combustion process. In a hybrid rocket, the mixing happens at the melting or evaporating surface. The mixing is not well-controlled and generally, a lot of propellant is left unburned, limiting the motor’s efficiency. On the other hand, liquid propellants are generally mixed with oxidizer by an injector at the top of the combustion chamber which directs many small streams of fuel and oxidizer into one another. Based on reasonable efficiencies of both liquid fuel and hybrid fuel processes, estimate the weight of fuel necessary to get a specific rocket of your choice to low Earth orbit if the fuel is liquid vs. hybrid. \\
\textbf{Target phenomena:} Two rockets (one with liquid fuel and one with hybrid fuel) reach low Earth orbit. \\
\textbf{Set-up conditions:} Two rockets (one with liquid fuel and the other with hybrid  - solid and gas or solid and liquid) conditions, Earth’s atmosphere conditions. \\
\textbf{Entities:} Fuel, propellants, molecules, pumps, valves, oxidizer. \\
\textbf{Activities:} Fuel and oxidizer mixing, combined substance burning, molecules compressing, combined substance ejecting. \\
\textbf{Properties:} State of matter (liquid, solid, gas) of fuel, amount of space taken up by the fuel, temperature or pressure of the molecules.  \\
\textbf{Organization:} Fuel and oxidizer starting out in separate containers, slowly combining, molecules coming out of the rocket. \\
\textbf{Chaining:} Liquid fuel slowly mixes with the oxidizer, and combusts to release hot gas as thrust for the rocket to move. [Not mentioned by the students] }}
\end{center}

\section{Human annotation procedure and IRR analysis}\label{human_annot}
Two human analysts collaborated to tailor Russ’ framework to a specific problem context -- a professor of mechanical engineering education and a postdoctoral researcher working at the intersection of machine learning and education research. The two analysts first established a general coding strategy applicable across all problems, and then calibrated their problem-specific application of that strategy using the snowmaking problem as a reference case. To obtain evidence of annotation reproducibility with the developed code, the two analysts conducted an inter-rater reliability (IRR) procedure on student text from this problem. After this calibration phase, the remaining problem codes were developed by the postdoctoral researcher. Below, we describe the general coding strategy details, as well as the IRR implementation and results. \\

\noindent \textbf{General coding strategy details.} For a specific problem, the first step in coding the students' talk for mechanistic reasoning is to clarify for ourselves the target phenomenon. We highlight the importance of this decision, as what is chosen to be the target phenomena defines the scale of objects we believe students will reason mechanistically about. For example, if the target phenomena is the amount of snow coverage as a result of the snowgun vs.~the amount of snow coverage after a day in the sun. In the first scenario, we expect evidence of mechanistic reasoning to look like a discussion about the inter-workings of the snowgun. In the second scenario, we expect evidence to looks like a discussion about how sunlight interacts with snow. 

In this snowmaking problem, the students are tasked with determining how long it takes to cover a ski resort with snow from a snow gun. Thus, the target phenomenon is production of snow to cover ski trails. Next we do a first read-through of the entire conversation to identify the specific instantiations of the remaining mechanistic elements for this particular problem. This step enables us to be internally consistent - within a specific conversation - regarding which student language we consider to refer to activities versus properties versus organization of the different entities they have identified. Finally, we code sentence-by-sentence, considering whether each sentence presents evidence of one or more of the seven categories. We apply a code when we find an \textit{act} of mechanistic reasoning in the conversational turn rather than only the linguistic features of mechanistic reasoning. For example, we code for entities when the student utterance is \textit{focused} on identifying the entities that are relevant to the problem, rather than \textit{mentioning} the entities to take the next identification step. For example, in the following utterance, the act is not identifying the entities of air and water (category \#3) but rather mentioning them in order to accomplish the higher level act of identifying an activity (category \#4). ``If we have to have the air and the water come out of the same nozzle, we shouldn't do - we shouldn't split it. They just do that so it covers more area."

\noindent \textbf{IRR implementation and results.} The two analysis agreed on the general coding strategy detailed above, and then selected the snowmaking problem, due to its large number of utterances compared to other problems (372 utterances), to calibrate their strategy for developing a problem-specific code and applying it to the text. The analysts conducted a full read-through of the student conversation, developed an initial code for the problem together, and then coded the full problem individually (372 utterances). As evidence of initial inter-rater reproducibility, we report the Cohen’s $\kappa$ and accuracy metrics for the individual coders across all categories in Table \ref{IRR1}. 

\begin{table}[t]
\centering
\begin{tabular}{lrr}
\toprule
Label & Accuracy & Cohen's $\kappa$ \\
\midrule
Target phenomena  & 98.59 & 0.6989 \\
Set-up conditions & 96.89 & 0.4073 \\
Entities  & 67.80 & 0.0842 \\
Activities  & 88.70 & 0.4024 \\
Properties & 81.64 & 0.2809 \\
Organization & 95.76 & 0.6716 \\
Chaining  & 98.87 & 0.4943 \\
No evidence & 81.92 & 0.5723 \\
Max evidence & 72.03 & 0.4262 \\
\bottomrule
\end{tabular}

\caption{Pre-discussion inter-rater reliability metrics for each code column and the maximum code evidence on the full snowmaking problem.}
\label{tab:irr_metrics}
\end{table}\label{IRR1}

\begin{table}[t]
\centering
\begin{tabular}{lrr}
\toprule
Column & Accuracy & Cohen's $\kappa$ \\
\midrule
Target phenomena  & 97.22 & 0.6032 \\
Set-up conditions & 99.44 & 0.8861 \\
Entities  & 91.11 & 0.3814 \\
Activities  & 89.44 & 0.4780 \\
Properties & 85.56 & 0.2439 \\
Organization & 98.33 & 0.9097 \\
Chaining  & 100.00 & NaN \\
No evidence & 86.67 & 0.6837 \\
Max evidence & 82.22 & 0.6380 \\
\bottomrule
\end{tabular}

\caption{Post-discussion inter-rater reliability metrics for each code column and the maximum code evidence on the first 181 utterances of the snowmaking problem.}
\label{tab:irr_metrics}
\end{table}\label{IRR2}

\begin{table}[t]
\centering
\begin{tabular}{lrr}
\toprule
Column & Accuracy & Cohen's $\kappa$ \\
\midrule
Target phenomena  & 99.45 & 0.9063 \\
Set-up conditions & 96.15 & 0.3480 \\
Entities  & 71.98 & 0.1749 \\
Activities  & 87.91 & 0.4701 \\
Properties & 82.42 & 0.0125 \\
Organization & 93.96 & 0.6847 \\
Chaining  & 99.45 & 0.0000 \\
No evidence & 85.71 & 0.6703 \\
Max evidence & 72.53 & 0.4577 \\
\bottomrule
\end{tabular}

\caption{Pre-discussion inter-rater reliability metrics for each code column and the maximum code evidence on the first 181 utterances of the snowmaking problem.}
\label{tab:irr_metrics}
\end{table}\label{IRR3}

We report the Cohen’s $\kappa$ and accuracy metrics for each category (e.g.~target phenomenon). The same student utterance can have multiple categorical labels, and so we include a separate metric for the inter-rater agreement of the maximum (max) evidence category in the hierarchy per utterance. This informs how often two raters observe the same max evidence in a given utterance. This validation is the most relevant to our usage of the data, as post the IRR procedure, decided to train the ML classifier feedback mechanism discussed in Sec.~\ref{feedback_mech} with only the max evidence labels. Throughout the main text and in the Appendix, when we refer to the number of positive labels for a category, we \emph{only} refer to the max evidence. As in, a student utterance is labeled as its max category, and it is then a positive example for that category. 

We observe relatively high accuracies across categories except for entities. Through discussion, it was learned that both raters needed to resolve ambiguity around what counted as an \emph{act} of reasoning about entities vs.~the mere \emph{mentioning} of entities. The Cohen's $\kappa$ values indicate mostly moderate agreement. We expect this to be lower than accuracy as $\kappa$ values take into account how likely a rater is to mark $0/1$ for a certain category based on their overall marking frequency. 

Post the IRR assessment, the analysts came together for a discussion to see whether they could resolve their disagreement and improve the language in the codebook. The analysts updated the code with the intention of making the code less ambiguous for a future user. After this update, the analysts re-coded nearly half (181 utterances) of the problem again separately. As evidence of reproducibility following codebook refinement, we report Cohen’s $\kappa$ and accuracy across all categories for the two coders in Table \ref{IRR2}. We also include the inter-rater agreement metrics on the first 181 utterances from the previous codings for an apples-to-apples comparison in Table \ref{IRR3}. 

We observe from comparing the values in Table \ref{IRR2} and Table \ref{IRR3} that all accuracies and $\kappa$ values increase post the discussion, except for the category target phenomena. This is evidence to suggest that the raters did not fully resolve more ambiguity for this category as a result of the discussion. Despite this, agreement on the max evidence category increased significantly -- from moderate to substantial-- in $\kappa$ value. We again emphasize this increase, as this is the annotation that we utilize in the ML portion of this work. 

Overall, the IRR assessment provides moderate to substantial evidence that annotations for the snowmaking problem using the code in Sec.~\ref{problems} can be reproduced. However, stronger evidence of reproducibility would require a third rater, who had not been a part of the original discussions, to code the problem and compare with the first two rater's results. We acknowledge this as a resource limitation. Furthermore, we recognize a key limitation of the overall IRR procedure is that all coding was done on the same problem. We utilized the time resources that we had to prioritize the fine-tuning of the codebook for one problem, so that learned considerations throughout the analyst's discussions could then be applied across problems. We acknowledge that an improved reproducibility procedure would include the step of the analysts developing a code for a new problem together (or separately for even stronger evidence), coding that entire problem individually, and then conducting another round of IRR validation. This would provide stronger evidence that the strategy of developing a specific code for a problem and then applying that code to the student text is reproducible. Due to limited resources, the postdoctoral researcher developed the remaining problem-specific codes independently using learnings from the discussions on the snowmaking problem, and then shared the codes with the mechanical engineering education professor for final validation. No further updates were made to the codebook. The postdoctoral researcher coded the rest of the student text using these codes. Only the labels produced by the postdoctoral researcher are used to train the feedback classifier in Sec.~\ref{feedback_mech}. In future work, we believe it's important to obtain more resources (e.g.~coding time, more raters) to increase the evidence of annotation reproducibility. If we cannot agree on what evidence of mechanistic reasoning looks like, we lack a well-defined ``ground truth" for the ML models. 

\section{Data summary}\label{data_summary}

See Tables \ref{utterances1A} - \ref{utterances5B} for the number of total student utterances in each problem and the distribution of utterances across students in each problem. In the raw data for problem 1B, two additional students walked by the group and exchanged a two or three casual utterances. None of these utterances contain evidence of mechanistic reasoning. As we did not believe it made sense for these students to be represented as entities, we removed these lines during the initial processing of the raw data. 

\begin{table}[htbp]
\centering
\normalsize 
\begin{tabular}{ccc} 
\toprule
{Pseudonym} & {Utterance counts}\\
\midrule
Cameron & 154 \\
Logan & 137 \\
Sam & 114 \\
Max & 79 \\
Total & 484 \\
\bottomrule
\end{tabular}
\caption{Problem 1A}
\label{utterances1A}
\end{table}

\begin{table}[htbp]
\centering
\normalsize 
\begin{tabular}{ccc} 
\toprule
{Pseudonym} & {Utterance counts}\\
\midrule
Cameron & 69 \\
Logan & 61 \\
Sam & 54 \\
Max & 53 \\
Total & 237 \\
\bottomrule
\end{tabular}
\caption{Problem 1B}
\label{utterances1B}
\end{table}

\begin{table}[htbp]
\centering
\normalsize 
\begin{tabular}{ccc} 
\toprule
{Pseudonym} & {Utterance counts}\\
\midrule
Jamie & 186 \\
Morgan & 186 \\
Total & 372 \\
\bottomrule
\end{tabular}
\caption{Problem 2A}
\label{utterances2A}
\end{table}

\begin{table}[htbp]
\centering
\normalsize 
\begin{tabular}{ccc} 
\toprule
{Pseudonym} & {Utterance counts}\\
\midrule
Jamie & 55 \\
Morgan & 54 \\
Total & 109 \\
\bottomrule
\end{tabular}
\caption{Problem 2B}
\label{utterances2B}
\end{table}

\begin{table}[htbp]
\centering
\normalsize 
\begin{tabular}{ccc} 
\toprule
{Pseudonym} & {Utterance counts}\\
\midrule
Drew & 189 \\
Alex & 151 \\
Charlie & 149 \\
Blake & 87 \\
Total & 576 \\
\bottomrule
\end{tabular}
\caption{Problem 3A}
\label{utterances3A}
\end{table}

\begin{table}[htbp]
\centering
\normalsize 
\begin{tabular}{ccc} 
\toprule
{Pseudonym} & {Utterance counts}\\
\midrule
Drew & 61 \\
Alex & 106 \\
Charlie & 94 \\
Blake & 67 \\
Total & 328 \\
\bottomrule
\end{tabular}
\caption{Problem 3B}
\label{utterances3B}
\end{table}

\begin{table}[htbp]
\normalsize 
\begin{tabular}{ccc} 
\toprule
{Pseudonym} & {Utterance counts}\\\midrule
Xian & 285 \\
Taylor & 270 \\
Andy & 43 \\
Parker & 10 \\
Total & 608 \\
\bottomrule
\end{tabular}
\caption{Problem 4A}
\label{utterances4A}
\end{table}

\begin{table}[htbp]
\centering
\normalsize 
\begin{tabular}{ccc} 
\toprule
{Pseudonym} & {Utterance counts}\\
\midrule
Xian & 178 \\
Taylor & 176 \\
Andy & 19 \\
Parker & 18 \\
Total & 391 \\
\bottomrule
\end{tabular}
\caption{Problem 4B}
\label{utterances4B}
\end{table}

\begin{table}[htbp]
\centering
\normalsize 
\begin{tabular}{ccc} 
\toprule
{Pseudonym} & {Utterance counts}\\
\midrule
Palmer & 318 \\
Golden & 310 \\
Dylan & 91 \\
Rowan & 74 \\
Total & 793 \\
\bottomrule
\end{tabular}
\caption{Problem 5A}
\label{utterances5A}
\end{table}

\begin{table}[htbp]
\centering
\normalsize 
\begin{tabular}{ccc} 
\toprule
{Pseudonym} & {Utterance counts}\\
\midrule
Palmer & 102 \\
Golden & 81 \\
Dylan & 79 \\
Rowan & 12 \\
Total & 274 \\
\bottomrule
\end{tabular}
\caption{Problem 5B}
\label{utterances5B}
\end{table}

\section{Probabilistic model factorization details}\label{factorization_details}

As discussed in Sec.~\ref{probabilistic_assumptions}, our model defines a joint distribution over all random variables: $p(s^{1:N}_{0:T}, z^{1:N,1:J}_{0:T}, x^{1:N,1:J}_{0:T}, \theta)$. Under the assumption that each generated sequence $i$ is drawn independently, we re-write the model in the following form: 
\begin{gather}
      \prod_{i=1}^N p(x_{0:T}^{i, 1:J}, z^{i, 1:J}_{0:T}, s_{0:T}^{i}|\theta)p(\theta) \\  = \prod_{i=1}^N
      p(s_{0:T}^{i}|\theta)p(x_{0:T}^{i, 1:J}, z^{i, 1:J}_{0:T} | s_{0:T}^{i}, \theta)p(\theta) \nonumber
\end{gather}\label{prob1}

From here, we incorporate the assumptions discussed in Sec.~\ref{probabilistic_assumptions} to obtain a further factorized form of the components above: 

\begin{gather}
p(s_{0:T}^{i}|\theta) = p(s^{i}_0|\theta_{s-init})  \prod^T_{t=1} p(s^{i}_t|s^{i}_{t-1}, x^{i, 1:J}_{t-1}, \theta_{ss}), \\
p(x_{0:T}^{i, 1:J}, z^{i, 1:J}_{0:T} | s_{0:T}^{i}, \theta) = \prod ^J_{j=1} p(z^{i, j}_0|\theta_{z-init}) \nonumber \\ p(x^{i, j}_0|\theta_{x-init})   \prod^T_{t=1} p(z^{i, j}_t|s^{i}_t, z^{i, j}_{t-1}, x^{i, j}_{t-1}, \theta_{es}) \nonumber \\ p(x^{i, j}_t|z^{i, j}_{t}, x^{i, j}_{t-1}, \theta_{ee}) \nonumber
\end{gather}

Each of these factorized distributions correspond to different components introduced in Sec.~\ref{probabilistic_assumptions}: system-level starting distribution, system-level transition distributions defined by $G$, entity-level starting distribution, entity-level transition distributions defined by $F$, emission at $t=0$, and emission for $t \geq0$. 

\section{Inductive bias via data feedback details}\label{feedback_app}

\subsection{Classifier selection details.}\label{classifier_selection}
We considered various pre-trained and need-to-train classifiers, and selected based on task alignment and the amount of computational resource requirements for deployment (e.g.~the number of parameters stored and used during inference). Task alignment refers to the extent to which the pre-training objective is congruent with, or relevant to, our current task. Secondly, for the trained adapted-HSRDM to be able to run on one's local device (e.g.~M1 chip with 32GB RAM), we need to be able to download and run the model for the feedback mechanism locally while also running variational inference. Thus, we need to minimize the number of parameters stored at inference time. 
 
\noindent \textbf{Re-purpose from prior work.} We first considered re-purposing an existing method from the literature. The authors in \cite{gili2025} fine-tune and validate the Bidirectional Encoder Representations from Transformers (BERT) \cite{BERT} encoder-only language model for the binary multi-label classification task of measuring evidence of students' mechanistic reasoning on self-contained responses to physics problems. The authors utilize domain labels that map to Russ' framework minus the category target phenomena. Re-purposing this classifier for our task eliminates the need to extract training data for the feedback mechanism. However, this method is not useful due to task mis-alignment. There is a difference between our interpretation of a student \emph{identifying} the items in Russ' framework if the data is an individual summary explanation vs.~a sequence of in-moment dialogue. In the former, we have no information about the process leading up to the explanation generation, and so we can reasonably make the assumption that the text represents a self-contained account of all of the student's intellectual work (either conducted in the moment of writing the explanation or in moments prior). Therefore, if a student \emph{mentions} one of the items (e.g.~an entity, activity, property), the annotator assumes this is \emph{identifying} one of these items. In contrast, when we know that students are engaging in dialogue to solve a problem, we can reasonably assume that the intellectual work is spread across different moments in time, and mentioning an item in one moment does not necessarily mean that a student is doing the intellectual work in that moment to identify that item; it could be referencing the item from intellectual work that was done in previous moments. As a more concrete example, consider the following two utterances from the snowmaking problem: \\

\noindent J: \emph{``Alright. Also I'm concerned. Is this a pure substance? Do we have to worry about that? The fact that it's air and water mixed together?"} \\

\noindent M: \emph{``I don't know the density of water in a snow gun."} \\

In the first utterance, the student identifies that the substance under consideration is composed of the entities air and water. In the second utterance, the student uses the word ``water'', but is not using it to identify water as an entity. Rather, in this moment, the student is reasoning about a property of that entity - the density. As the intellectual work of identifying ``water'' was done in the first utterance, we would want to mark a $1$ for the entity domain. As this does not appear to be the intellectual work done in the second utterance, we would want to mark this a $0$ despite the term ``water'' being used. Given that the classifier in \cite{gili2025} was never trained on student responses that \emph{mentioned} items without \emph{doing} the intellectual work (i.e.~the response should receive a label $0$ for that item rather than a $1$), we concluded that this classifier would do poorly on our data, which includes these characteristics in a large portion of the responses. 

\noindent \textbf{Transformer decoder for in-context learning.} We considered using a pre-trained decoder-only language model, where one can explicitly prompt the model to return the labels associated with the provided utterance. These labels can then be mapped to a binary vector. Recent work \cite{weller2025seqvsseqopen} provides empirical evidence that using decoder-only language models with more parameters for classification tasks performs \emph{worse} than using encoder-only models for these tasks with fewer parameters. Choosing such a method requires storing a $\sim$ billion parameter model and running this model at each timestep in the adapted-HSRDM. We prioritized a feedback method that keeps the number of model parameters low rather than needing to download a $\sim$ billion parameter model. 

\noindent \textbf{Transformer encoder for transfer learning.} Lastly, we considered training a foundation language encoder-only model for our specific downstream classification task. In an attempt to not retrain foundation model weights, which demands more computational resources -- especially when optimizing over various hyper-parameters, we decided to freeze the weights of the EmbeddingGemma encoder and train a small neural network architecture to classify each generated embedding. Unlike decoder-only models, the classification outputs are deterministic for each embedding, and the number of parameters that need to be stored locally and used at each timestep is significantly fewer (millions instead of billions). 

\subsection{Training data selection.}\label{FM training details}

We first selected 2 problems (2A and 2B) as a test set for the adapted-HSRDM, which left 8 problems for training the supervised feedback mechanism classifier and the CAVI training. As we did not want our feedback mechanism to be trained on the exact same data as the CAVI training, we used the smallest sub-set believed possible for training the classifier (4 problems). To once again prevent data leakage across students, we chose 4 problems from two student groups. As we only have one problem outside of our test set that contains examples of chaining (problem 4A), we included the data from problems 4A and 4B in our training. To select the second student group, we picked the pair of problems that had the fewest number of positive labels in the hierarchy past the entity domain. The reason for this is that problems 4A and 4B contain the most positive labels in this sub-section (entity--chaining) of the hierarchy out of all of the training data, and we would like to ensure that there is enough non-overlapping positive labels in this sub-section between the supervised training and the unsupervised training loss. Thus, we picked problems 1A and 1B. See Table  \ref{tab:positive_label_counts} for all positive label counts present in the dataset and in the train/test splits. 

\begin{table*}[htbp]
\centering
\small
\begin{tabular}{ccccccccc}
\toprule
{Problem} & {No evidence} &  \makecell{{Target}\\{phenomena}} &  \makecell{{Set-up} \\{conditions}} &  {Entities} &  {Activities} &  {Properties} &  {Organization} &  {Chaining} \\
\midrule
1A &  415 & 3 & 60 &  5 &           0 &           2 &             0 &         0 \\
                      1B &                            209 &                 1 &                 17 &         9 &           1 &           0 &             0 &         0 \\
                      2A &                            298 &                 6 &                  8 &         5 &          13 &          21 &            21 &         1 \\
                      2B &                            102 &                 0 &                  6 &         1 &           0 &           0 &             0 &         0 \\
                      3A &                            513 &                 1 &                 51 &        10 &           0 &           1 &             0 &         0 \\
                      3B &                            267 &                 7 &                 36 &        12 &           4 &           0 &             2 &         0 \\
                      4A &                            505 &                 1 &                  6 &        39 &          18 &          27 &            10 &         2 \\
                      4B &                            323 &                 0 &                 68 &         0 &           0 &           0 &             0 &         0 \\
                      5A &                            761 &                 3 &                 10 &         2 &           5 &          11 &             1 &         0 \\
                      5B &                            243 &                 0 &                 29 &         1 &           0 &           1 &             0 &         0 \\
                   Total &                           3636 &                22 &                291 &        84 &          41 &          63 &            34 &         3 \\
\midrule \\
{Supervised train} & 1452 &                 5 &                151 &        53 &          19 &          29 &            10 &         2 \\
{Unsupervised train} &                           3338 &                16 &                277 &        78 &          28 &          42 &            13 &         2 \\
{Semi-supervised test} &                            400 &                 6 &                 14 &         6 &          13 &          21 &            21 &         1 \\
\bottomrule
\end{tabular}
\centering
\caption{Positive label counts by problem}
\label{tab:positive_label_counts}
\end{table*}

\subsection{Method details.}\label{FM method details}

\textbf{Method justification.} To avoid needing a validation set, we utilized a recent Bayesian variational inference approach put forth in \cite{harvey2025} for model selection, where one can minimize a negative data-emphasized ELBO (DE-ELBO) as a method of implicitly choosing the model with the optimal prior parameters over neural network parameters (i.e.~the weight decay hyper-parameter), as well as the model that most likely generates the training data. By definition, minimizing the negative DE-ELBO mitigates over-fitting behavior if the assumed prior and likelihood distributions for the data are reasonably specified. 

The data-emphasized term simply adjusts the amount of importance that is placed on the data relative to the prior term in the ELBO, which may be necessary when the number of model parameters $n_p$ is greater than the amount of data $N$ for training ($n_p {=} 17544 > N {=} 1721$ in our situation). The demonstrations in Ref.~\cite{harvey2025} provide evidence that this method achieves just as high or better performance on test data for multi-label text classification tasks. In utilizing this method, we were able to use a 4 problem training set for the classifier, rather than 6 problems (the extra two problems would have served as our validation set). 

\textbf{Training details.} We trained a classifier that takes EmbeddingGemma encoded utterance vectors $x_i \in \mathbb{R}^{128}$ as input and predicts the maximum mechanistic-evidence class $y_i \in \{0, 2 ,...7\}$ under Russ’ taxonomy (Sec.~\ref{Background}). The zeroth class is for no-evidence, whereas classes $1-7$ quantify how much mechanistic reasoning evidence is present in the utterance. We utilized a 2-layer neural network (NN) with a ReLu activation function as our model, and assumed an isotropic gaussian distribution for our prior over these parameters $\theta$ with a learnable weight decay $\lambda$. We set the hidden layer of the NN to our embedding dimension size $D=128$. We assumed a categorical distribution for our probabilistic likelihood function, which is computed as the softmax over per-example class logits $z_i \in \mathbb{R^8}$ from the output of the 2-layer NN. We define our probabilistic likelihood $p(y_i|x_i, \theta, M)$ and prior $p(\theta)$ density in the following form: 
\begin{gather}
    p(y_i|x_i, \theta) = \text{Cat}(y_i| \text{softmax}(z_i)), \\
    p(\theta) = \mathcal{N}(\theta |0, \lambda^{-1}\mathcal{I})
\end{gather}
For our approximate posterior, we also assumed a parameterized $\phi$ isotropic gaussian distribution, with a mean based on the current parameter mean $\bar\theta$ and a trainable parameter for the variance $\rho^2$. We define our approximate posterior in the following form:  
\begin{equation}
    q_{\phi}(\theta) = \mathcal{N}(\theta | \bar\theta, \rho^2\mathcal{I})
\end{equation}
The approximate posterior is trained to match the true posterior, such that $ q_{\phi}(\theta) \approx p(\theta|x_{1:N}, y_{1:N})$. The parameters of the true distribution and the approximate posterior can be learned via an optimization procedure that minimizes the per-datapoint negative (DE-ELBO):
\begin{gather}\label{feedback_ELBO}
   -\text{ELBO}= -\frac{1}{N}\sum^N_{i=1}[\frac{n_p}{N} \mathbb{E}_{\theta \sim q(\theta)}[ \log p(y_i|x_i, \theta)] \\  - KL(q(\theta) || p(\theta) ] \nonumber
\end{gather}
Note that the negative-log-likelihood (NLL) term (left) prioritizes fitting the training data well and the Kullback–Leibler (KL) divergence term (right) prioritizes obtaining a posterior distribution over parameters that remains close to the prior. The $\frac{n_p}{N}$ term allows one to up-weight the data-fit term when a large amount of parameters might cause the prior term to dominate. The data-fit term will be maximized when the correct label $y_i$ is generated from the model. Due to the class imbalance over our $K{=}8$ classes, we weighted the importance of each class in the training by $\frac{N}{8\times n_k}$, where $n_k$ is the number of examples belonging to the class $k$. Training data selection details can be found in Appendix Sec.~\ref{FM training details}.

\noindent \textbf{Results.}~The weight decay hyper-parameter $\lambda$ was learned implicitly via the variational approach, and we conducted a grid search to optimize for the learning rate (LR) $\alpha$ over different parameter initializations that produces the lowest negative ELBO over $1000$ training epochs. See Table \ref{tab:final-losses} for all final negative ELBO results over five different parameter initialization seeds and fourteen different learning rates $\{\alpha\}$. The best model is in bold ($seed {=} 7, \alpha {=} 2.9$). See Table \ref{class_accuracy} for the per-class accuracy scores computed from the model's predictions on the training data. We conducted three additional checks, outlined below, to verify that our model choice is reasonable. 

\begin{table}[htbp]
\centering
\caption{Final losses by initialization and LR (Part 1).}
\label{tab:final-losses-part1}
\small
\begin{tabular}{ccc}
\toprule
Init seed & LR & Final loss \\
\midrule
7 & 3.3 & 0.0096656819805502 \\
7 & 2.9 & 0.009665491990745 \\
7 & 2.5 & 0.0096656940877437 \\
7 & 2.1 & 0.0096656968817114 \\
7 & 1.7 & 0.0096655311062932 \\
7 & 1.3 & 0.0096656577661633 \\
7 & 0.9 & 0.0096659399569034 \\
7 & 0.5 & 0.0096686063334345 \\
7 & 0.1 & 0.0131261078640818 \\
7 & 0.01 & 0.0179602280259132 \\
7 & 0.001 & 0.0886219218373298 \\
7 & 0.0001 & 5.237014293670654 \\
7 & 1e-05 & 6.147874355316162 \\
7 & 1e-06 & 6.2400665283203125 \\
123 & 3.3 & 0.0096749262884259 \\
123 & 2.9 & 0.0096757467836141 \\
123 & 2.5 & 0.0096768550574779 \\
123 & 2.1 & 0.0096779707819223 \\
123 & 1.7 & 0.0096803484484553 \\
123 & 1.3 & 0.0096833296120166 \\
123 & 0.9 & 0.0096903471276164 \\
123 & 0.5 & 0.0097132893279194 \\
123 & 0.1 & 0.014538419432938 \\
123 & 0.01 & 0.0212347786873579 \\
123 & 0.001 & 0.0887199118733406 \\
123 & 0.0001 & 5.237014293670654 \\
123 & 1e-05 & 6.147874355316162 \\
123 & 1e-06 & 6.240066051483154 \\
213 & 3.3 & 0.0096758734434843 \\
213 & 2.9 & 0.0096771800890564 \\
213 & 2.5 & 0.0096781142055988 \\
213 & 2.1 & 0.0096795102581381 \\
213 & 1.7 & 0.0096811978146433 \\
213 & 1.3 & 0.0096835847944021 \\
213 & 0.9 & 0.0096878539770841 \\
213 & 0.5 & 0.0097013833001255 \\
213 & 0.1 & 0.0130045237019658 \\
213 & 0.01 & 0.0172159876674413 \\
213 & 0.001 & 0.0886680483818054 \\
213 & 0.0001 & 5.237014293670654 \\
213 & 1e-05 & 6.147874355316162 \\
213 & 1e-06 & 6.240066051483154 \\
\bottomrule
\end{tabular}
\end{table}

\begin{table}[htbp]
\centering
\caption{Final losses by initialization and LR (Part 2).}
\label{tab:final-losses-part2}
\small
\begin{tabular}{ccc}
\toprule
Init seed & LR & Final loss \\
\midrule
512 & 3.3 & 0.0096786580979824 \\
512 & 2.9 & 0.0096797160804271 \\
512 & 2.5 & 0.009681187570095 \\
512 & 2.1 & 0.0096823880448937 \\
512 & 1.7 & 0.0096843177452683 \\
512 & 1.3 & 0.0096874348819255 \\
512 & 0.9 & 0.0096928542479872 \\
512 & 0.5 & 0.009710619226098 \\
512 & 0.1 & 0.0133584877476096 \\
512 & 0.01 & 0.0178864225745201 \\
512 & 0.001 & 0.0886841490864753 \\
512 & 0.0001 & 5.2370147705078125 \\
512 & 1e-05 & 6.14787483215332 \\
512 & 1e-06 & 6.2400665283203125 \\
637 & 3.3 & 0.0096724024042487 \\
637 & 2.9 & 0.0096732759848237 \\
637 & 2.5 & 0.0096739986911416 \\
637 & 2.1 & 0.0096749421209096 \\
637 & 1.7 & 0.0096763456240296 \\
637 & 1.3 & 0.0096787083894014 \\
637 & 0.9 & 0.0096830110996961 \\
637 & 0.5 & 0.0096974587067961 \\
637 & 0.1 & 0.0137728871777653 \\
637 & 0.01 & 0.0196864567697048 \\
637 & 0.001 & 0.0886529237031936 \\
637 & 0.0001 & 5.237014293670654 \\
637 & 1e-05 & 6.147874355316162 \\
637 & 1e-06 & 6.2400665283203125 \\
\bottomrule
\end{tabular}
\end{table}

\begin{table}[t]
\centering
\begin{tabular}{lc}
\toprule
{Class} & {Accuracy} \\
\midrule
0 & 0.9074 \\
1 & 1.0000 \\
2 & 0.8409 \\
3 & 0.8099 \\
4 & 0.9603 \\
5 & 0.8047 \\
6 & 0.9885 \\
 7 & 1.0000 \\
\bottomrule
\end{tabular}\caption{\justifying The per-class accuracy scores of the classifier predictions on the training data.}
\end{table}\label{class_accuracy}

\begin{figure*}[htbp]
    \centering
    \begin{subfigure}[t]{0.32\textwidth}
        \centering
        \includegraphics[width=\linewidth]{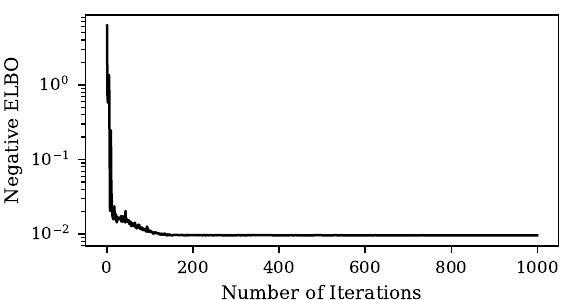} 
        \caption{Negative ELBO}
    \end{subfigure}
    \hfill
    \begin{subfigure}[t]{0.32\textwidth}
        \centering
        \includegraphics[width=\linewidth]{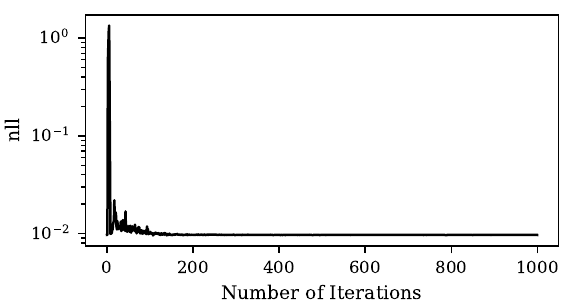}
        \caption{Negative-log likelihood}
    \end{subfigure}
    \hfill
    \begin{subfigure}[t]{0.32\textwidth}
        \centering
        \includegraphics[width=\linewidth]{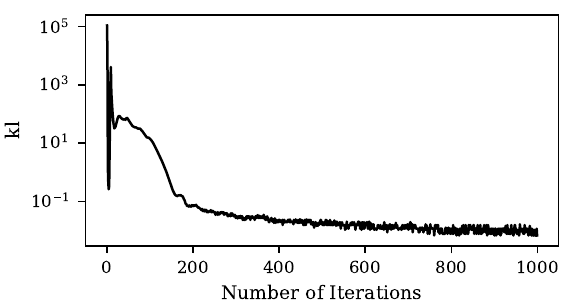}
        \caption{Kullback–Leibler divergence}
    \end{subfigure}

    \caption{Log-scale convergence behavior for the optimal model (seed: 7, LR: 2.9). Left: full negative ELBO. Middle: the negative log-likelihood or data term in the ELBO. Right: the KL term in the ELBO between the prior and approximate posterior.}
    \label{feedback_convergence}
\end{figure*}

\noindent\textbf{Negative ELBO, NLL, and KL behavior.} We checked whether the individual negative ELBO, NLL and KL terms in Eq.~\ref{feedback_ELBO} computed over $1000$ training epochs also descend and converge to minimum values. The NLL term represents the fit of our model to our training data and the KL terms represents the  regularization of our approximate posterior distribution to our prior distribution. While we expect the negative ELBO to be the primary guidepost, if the NLL term is low and the KL term is high, this is a signal of overfitting. Thus, for our best model ($seed {=} 7, \alpha{=} 2.9$), we ensured that both terms computed over epochs show reasonable behavior (see Fig.~\ref{feedback_convergence}). 

\noindent \textbf{Learning rate behavior.} In our initial training procedure, we only included learning rates up to $\alpha = 0.5$, and found the model associated with this rate to be optimal. Thus, we extended our grid search to values that iteratively $+0.4$ up to $\alpha {=} 3.3$. We then found that $\alpha {=} 2.9$ led to the optimal model. However, we noticed that the lowest negative ELBO value was similar among all high learning rates, indicating similar convergence paths towards local minima. Thus, to do a final check that model training ran as expected, we zoomed in on the first 20 steps of training for each learning rate and looked for similar descents happening in fewer iterations. We observed that increasing the LR led to earlier descents. 

\noindent \textbf{Parameter initialization stability.} Oftentimes, parameter initialization can impact model performance, and different initializations then require different learning rates to do well. We checked the training stability to observe whether the initialization parameters played a critical role in the model selection. The results in Table \ref{tab:final-losses} show that model performance is stable across initializations for different learning rates. We used this as evidence to conclude that searching over more starting points would not necessarily lead to a better model. 
\vspace{-2em}
\section{ML training details}\label{ml_method_details}

\subsection{Training Algorithm Overview}\label{vi_alg_overview}

the HSRDM is trained in a fully unsupervised fashion via Bayesian variational inference techniques \cite{beal2003variational, Blei_2017}. One can utilize a structured distribution $q_{\phi_{i}}(z^{i,j}_{0:T},s^{i}_{0:T})$ parameterized by $\phi_{i}$ that can be trained to approximate the true posterior distribution $p(z^{i,1:J}_{0:T}, s_{0:T}^{i}| x_{0:T}^{i, 1:J}, \theta)$ over latent variables in sequence $i$. This can be done by maximizing the per-datapoint Evidence Lower Bound (ELBO): 
\begin{gather}
      \text{ELBO} = \frac{1}{N}\sum^N_{i=1}{\mathbb{E}_{q_{\phi_i}}} \log p(x_{0:T}^{i, 1:J}, z^{i, 1:J}_{0:T}, s_{0:T}^{i}|\theta) \\  - {\mathbb{E}_{q_{\phi_i}}} \log {q_{\phi_i}(z^{i, 1:J}_{0:T}, s_{0:T}^{i})} \nonumber
\end{gather}\label{ELBO_equation}
For the interested reader, see Appendix Sec.~\ref{elbo_derivation}, for a detailed derivation of the ELBO from our joint distribution and factorization assumptions in Sec.\ref{probabilistic_assumptions}, which offers a theoretical explanation as to how maximizing the ELBO allows one to approximate the assumed true posterior. As done in Ref.~\cite{wojnowicz2024discovering}, to maximize the ELBO in practice, we make a mean field assumption to factorize $q_{\phi_{i}}(z^{i,j}_{0:T},s^{i}_{0:T}) = {q_{\phi_{i_1}}}(z^{i,j}_{0:T}), {q_{\phi_{i_2}}}(s^{i}_{0:T})$, an additional independence assumption that allows the parameters $\phi_{i} = \{\phi_{i_1}, \phi_{i_2}\}$ of each distribution to be computed and updated separately. We implement coordinate ascent variational inference (CAVI) \cite{Blei_2017} to train all model parameters, which includes three separate steps:  
\begin{gather}
\text{Variational expectation system (VES)}\\
      q_{\phi_{i_2}}^*(s_{0:T}^{i})  \propto \exp \mathbb{E}_{q_{\phi_{i_1}}}[\log p(x_{0:T}^{i, 1:J}, z^{i, 1:J}_{0:T}, s_{0:T}^{i}|\theta)] \nonumber
\end{gather}
\begin{gather}
\text{Variational expectation entity (VEZ)}\\
      q_{\phi_{i_1}}^*(z^{i, 1:J}_{0:T})  \propto \exp \mathbb{E}_{q_{\phi_{i_2}}}[\log p(x_{0:T}^{i, 1:J}, z^{i, 1:J}_{0:T}, s_{0:T}^{i}|\theta)] \nonumber 
\end{gather}
\begin{gather}
\text{Maximization step (M-step)}\\
      \theta^* = \text{argmax}_{\theta}  \sum^N_{i}{\mathbb{E}_{q_{\phi_i}}} \log p(x_{0:T}^{i, 1:J}, z^{i, 1:J}_{0:T}, s_{0:T}^{i}|\theta)\nonumber
\end{gather}
In the VES step, the optimal parameters of each approximate system posterior distribution $q_{\phi_{i_2}}^*$ are computed via a forwards-backwards dynamic programming algorithm \cite{rabinerTutorialHiddenMarkov1989}, given a fixed $\theta$ and the current approximate entity posterior distribution $q_{\phi_{i_1}}^*$. In the VEZ step, the same procedure is conducted, but the system posterior parameters are fixed along with $\theta$. Both VES and VEZ steps combined form the E-step. In the M-step, given the current optimal system and entity approximate posterior distribution parameters, $\phi_{i_1}, \phi_{i_2}$, the other model parameters (i.e.~those of the true distributions) $\theta$ are maximized via gradient-ascent or a closed-form solution when available. These steps offer an efficient algorithm to update all of the parameters in the model to maximize the ELBO in Eq.~\ref{ELBO_equation}. For the interested reader, see Appendix Sec. \ref{cavi_update_derivations} for a detailed derivation of the CAVI updates for our problem and see Appendix Sec.~\ref{cavi_update_algorithms} for details regarding the algorithmic implementations for these updates, including the forwards-backwards inference procedure. 

\subsection{ELBO derivation}\label{elbo_derivation}
Starting from our joint distribution over our observed data $\{x_{0:T}^{i, 1:J}\}_{i=1}^N$ and our latent variables $\{z^{i, 1:J}_{0:T}\}_{i=1}^N$, $\{s_{0:T}\}_{i=1}^N$, given our parameters $\theta$, we derive the Evidence Lower Bound (ELBO) for unsupervised training of the model. First, we factorize the joint into the following form: 
\begin{gather*}
\prod_{i=1}^N 
p(x_{0:T}^{i,1:J}, z^{i,1:J}_{0:T}, s_{0:T}^{i} \mid \theta) \\[6pt]
= \prod_{i=1}^N 
p(x_{0:T}^{i,1:J} \mid z^{i,1:J}_{0:T}, s_{0:T}^{i}, \theta) \,
p(z^{i,1:J}_{0:T}, s_{0:T}^{i}, \theta) \\[6pt]
= \prod_{i=1}^N 
p(z^{i,1:J}_{0:T}, s_{0:T}^{i} \mid x_{0:T}^{i,1:J}, \theta) \,
p(x_{0:T}^{i,1:J} \mid \theta)
\end{gather*}
From here, we factorize and re-position some terms to obtain the latent variable posterior: 
\begin{gather}
   \prod^N_{i=1} p(z^{i,1:J}_{0:T}, s_{0:T}^{i}| x_{0:T}^{i, 1:J}, \theta)\\  = \frac{\prod^N_{i=1} p(x_{0:T}^{i, 1:J}|z^{i, 1:J}_{0:T}, s_{0:T}, \theta) p(s_{0:T}^{i},z^{i, 1:J}_{0:T}|\theta)}{\prod^N_{i=1} p(x^{i, 1:J}_{0:T}|\theta)} \nonumber
\end{gather}
Rather than learning the latent variables and model parameters $\theta$ that maximize the posterior distribution, we aim to learn the posterior distribution that maximizes the evidence term ${\prod^N_{i=1} p(x^{i, 1:J}_{0:T}|\theta)}$. However to maximize this evidence term, we would have to compute a highly intractable integral -- marginalizing over all hidden variables for each entity and across time points. Thus, instead, we maximize the \emph{lower bound} of the $log$ evidence as a proxy with an approximate distribution for the posterior. For example, the evidence can be written as: 
\begin{gather}
    {\prod^N_{i=1} p(x^{i, 1:J}_{0:T}| \theta)} = \frac{\prod^N_{i=1} p(x_{0:T}^{i, 1:J}|z^{i, 1:J}_{0:T}, s_{0:T}^{i}, \theta)p(s_{0:T}^{i},z^{i, 1:J}_{0:T}|\theta)}{\prod^N_{i=1} p(z^{i, 1:J}_{0:T}, s_{0:T}^{i}| x_{0:T}^{i,1:J}, \theta)} 
\end{gather}
The denominator and numerator are combined so that we have: 
\begin{gather}
    {\prod^N_{i=1} p(x^{i,1:J}_{0:T}| \theta)} \\ = \prod^N_{i=1} \sum_{z,s} p(x_{0:T}^{i, 1:J}|z^{i, 1:J}_{0:T}, s_{0:T}^{i},\theta) p(z^{i, 1:J}_{0:T}, s_{0:T}^{i}|\theta) \nonumber
\end{gather}
By taking the $log$ of both sides, we obtain: 
\begin{gather}
    {\sum^N_{i=1} \log p(x^{i, 1:J}_{0:T}| \theta)}\\ = \sum^N_{i=1} \log \sum_{z,s} p(x_{0:T}^{i, 1:J}|z^{i, 1:J}_{0:T}, s_{0:T}^{i},\theta) p(z^{i, 1:J}_{0:T}, s_{0:T}^{i}|\theta)  \nonumber
\end{gather} 
We assume some structured auxiliary probability distribution over our latent variables $q_{\phi_i}(z^{i, 1:J}_{0:T}, s_{0:T}^{i})$ that we are going to \emph{learn} the $\phi$ parameters of alongside our parameters $\theta$ for all $N$ sequences. 

We add a $\frac{q_{\phi_i}(z^{i,1:J}_{0:T}, s_{0:T}^{i})}{q_{\phi_i}(z^{i, 1:J}_{0:T}, s_{0:T}^{i})}$ to the right side of the equation: 
\begin{gather}
     {\sum^N_{i=1} \log p(x^{i, 1:J}_{0:T}| \theta)} = \sum^N_{i=1} \log \sum_{z,s} p(x_{0:T}^{i, 1:J}|z^{i, 1:J}_{0:T}, s_{0:T}^{i},\theta)\\  p(z^{i, 1:J}_{0:T}, s_{0:T}^{i}|\theta)\frac{q_{\phi_i}(z^{i,1:J}_{0:T}, s_{0:T}^{i})} {q_{\phi_i}(z^{i,1:J}_{0:T}, s_{0:T}^{i})} \nonumber
\end{gather}
We combine the terms on top and transform the sum over hidden variables into the expectation value with respect to latent variables sampled from our distribution $q_{\phi}(z^{i,1:J}_{0:T}, s_{0:T}^{i})$: 
\begin{gather}
     {\sum^N_{i=1} \log p(x^{i,1:J}_{0:T}| \theta)} = \sum^N_{i=1} \log {\mathbb{E}_{q_{\phi}}} [\frac{p(x_{0:T}^{i,1:J}, z^{i,1:J}_{0:T}, s_{0:T}^{i}|\theta)} {{q_{\phi_i}(z^{i,1:J}_{0:T}, s_{0:T}^{i})}} ]
\end{gather}
By using Jenson's Inequality $\log \mathbb{E}[f(x)] >= \mathbb{E}[\log f(x)]$, we obtain: 
\begin{gather}
     {\sum^N_{i=1} \log p(x^{i,1:J}_{0:T}| \theta)} = \sum^N_{i=1} \log {\mathbb{E}_{q_{\phi_i}}} [\frac{p(x_{0:T}^{i,1:J}, z^{i,1:J}_{0:T}, s_{0:T}^{i}|\theta)} {{q_{\phi}(z^{i,1:J}_{0:T}, s_{0:T}^{i})}} ] \\ >= \sum^N_{i=1} {\mathbb{E}_{q_{\phi_i}}} \log  [\frac{p(x_{0:T}^{i,1:J}, z^{1:J}_{0:T}, s_{0:T}^{i}|\theta)} {{q_{\phi_i}(z^{i,1:J}_{0:T}, s_{0:T}^{i})}} ] \nonumber
\end{gather}
We separate the $\log$ in the right-most expression into a two term subtraction to end up with a joint term and an entropy term.
\begin{gather}
     {\sum^N_{i=1} \log p(x^{i,1:J}_{0:T}| \theta)} \\ >= \sum^N_{i=1}  {\mathbb{E}_{q_{\phi_i}}} \log p(x_{0:T}^{i,1:J}, z^{i,1:J}_{0:T}, s_{0:T}^{i}|\theta) - {\mathbb{E}_{q_{\phi_i}}} \log {q_{\phi_i}(z^{i,1:J}_{0:T}, s_{0:T}^{i})} \nonumber
\end{gather}
The term to the right is our per datapoint variational ELBO: 
\begin{gather}
      \sum^N_{i=1}ELBO(q_{\phi_{i}}, \theta) \\ = \sum^N_{i=1}{\mathbb{E}_{q_{\phi_i}}} \log p(x_{0:T}^{i, 1:J}, z^{i, 1:J}_{0:T}, s_{0:T}^{i}|\theta) \nonumber - \\ {\mathbb{E}_{q_{\phi_i}}} \log {q_{\phi_i}(z^{i, 1:J}_{0:T}, s_{0:T}^{i})} \nonumber
\end{gather}
This is the training loss for the HSRDM. This loss is maximized via Coordinate Ascent Variational Inference (CAVI) updates \cite{Blei_2017}, which we derive in the following section.  
\subsection{CAVI update rule derivations}\label{cavi_update_derivations}

To utilize the CAVI updates proposed in \cite{Blei_2017}, we make a \emph{mean field approximation}, meaning that $q_{\phi_i}(z^{i, 1:J}_{0:T}, s_{0:T}^{i}) = q_{\phi_{i_1}}(z^{i, 1:J}_{0:T})q_{\phi_{i_2}}(s_{0:T}^{i})$. This allows us to take derivatives of the ELBO with respect to $q_{\phi_{i_1}}(z^{i, 1:J}_{0:T}) \text{ and } q_{\phi_{i_2}}(s_{0:T}^{i})$ separately. First, we disentangle these two posterior terms in the ELBO. The first ELBO term is written as: 
\begin{gather}
\mathbb{E}_{q_{\phi_{i}}} 
  \log p(x_{0:T}^{i,1:J}, z^{i,1:J}_{0:T}, s_{0:T}^{i} \mid \theta) \\ \nonumber
  = \sum_{q_{\phi_{i_1}}} \sum_{q_{\phi_{i_2}}} 
       \log p(x_{0:T}^{i,1:J}, z^{i,1:J}_{0:T}, s_{0:T}^{i} \mid \theta) \, 
       q(z^{i,1:J}_{0:T}) \,
       q_{\phi_{i_2}}(s_{0:T}^{i}) \\ \nonumber
  = \sum_{q_{\phi_{i_1}}} 
       \mathbb{E}_{q_{\phi_{i_2}}} \Big[
         \log p(x_{0:T}^{i,1:J}, z^{i,1:J}_{0:T}, s_{0:T}^{i} \mid \theta)
       \Big] \,
       q_{\phi_{i_1}}(z^{i,1:J}_{0:T}) \nonumber
\end{gather}
We split the second term into two parts -- one part that depends on $q_{\phi_{i_2}}(s_{0:T}^{i})$ and another that depends on $q_{\phi_{i_2}}(z^{i, 1:J}_{0:T})$. 
\begin{gather}
\mathbb{E}_{q_{\phi_{i}}} \log q_{\phi_i}(z^{i,1:J}_{0:T}, s_{0:T}^{i}) \\ \nonumber
= \sum_{q_{\phi_{i_{2}}}} \sum_{q_{\phi_{i_{1}}}}
       \log \!\Big[
         q_{\phi_{i_{1}}}(z^{i,1:J}_{0:T})
         q_{\phi_{i_{2}}}(s_{0:T}^{i})
       \Big] \, 
       q_{\phi_{i_{1}}}(z^{i,1:J}_{0:T})
       q_{\phi_{i_{2}}}(s_{0:T}^{i}) \\ \nonumber
= \sum_{q_{\phi_{i_{2}}}} \sum_{q_{\phi_{i_{1}}}}
       \Big[
         \log q_{\phi_{i_{1}}}(z^{i,1:J}_{0:T})
         + \log q_{\phi_{i_{2}}}(s_{0:T})^{i}
       \Big] \, \\ \nonumber
       q_{\phi_{i_{1}}}(z^{i,1:J}_{0:T})
       q_{\phi_{i_{2}}}(s_{0:T}^{i}) \\ \nonumber
= \sum_{q_{\phi_{i_{2}}}} \sum_{q_{\phi_{i_{1}}}}
       \log \!\Big[
         q_{\phi_{i_{1}}}(z^{i,1:J}_{0:T})
         q_{\phi_{i_{1}}}(z^{i,1:J}_{0:T})
         q_{\phi_{i_{2}}}(s_{0:T}^{i})
       \Big] \\ \nonumber
  \quad + \sum_{q_{\phi_{i_{2}}}} \sum_{q_{\phi_{i_{1}}}}
       \log q_{\phi_{i_{2}}}(s_{0:T}^{i}) \,
       q_{\phi_{i_{1}}}(z^{i,1:J}_{0:T})
       q_{\phi_{i_{2}}}(s_{0:T}^{i}) \\ \nonumber
  = \sum_{q_{\phi_{i_{1}}}}
       q_{\phi_{i_{1}}}(z^{i,1:J}_{0:T})
       \log q_{\phi_{i_{1}}}(z^{i,1:J}_{0:T}) \\
  \quad + \sum_{q_{\phi_{i_{2}}}}
       q_{\phi_{i_{2}}}(s_{0:T}^{i})
       \log q_{\phi_{i_{2}}}(s_{0:T}^{i}) \nonumber
\end{gather}
We obtain the final result by using the rules that he $\sum$ operator is linear and the summation over a probability density is constrained to be one - e.g.~ $\sum_{(s_{0:T}^{i})}q_{\phi_{i_{2}}}(s_{0:T}^{i})= 1$. Now, we take the derivative with respect to an individual $q_{\phi_{i_{2}}}(s_{0:T}^{i})$ or $q_{\phi_{i_{1}}}(z^{i, 1:J}_{0:T})$ and set equal to zero to obtain the optimal value for each. We show the full example for the entity level update $q_{\phi_{i_{1}}}(z^{i, 1:J}_{0:T})$. 

First, we take the derivative with respect to $q_{\phi_{i_{1}}}(z^{i, 1:J}_{0:T})$: 
\begin{gather}
\frac{\mathrm{d}\, ELBO(q_{\phi_i}, \theta)}
     {\mathrm{d}q_{\phi_{i_1}}(z^{i,1:J}_{0:T})}
  = \sum_{i=1}^N \Bigg\{ 
       \frac{\mathrm{d}}{\mathrm{d}q_{\phi_{i_1}}(z^{i,1:J}_{0:T})}
        \\ \nonumber \sum_{q_{\phi_{i_1}}} 
           \mathbb{E}_{q_{\phi_{i_2}}} \Big[
             \log p(x^{i,1:J}_{0:T}, z^{i,1:J}_{0:T}, s^{i}_{0:T} \mid \theta)
           \Big] \,
           q_{\phi_{i_1}}(z^{i,1:J}_{0:T}) \\ \nonumber 
  \quad - 
       \frac{\mathrm{d}}{\mathrm{d}q_{\phi_{i_1}}(z^{i,1:J}_{0:T})}
         \sum_{q_{\phi_{i_1}}}
           q_{\phi_{i_1}}(z^{i,1:J}_{0:T})
           \log q_{\phi_{i_1}}(z^{i,1:J}_{0:T}) \\ \nonumber 
  \quad +
       \frac{\mathrm{d}}{\mathrm{d}q_{\phi_{i_1}}}
         \sum_{q_{\phi_{i_2}}}
           q_{\phi_{i_2}}(s^{i}_{0:T})
           \log q_{\phi_{i_2}}(s^{i}_{0:T})
     \Bigg\} \\ \nonumber 
  = 0
\end{gather}
The last term ends up going to zero. For the other two terms, we need to take the derivative for each component $z=z'$ in the distribution $q_{\phi_{i_{1}}}(z^{i, 1:J}_{0:T} = z')$, $z' \in \mathbb{R}^{T \times J \times K}$. The first term ends up being ${\mathbb{E}_{q_{\phi_{i_{2}}}}}[\log p(x_{0:T}^{i, 1:J}, z^{i, 1:J}_{0:T}, s_{0:T}^{i}|\theta)]$ for each $q_{\phi_{i_{1}}}(z^{i, 1:J}_{0:T} = z')$. 

The second term ends up being $-(\log(q_{\phi_{i_{1}}}(z^{i, 1:J}_{0:T}))+ 1)$ for all $q_{\phi_{i_{1}}}(z^{i, 1:J}_{0:T} = z')$. We add in a third term as a \emph{Lagrange multiplier} that imposes a normalization constraint on the objective: $\sum_{q_{\phi_1}}q_{\phi_{i_{1}}}(z^{i, 1:J}_{0:T}) = 1$. We add zero to the function by adding $\lambda(\sum_{q_{\phi_1}}q_{\phi_{i_{1}}}(z^{i, 1:J}_{0:T}) -1)$, and the derivative of this is simply a constant $\lambda$. As such, we obtain the following proportion: 
\begin{gather}
      \log q_{\phi_{i_1}}^*(z^{i, 1:J}_{0:T})  \propto \mathbb{E}_{q_{\phi_{i_2}}}[\log p(x_{0:T}^{i, 1:J}, z^{i, 1:J}_{0:T}, s_{0:T}^{i}|\theta)] 
\end{gather}
As such, we get the following expression for the optimal $q_{\phi_{i_1}}^*(z^{i, 1:J}_{0:T})$: 
\begin{gather}
      q_{\phi_{i_1}}^*(z^{i, 1:J}_{0:T})  \propto \exp \mathbb{E}_{q_{\phi_{i_2}}}[\log p(x_{0:T}^{i, 1:J}, z^{i, 1:J}_{0:T}, s_{0:T}^{i}|\theta)] 
\end{gather}
By the same set of rules, we obtain the expression for the optimal $q_{\phi_{i_2}}^*(s_{0:T}^{i})$: 
\begin{gather}
      q_{\phi_{i_2}}^*(s_{0:T}^{i})  \propto \exp \mathbb{E}_{q_{\phi_{i_1}}}[\log p(x_{0:T}^{i, 1:J}, z^{i, 1:J}_{0:T}, s_{0:T}^{i}|\theta)] 
\end{gather}
These are the CAVI updates for the VEZ and VES steps in the three-step training algorithm. As for the M-step, we take the derivative of the ELBO with respect to $\theta$. By doing this, we see that the second term of the ELBO, which does not depend on $\theta$ goes to zero, and only the first term is maximized. 
\begin{gather}
      \theta^* = \text{argmax}_{\theta}  \sum^N_{i=1}{\mathbb{E}_{q_{\phi_i}}} \log p(x_{0:T}^{i, 1:J}, z^{i, 1:J}_{0:T}, s_{0:T}^{i}|\theta)  
\end{gather}
Each of these CAVI updates is computed separately. The VEZ and VES updates are computed via the forward-backward inference algorithm outline in the following section. The M-step update is implemented either with gradient-ascent or with a closed-form solution if the probability distributions are assumed gaussian, also outlined in the following section. 
\subsection{CAVI update rule algorithms}\label{cavi_update_algorithms}

\textbf{VEZ and VES forward-backwards algorithm.} 
Recall that we have some optimal $q^*_{\phi_{i}}$ functions that we can compute directly. 
\begin{gather}
      q_{\phi_{i_1}}^*(z^{i, 1:J}_{0:T})  \propto \exp \mathbb{E}_{q_{\phi_{i_2}}}[\log p(x_{0:T}^{i, 1:J}, z^{i, 1:J}_{0:T}, s_{0:T}^{i}|\theta)] 
\end{gather}
\begin{gather}
      q_{\phi_{i_2}}^*(s_{0:T}^{i})  \propto \exp \mathbb{E}_{q_{\phi_{i_1}}}[\log p(x_{0:T}^{i, 1:J}, z^{i, 1:J}_{0:T}, s_{0:T}^{i}|\theta)] 
\end{gather}
To compute each function, we hold the other distribution fixed. We start with holding $q_{\phi_{i_2}}^*(s_{0:T}^{i})$ fixed and take the expectation with respect to that distribution. Our $\theta$ values are also held fixed. Then, we write the expression in the factorized form from above: 
\begin{gather}
\exp \, \mathbb{E}_{q_{\phi_{i_2}}}
  \big[ \log p(x_{0:T}^{i, 1:J}, z^{i, 1:J}_{0:T}, s_{0:T}^{i} \mid \theta) \big]
  \\ \nonumber
  = \exp \, \mathbb{E}_{q_{\phi_{i_2}}} \Bigg\{ \log \Bigg[ 
        p(s^{i}_0 \mid \theta) \,
        \prod_{t=1}^T p(s^{i}_t \mid s^{i}_{t-1}, x^{i,1:J}_{t-1}, \theta) \\ \nonumber
  \qquad \times 
        \prod_{j=1}^J \Bigg(
            p(z^{i,j}_0 \mid s^{i}_0, \theta) \,
            p(x^{i,j}_0 \mid z^{i,j}_0, \theta) \\ \nonumber
  \qquad\quad \times
            \prod_{t=1}^T p(z^{i,j}_t \mid s^{i}_t, z^{i,j}_{t-1}, x^{i,j}_{t-1}, \theta) \,
            p(x^{i,j}_t \mid z^{i,j}_t, x^{i,j}_{t-1}, \theta)
        \Bigg) \Bigg] \Bigg\}
\end{gather}
To compute $q_{\phi_{i_1}}^*(z^{i, 1:J}_{0:T})$, we drop all of the terms that do not depend on $q_{\phi_{i_1}}^*(z^{i, 1:J}_{0:T})$ as these will just be absorbed into the constant (hence the $\propto$ term). Now what we have left is: 
\begin{gather}
\exp \mathbb{E}_{q_{\phi_{i_2}}}\{\log[ \prod ^J_{j=1} p(z^{i, j}_0|s^{i}_0, \theta) p(x^{i, j}_0|z^{i, j}_0, \theta) \\  \prod^T_{t=1} p(z^{i, j}_t|s^{i}_t, z^{i, j}_{t-1}, x^{i, j}_{t-1}, \theta) p(x^{i, j}_t|z^{i, j}_{t}, x^{i, j}_{t-1}, \theta)]\} \nonumber
\end{gather}
We move the $\exp \mathbb{E}_{q_{\phi_{i_1}}}\log$ operations throughout the expression and obtain: 
\begin{gather}
\prod ^J_{j=1} p(x^{i, j}_0|z^{i, j}_0, \theta)  \prod^L_{l = 1}p(z^{i, j}_0|s^{i}_0, \theta)^{(s_0^i = l)} \\  \prod^T_{t=1} p(x^{i, j}_t|z^{i, j}_{t}, x^{i, j}_{t-1}, \theta) \prod^L_{l = 1} p(z^{i, j}_t|s^{i}_t, z^{i, j}_{t-1}, x^{i, j}_{t-1}, \theta)^{(s_t^i = l)} \nonumber
\end{gather}

This expression comes from a few rules: (1) The expectation does not change a function that does not depend on ${q_{\phi_{i_2}}}$, so we just have a $\exp \log$ which does not change anything. (2) We can use the properties $\exp \sum^N = \prod^N$ and $\exp(x\log(y)) = y^x$ to obtain the terms like $\prod^L_{l = 1}p(z^{i, j}_0|s^{i}_0, \theta)^{(s_0^i = l)}$. This functional form takes that of a discrete latent variable \emph{Hidden Markov Model} for $J$ independent entities, where we can use a forwards-backwards dynamic programming method to compute the marginals \emph{and} pairwise marginals for each $q_{\phi_{i_1}}^*(z^{i, j}_{0:T})$, which is all we need for downstream computations due to our Markov assumptions. 

So for each entity $j$, we have the following expression: 
\begin{gather}
p(x^{i, j}_0|z^{i, j}_0, \theta)  \prod^L_{l = 1}p(z^{i, j}_0|s^{i}_0, \theta)^{(s_0^i = l)}  \\ \prod^T_{t=1} p(x^{i, j}_t|z^{i, j}_{t}, x^{i, j}_{t-1}, \theta) \prod^L_{l = 1} p(z^{i, j}_t|s^{i}_t, z^{i, j}_{t-1}, x^{i, j}_{t-1}, \theta)^{(s_t^i = l)} \nonumber
\end{gather}

We break this into \emph{initial}, \emph{transition}, and \emph{emission} probability distributions, where for each term where we compute $\prod^L_{l = 1}$, we can normalize to ensure that our distributions sum to one. For example: 
\begin{gather}
\text{Initial: } \\ \nonumber
\tilde \pi_0 = \frac{\prod^L_{l = 1}p(z^{i, j}_0|s^{i}_0, \theta)^{(s_0^i = l)} }{\sum^K_{k_j = 1} \prod^L_{l = 1}p(z^{i, j}_0 = k_j|s^{i}_0, \theta)^{(s_0^i = l)} }
\end{gather}
\begin{gather}
\text{Transition: }\\ \nonumber \tilde \Psi_t = \frac{\prod^L_{l = 1} p(z^{i, j}_t|s^{i}_t, z^{i, j}_{t-1}, x^{i, j}_{t-1}, \theta)^{(s_t^i = l)}}{\sum^K_{k_j = 1} \prod^L_{l = 1} p(z^{i, j}_t = k_j|s^{i}_t, z^{i, j}_{t-1}, x^{i, j}_{t-1}, \theta)^{(s_t^i = l)} }
\end{gather}
Here, $k_{j}$ indicates the current state $p(z^{i, j}_t = k_j)$ at time $t$, and $k_{i}$ indicates the previous state $p(z^{i, j}_{t-1} = k_i)$ at time $t-1$. $\tilde \pi \in (0,1)^K$ represents the initial probability distribution of the latent variable, whereas $\tilde \Psi_t \in (0,1)^{K \times K}$ represents the transition probability distribution for a current latent state given the previous latent state. The term $\phi_t = p(x^{i, j}_t|z^{i, j}_{t}, x^{i, j}_{t-1}, \theta)$ represents each emission probability density for an observation given the current latent state. \\

Forwards-backwards inference allows us to compute the following forward probability via recursion for each state $k_j$: 
\begin{gather}
\alpha_0(k_j) = \tilde\pi_0(z^{i, j}_0 = k_j)\phi_0(z^{i, j}_0 = k_j)
\end{gather}
\begin{gather}
\alpha_t(k_j) = \phi_t(z^{i, j}_t = k_j)\sum^K_{k_i = 1} \alpha_{t-1}(z^{i, j}_{t-1} = k_i) \\ \tilde\Psi_t(z^{i, j}_{t-1} = k_i, z^{i, j}_{t} = k_j) \nonumber
\end{gather}
Now, $\alpha_t(k_j) = p(x^{i,j}_{0:t}, z^{i,j}_t = k_j|s_t, \theta)$. To obtain the latent marginal and pairwise marginals, we implement the backward step. 
\begin{gather}
\beta_T(k_j) = 1
\end{gather}
\begin{gather}
\beta_{t-1}(k_i) = \sum^K_{k_j=1} \phi_{t}(z^{i, j}_{t} = k_j) \\ \tilde \Psi_t(z^{i, j}_{t-1}= k_i, z^{i, j}_{t} = k_j)\beta_t(z^{i,j}_t = k_j) \nonumber
\end{gather}

Now, $\beta_t(k_j) = p(x^{i,j}_{t+1:T}|z^{i,j}_t = k_j, s_t, \theta)$. To obtain the marginal probability, we compute the following: 

\begin{gather}
p(z^{i,j}_t = k_j \mid x_{0:T}^{i,j}, s_t, \theta) 
  = \frac{\alpha_t(k_j)\beta_t(k_j)}
          {\sum_{k_j=1}^K \alpha_t(k_j)\beta_t(k_j)} \\ \nonumber
  = \frac{p(x^{i,j}_{t+1:T} \mid z^{i,j}_t = k_j, s_t, \theta) \,
           p(x^{i,j}_{0:t}, z^{i,j}_t = k_j \mid s_t, \theta)}
          {\sum_{k_j=1}^K p(x^{i,j}_{t+1:T} \mid z^{i,j}_t = k_j, s_t, \theta) \,
           p(x^{i,j}_{0:t}, z^{i,j}_t = k_j \mid s_t, \theta)} \\ \nonumber
  = \frac{p(x^{i,j}_{0:T}, z^{i,j}_t = k_j \mid s_t, \theta)}
          {p(x^{i,j}_{0:T} \mid s_t, \theta)} 
\end{gather}

Now, for the pairwise marginal probability, we compute: 
\begin{gather}
p(z^{i,j}_{t-1} = k_i,z^{i,j}_{t} = k_j|x_{0:T}^{i,j}, s_t, \theta) \\ \nonumber  =  \frac{\alpha_{t-1}(k_i)\tilde \Psi_t(k_i,k_j) \phi_t(k_j)\beta_t(k_j)}{\sum_{{k_i}, {k_j}}\alpha_{t-1}(k_i)\tilde \Psi_t(k_i,k_j) \phi_t(k_j)\beta_t(k_j)}
\end{gather}
We do these computations for all $K$ states to end up with the optimal probability distributions $q^*(z^{i,j}_t)$ and $q^*(z^{i,j}_{t-1},z^{i,j}_t)$ for each entity, which is all we need for computing the other EM steps. Similar to this step, for computing $q^*(s^{i}_t)$ and $q^*(s^{i}_{t-1},s^{i}_t)$, we only need $q^*(z^{i,j}_t)$. As we show below, for the maximization step, we will only need the pairwise marginals for both. To compute the full forward-backward inference to obtain the optimal distributions for all entities, it takes $\mathcal{O}(K^2TLJ)$ steps ($K$ multiplications for $T$ time steps per $K$ states for each of the $J$ entities). 

\noindent \textbf{M-step optimization.} Given the optimal variational distributions, we  compute the following: 
\begin{gather}
\theta^* = \text{argmax}_{\theta}  \sum^N_{i=1}{\mathbb{E}_{q_{\phi_i}}}[ \log p(x_{0:T}^{i, 1:J}, z^{i, 1:J}_{0:T}, s_{0:T}^{i}|\theta)] 
\end{gather}
We factor the data likelihood and spread out the terms in the $\log$ function: 
\begin{gather}
\theta^* = \text{argmax}_{\theta} 
  \sum_{i=1}^N \mathbb{E}_{q_{\phi_i}} \Bigg\{ 
     \log p(s^{i}_0 \mid \theta_{s-init}) \\ \nonumber
    + \sum_{t=1}^T \log p(s^{i}_t \mid s^{i}_{t-1}, x^{i, 1:J}_{t-1}, \theta_{ss}) 
       \\ \nonumber
    + \sum_{j=1}^J \Big[ 
        \log p(z^{i, j}_0 \mid s^{i}_0, \theta_{z-init})  \\ \nonumber
    \quad + \log p(x^{i, j}_0 \mid z^{i, j}_0, \theta_{x-init})  \\ \nonumber
    \quad + \sum_{t=1}^T \log p(z^{i, j}_t \mid s^{i}_t, z^{i, j}_{t-1}, x^{i, j}_{t-1}, \theta_{es}) 
              \\ \nonumber
    \quad + \log p(x^{i, j}_t \mid z^{i, j}_t, x^{i, j}_{t-1}, \theta_{ee}) 
      \Big] 
  \Bigg\}
\end{gather}
We take the derivative with respect to each distinct $\theta$ independently for each distribution. When we assume gaussian distributions, we obtain closed-form solutions. Otherwise, we implement gradient-ascent.  

\vspace{-2em}
\section{ML training results}\label{ML_results}
\subsection{K-means initializations}\label{k-means}

We include the adapted-HSRDM training results across different random k-means initializations. We used the k-means algorithm as implemented in scikit-learn \cite{scikit-learn} to cluster the EmbeddingGemma vectors into $K{=}4$ groups. Then, using only the data assigned to each cluster, we fit an autoregressive linear model over time to estimate the $A_k, b_k, Q_k$ parameters for that cluster. Immediately following this initialization, we clamp the parameters related to silent entity states $\SSN$ or $\SSY$ to a silent embedding mean and a $0.001$ variance. Thus, the difference between the informed initialization and k-means is in the parameters for states $\TTN, \TTY$. Only the informed settings contain prior knowledge of which states align with utterances containing evidence of mechanistic reasoning. K-means does rely on a random seed, so for that procedure we keep the best run of $5$ seeds. 
We define ``best'' as the highest speaker correlation metric results on the training set. While not relevant for model selection, we report all metric results for the model's performance on training data with the informed initialization under various ablations (e.g.~removing the classifier feedback mechanism). 

We include the results across seeds for investigating all four hypotheses regarding the desired model behavior: H(i) results in Table \ref{kmeans_speaker_corr}, H(ii) results in Table \ref{kmeans_speaker_means}, H(iii) non-speaker to non-speaker results in Table \ref{nonspeaker_k1}, and H(iii) non-speaker to speaker results \ref{nonspeaker_k3}. We selected best model for test based on the seed with the highest speaker correlation between the human annotations (seed 17) and the $\SSY$ state (the bold column in Table). 
\FloatBarrier
\justifying
\begin{table}[H]
\centering
\begin{tabular}{cccccc}

\toprule
Seed & Entity & $\SSN$ & $\boldsymbol{\SSY}$ & $\TTN$ & $\TTY$ \\
\midrule

\addlinespace
\multirow[c]{4}{*}{17} & 0 & -0.157 & \textbf{0.323} & -0.049 & -0.028 \\
 & 1 & -0.130 & \textbf{0.295} & -0.055 & -0.019 \\
 & 2 & -0.138 & \textbf{0.409} & -0.107 & -0.051 \\
 & 3 & -0.211 & \textbf{0.478} & -0.108 & -0.041 \\
\addlinespace
\multirow[c]{4}{*}{41} & 0 & 0.182 & \textbf{-0.093} & -0.028 & -0.049 \\
 & 1 & 0.172 & \textbf{-0.087} & -0.019 & -0.055 \\
 & 2 & 0.165 & \textbf{-0.021} & -0.051 & -0.107 \\
 & 3 & 0.200 & \textbf{-0.055} & -0.041 & -0.108 \\
\addlinespace
\multirow[c]{4}{*}{76} & 0 & 0.010 & \textbf{0.075} & -0.041 & -0.037 \\
 & 1 & 0.028 & \textbf{0.060} & -0.039 & -0.041 \\
 & 2 & 0.007 & \textbf{0.214} & -0.081 & -0.081 \\
 & 3 & -0.030 & \textbf{0.217} & -0.075 & -0.082 \\
\addlinespace
\multirow[c]{4}{*}{126} & 0 & -0.157 & \textbf{0.323} & -0.049 & -0.028 \\
 & 1 & -0.130 & \textbf{0.295} & -0.055 & -0.019 \\
 & 2 & -0.138 & \textbf{0.409} & -0.107 & -0.051 \\
 & 3 & -0.211 & \textbf{0.478} & -0.108 & -0.041 \\
\addlinespace
\multirow[c]{4}{*}{166} & 0 & 0.182 & \textbf{-0.093} & -0.028 & -0.049 \\
 & 1 & 0.172 & \textbf{-0.087} & -0.019 & -0.055 \\
 & 2 & 0.165 & \textbf{-0.021} & -0.051 & -0.107 \\
 & 3 & 0.200 & \textbf{-0.055} & -0.041 & -0.108 \\
\bottomrule
\end{tabular}
\caption{\textbf{H(i): Pearson $r$ correlation experiments on the training data with k-means initializations.}
For each k-means random initialization, we examine turns when an entity speaks, 
and assess correlation between the posterior probability mass of the next state (columns) and the human-annotated evidence strength present in the utterance.
For good models, the desired behavior is a strong positive correlation for state $\SSY$ (bold). We observe the correlation decrease as we ablate inductive bias components in the model architecture. }\label{kmeans_speaker_corr}
\end{table} 
\justifying
\begin{table}[H]
\centering
\begin{tabular}{cccc}

\toprule
 &  & Mean  & Std. Dev  \\
Seed & Evidence & Posterior & Posterior \\
\midrule

\addlinespace
\multirow[c]{2}{*}{17} & Yes & 0.796 & 0.159 \\
 & No & 0.438 & 0.224 \\
\addlinespace
\multirow[c]{2}{*}{41} & Yes & 0.213 & 0.214 \\
 & No & 0.095 & 0.144 \\
\addlinespace
\multirow[c]{2}{*}{76} & Yes & 0.430 & 0.194 \\
 & No & 0.166 & 0.173 \\
\addlinespace
\multirow[c]{2}{*}{126} & Yes & 0.796 & 0.159 \\
 & No & 0.438 & 0.224 \\
\addlinespace
\multirow[c]{2}{*}{166} & Yes & 0.213 & 0.214 \\
 & No & 0.095 & 0.144 \\
\bottomrule
\end{tabular}
\caption{\textbf{H(ii): Mean probability of entity state $\SSY$, for previous speaker turns with and without evidence of MR on the training data with k-means.}
For all k-mean initializations, we report the mean and standard deviation of the previous speaker’s posterior probability mass assigned to state $\SSY$, conditioned on whether their utterance contains evidence of mechanistic reasoning. We expect for this state to have a larger probability when evidence is present than not.}\label{kmeans_speaker_means}
\end{table} 
\justifying
\begin{table}[!htbp]
\centering
\begin{tabular}{cccc}

\toprule
 &  & Mean  & Std. Dev  \\
Seed & Evidence & Posterior & Posterior \\
\midrule

\addlinespace
\multirow[c]{2}{*}{17} & Yes & 0.362 & 0.038 \\
 & No & 0.363 & 0.038 \\
\addlinespace
\multirow[c]{2}{*}{41} & Yes & 0.846 & 0.215 \\
 & No & 0.832 & 0.225 \\
\addlinespace
\multirow[c]{2}{*}{76} & Yes & 0.633 & 0.123 \\
 & No & 0.626 & 0.127 \\
\addlinespace
\multirow[c]{2}{*}{126} & Yes & 0.362 & 0.038 \\
 & No & 0.363 & 0.038 \\
\addlinespace
\multirow[c]{2}{*}{166} & Yes & 0.846 & 0.215 \\
 & No & 0.832 & 0.225 \\
\bottomrule
\end{tabular}
\caption{\justifying\textbf{H(iii): Mean probability of entity state $\SSY$, for silent-to-silent turns with and without evidence of MR, on the training data with k-means.}
For all k-mean initializations, we report the mean and standard deviation of the current non-speaker's posterior probability mass assigned to state $\SSY$, conditioned on whether the speaker's previous utterance contains evidence of mechanistic reasoning. We expect for this state to have \emph{noticeably} higher probability when evidence is present than not.}\label{nonspeaker_k1}
\end{table}
\justifying 
\begin{table}[!htbp]
\centering
\begin{tabular}{cccc}

\toprule
 &  & Mean  & Std. Dev  \\
Seed & Evidence & Posterior & Posterior \\
\midrule

\addlinespace
\multirow[c]{2}{*}{17} & Yes & 0.255 & 0.435 \\
 & No & 0.163 & 0.369 \\
\addlinespace
\multirow[c]{2}{*}{41} & Yes & 0.747 & 0.434 \\
 & No & 0.837 & 0.369 \\
\addlinespace
\multirow[c]{2}{*}{76} & Yes & 0.522 & 0.494 \\
 & No & 0.539 & 0.495 \\
\addlinespace
\multirow[c]{2}{*}{126} & Yes & 0.255 & 0.435 \\
 & No & 0.163 & 0.369 \\
\addlinespace
\multirow[c]{2}{*}{166} & Yes & 0.747 & 0.434 \\
 & No & 0.837 & 0.369 \\
\bottomrule
\end{tabular}
\caption{\justifying \textbf{H(iii): Mean probability of entity state $\TTY$, for silent-to-speaker turns with and without evidence of MR, on the training data with k-means.}
For all k-mean initializations, we report the mean and standard deviation of the current speaker’s posterior probability mass assigned to state $\TTY$, conditioned on whether the speaker's previous utterance contains evidence of mechanistic reasoning. We expect for this state to have \emph{noticeably} higher probability when evidence is present than not.}\label{nonspeaker_k3}
\end{table}
\justifying
\FloatBarrier
\subsection{Model ablation study with training data}
We report adapted-HSRDM training results across multiple ablations using the informed initialization from Sec. \ref{gauss_init}, and compare them to the best-performing k-means model. We include the following metrics: H(i) results in Table \ref{train_speaker_corr}, H(ii) results in Table \ref{train_speaker_mean}, H(iii) silent to silent results in Table \ref{no_speakerk1_train}, and H(iii) silent to speaker results \ref{no_speakerk3_train}.
\FloatBarrier
\justifying
\begin{table}[H]
\centering
\begin{tabular}{cccccc}

\toprule
Init + Feedback & Entity & $\SSN$ & $\boldsymbol{\SSY}$ & $\TTN$ & $\TTY$ \\
\midrule

\addlinespace
\multirow[c]{4}{*}{K-means + human} & 0 & 0.197 & \textbf{-0.126} & -0.042 & -0.036 \\
 & 1 & 0.177 & \textbf{-0.109} & -0.048 & -0.030 \\
 & 2 & 0.194 & \textbf{-0.076} & -0.098 & -0.062 \\
 & 3 & 0.267 & \textbf{-0.142} & -0.098 & -0.056 \\

\addlinespace
\multirow[c]{4}{*}{Informed + Human} & 0 & -0.188 & \textbf{0.477} & -0.049 & -0.027 \\
 & 1 & -0.137 & \textbf{0.414} & -0.056 & -0.018 \\
 & 2 & -0.145 & \textbf{0.618} & -0.115 & -0.038 \\
 & 3 & -0.235 & \textbf{0.654} & -0.101 & -0.053 \\
\addlinespace
\multirow[c]{4}{*}{K-means + Classifier} & 0 & -0.157 & \textbf{0.323} & -0.049 & -0.028 \\
 & 1 & -0.130 & \textbf{0.295} & -0.055 & -0.019 \\
 & 2 & -0.138 & \textbf{0.409} & -0.107 & -0.051 \\
 & 3 & -0.211 & \textbf{0.478} & -0.108 & -0.041 \\
\addlinespace
\multirow[c]{4}{*}{Informed + Classifier} & 0 & -0.180 & \textbf{0.346} & -0.049 & -0.027 \\
 & 1 & -0.139 & \textbf{0.310} & -0.056 & -0.018 \\
 & 2 & -0.151 & \textbf{0.419} & -0.115 & -0.040 \\
 & 3 & -0.232 & \textbf{0.491} & -0.100 & -0.053 \\
\addlinespace
\multirow[c]{4}{*}{Informed + No System} & 0 & -0.125 & \textbf{0.255} & -0.050 & -0.026 \\
 & 1 & -0.108 & \textbf{0.222} & -0.056 & -0.017 \\
 & 2 & -0.090 & \textbf{0.249} & -0.117 & -0.036 \\
 & 3 & -0.165 & \textbf{0.348} & -0.102 & -0.051 \\
\addlinespace
\multirow[c]{4}{*}{Informed + None} & 0 & 0.059 & \textbf{0.057} & -0.049 & -0.027 \\
 & 1 & 0.060 & \textbf{0.060} & -0.056 & -0.018 \\
 & 2 & 0.131 & \textbf{0.124} & -0.115 & -0.039 \\
 & 3 & 0.123 & \textbf{0.117} & -0.100 & -0.053 \\
\bottomrule
\end{tabular}
\caption{\textbf{H(i): Pearson $r$ correlation experiments on the training data.} For all model ablations on training data, we examine turns when an entity speaks, 
and assess correlation between the posterior probability mass of the next state (columns) and the human-annotated evidence strength present in the utterance.
For good models, the desired behavior is a strong positive correlation for state $\SSY$ (bold). We observe the correlation decrease as we ablate inductive bias components in the model architecture.}\label{train_speaker_corr}
\end{table} 
\justifying
\begin{table}[H]
\centering
\begin{tabular}{cccc}

\toprule
 & & Mean  & Std. Dev  \\
Init + Feedback & Evidence & Posterior & Posterior  \\
\midrule

\addlinespace
\multirow[c]{2}{*}{K-means + Human} & Yes & 0.043 & 0.068 \\
 & No & 0.009 & 0.016 \\
\addlinespace
\multirow[c]{2}{*}{Informed + Human} & Yes & 0.745 & 0.224 \\
 & No & 0.092 & 0.103 \\
\addlinespace
\multirow[c]{2}{*}{K-means + Classifier} & Yes & 0.796 & 0.159 \\
 & No & 0.438 & 0.224 \\
\addlinespace
\multirow[c]{2}{*}{Informed + Classifier} & Yes & 0.881 & 0.210 \\
 & No & 0.395 & 0.257 \\
\addlinespace
\multirow[c]{2}{*}{Informed + No System} & Yes & 0.832 & 0.220 \\
 & No & 0.761 & 0.262 \\
\addlinespace
\multirow[c]{2}{*}{Informed + None} & Yes & 0.496 & 0.024 \\
 & No & 0.501 & 0.022 \\
\bottomrule
\end{tabular}\caption{\textbf{H(ii): Mean probability of entity state $\SSY$, for turns with and without evidence of MR, on training data.}
For all model ablations, we report the mean and standard deviation of the previous speaker’s posterior probability mass assigned to state $\SSY$, conditioned on whether their utterance contains evidence of mechanistic reasoning.
We expect for this state to have a larger probability when evidence is present than not.}\label{train_speaker_mean}
\end{table} 
\justifying
\begin{table}[H]
\centering
\begin{tabular}{cccc}

\toprule
 & & Mean  & Std. Dev  \\
Init + Feedback & Evidence & Posterior & Posterior  \\
\midrule

\addlinespace
\multirow[c]{2}{*}{K-means + Human} & Yes & 0.767 & 0.292 \\
 & No & 0.747 & 0.303 \\
\addlinespace
\multirow[c]{2}{*}{Informed + Human} & Yes & 0.357 & 0.074 \\
 & No & 0.340 & 0.071 \\
\addlinespace
\multirow[c]{2}{*}{K-means + Classifier} & Yes & 0.362 & 0.038 \\
 & No & 0.363 & 0.038 \\
\addlinespace
\multirow[c]{2}{*}{Informed + Classifier} & Yes & 0.366 & 0.083 \\
 & No & 0.356 & 0.061 \\
\addlinespace
\multirow[c]{2}{*}{Informed + No System} & Yes & 0.082 & 0.181 \\
 & No & 0.092 & 0.189 \\
\addlinespace
\multirow[c]{2}{*}{Informed + None} & Yes & 0.498 & 0.023 \\
 & No & 0.498 & 0.021 \\
\bottomrule
\end{tabular}\caption{\textbf{H(iii): Mean probability of entity state $\SSY$, for silent-to-silent turns with and without evidence of MR, on the training data.}
For all model ablations, we report the mean and standard deviation of the current non-speaker's posterior probability mass assigned to state $\SSY$, conditioned on whether the speaker's previous utterance contains evidence of mechanistic reasoning. We expect for this state to have \emph{noticeably} higher probability when evidence is present than not.}\label{no_speakerk1_train}
\end{table}
\justifying
\begin{table}[H]
\centering
\begin{tabular}{cccc}

\toprule
 & & Mean  & Std. Dev  \\
Init + Feedback & Evidence & Posterior & Posterior  \\
\midrule

\addlinespace
\multirow[c]{2}{*}{K-means + Human} & Yes & 0.386 & 0.485 \\
 & No & 0.295 & 0.455 \\
\addlinespace
\multirow[c]{2}{*}{Informed + Human} & Yes & 0.340 & 0.473 \\
 & No & 0.154 & 0.359 \\
\addlinespace
\multirow[c]{2}{*}{K-means + Classifier} & Yes & 0.255 & 0.435 \\
 & No & 0.163 & 0.369 \\
\addlinespace
\multirow[c]{2}{*}{Informed + Classifier} & Yes & 0.340 & 0.472 \\
 & No & 0.160 & 0.364 \\
\addlinespace
\multirow[c]{2}{*}{Informed + No System} & Yes & 0.328 & 0.468 \\
 & No & 0.141 & 0.346 \\
\addlinespace
\multirow[c]{2}{*}{Informed + None} & Yes & 0.334 & 0.470 \\
 & No & 0.159 & 0.364 \\
\bottomrule
\end{tabular}
\caption{\textbf{H(iii): Mean probability of entity state $\TTY$, for silent-to-speaker turns with and without evidence of MR, on the training data}
For all model ablations, we report the mean and standard deviation of the current speaker’s posterior probability mass assigned to state $\TTY$, conditioned on whether the previous speaker's utterance contains evidence of mechanistic reasoning. We expect for this state to have \emph{noticeably} higher probability when evidence is present than not.}\label{no_speakerk3_train}
\end{table}
\justifying
\FloatBarrier
\subsection{ELBO results} 
\begin{figure}[H]
  \includegraphics[width=1\linewidth]{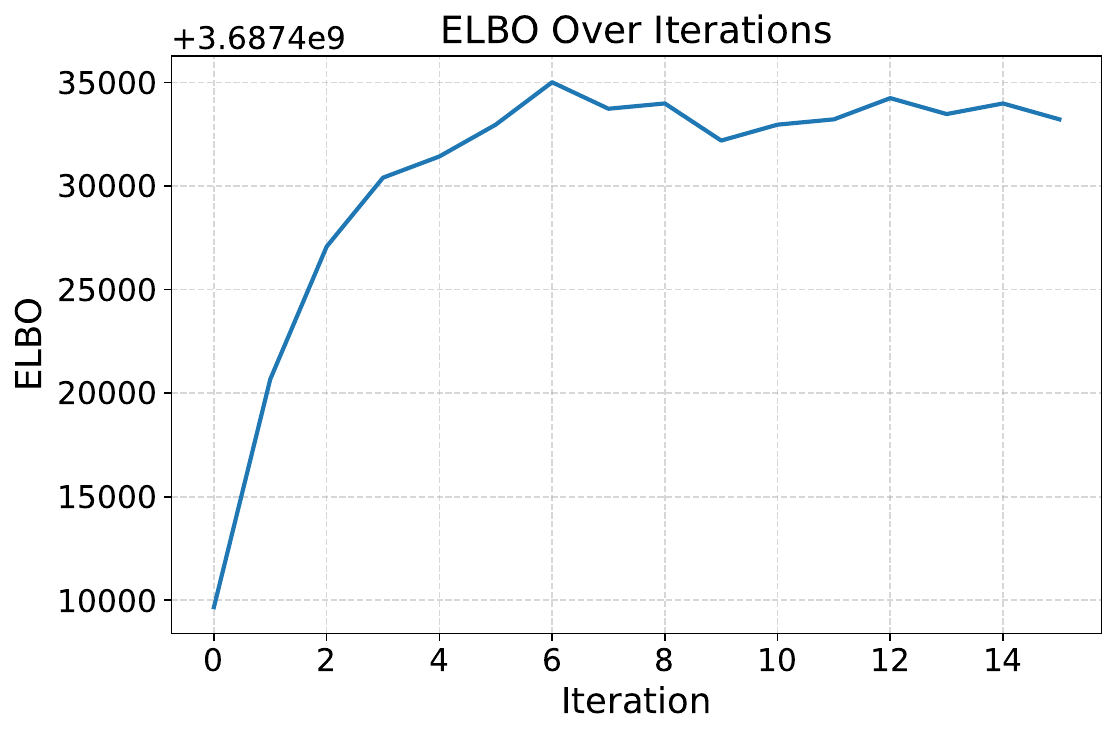}
  \caption{\justifying The ELBO training curve over $15$ CAVI iterations.}
  \label{ELBO_fig}
\end{figure}

We include the ELBO training plot in Fig.~\ref{ELBO_fig} for the best seed 17 based on the speaker correlation metric for the k-means initialization. We observe the ELBO increase over the $15$ CAVI iterations. \\

\end{document}